\documentclass[lettersize,journal]{IEEEtran}
\usepackage[cmex10]{amsmath}
\usepackage{amssymb}
\usepackage{amsthm}
\usepackage{multirow}
\usepackage{inputenc}
\usepackage{enumerate} 
\usepackage{textcomp}
\usepackage{listings}
\usepackage{accents}
\usepackage{array}
\usepackage[switch,mathlines]{lineno}
\usepackage{soul}
\usepackage{xfrac}
\usepackage{graphicx}
\usepackage{xcolor}
\usepackage{colortbl}
\usepackage{cite}
\usepackage{algorithm}
\usepackage{algorithm,algpseudocode}% http://ctan.org/pkg/{algorithms,algorithmx}
\algnewcommand{\Inputs}[1]{%
  \State \textbf{Inputs:}
  \Statex \hspace*{\algorithmicindent}\parbox[t]{.8\linewidth}{\raggedright #1}
}
\algnewcommand{\Initialize}[1]{%
  \State \textbf{Initialize:}
  \Statex \hspace*{\algorithmicindent}\parbox[t]{.8\linewidth}{\raggedright #1}
}

\usepackage{algpseudocode}

\usepackage{cancel}
\usepackage{cite}
\usepackage{bm}
\usepackage{subfig}
\usepackage{float}
\begin{document}

\title{A Robust Data-driven Process Modeling Applied to Time-series Stochastic Power Flow}

\author{Pooja Algikar,~\IEEEmembership{Member,~IEEE,}, Yijun Xu,~\IEEEmembership{Senior Member,~IEEE,} Somayeh Yarahmadi, ~\IEEEmembership{Member,~IEEE,} Lamine Mili, ~\IEEEmembership{Life Fellow,~IEEE.}
        % <-this % stops a space

\thanks{This work is supported, in part, by NSF 1917308 and by the Research Startup Fund of Southeast University in China under Grant 3216002206A1.}
\thanks{P. Algikar,  L. Mili are with the  Electrical Engineering Department, Virginia Tech, Falls Church, VA 22043, USA. (e-mail:\{apooja19, syarahmadi, lmili\}@vt.edu). }
\thanks{Y. Xu is with the  Southeast University, Nanjing, Jiangsu, China. (e-mail:yijunxu@seu.edu.cn). }
}

\markboth{IEEE Transactions on Power Systems}
{Algikar \MakeLowercase{\textit{\textit{et al.}}}: A Robust Data-driven Modeling Applied to Time-Series Stochastic Power Flow}


\maketitle

\begin{abstract}
In this paper, we propose a robust data-driven process model whose hyperparameters are robustly estimated using the Schweppe-type generalized maximum likelihood estimator. The proposed model is trained on recorded time-series data of voltage phasors and power injections to perform a time-series stochastic power flow calculation. Power system data are often corrupted with outliers caused by large errors, fault conditions, power outages, and extreme weather, to name a few. The proposed model downweights vertical outliers and bad leverage points in the measurements of the training dataset. The weights used to bound the influence of the outliers are calculated using projection statistics, which are a robust version of Mahalanobis distances of the time series data points. The proposed method is demonstrated on the IEEE $33$-Bus power distribution system and a real-world unbalanced $240$-bus power distribution system heavily integrated with renewable energy sources. Our simulation results show that the proposed robust model can handle up to $25\%$ of outliers in the training data set.
\end{abstract}


\section{Introduction}
A power system, as it stands currently, involves real-time operational and control actions based on the information provided by the state estimator. The latter processes a set of measurements at periodic time intervals consisting of real and reactive power flows and power injections and voltage magnitudes at selected lines and buses. They are collected from metered devices such as SCADA measurements, phasor measurement units (PMUs), and intelligent electronic devices (IEDs), among others  \cite{liacco1982role}. {They facilitate the time-series power flow analysis to forecast load duration curves and hence, to determine overload conditions in power distribution systems.} It is well known that these measurements are often corrupted with outliers. For instance, during fault conditions, the interference of inrush current in switchgear temporarily causes errors in the measurements. The communication methods used in power distribution systems are often exposed to heavy electromagnetic interference, resulting in corrupted data \cite{ZhangNext-generationCenter}. Furthermore, asynchronous sample time of PMUs \cite{Allen2013AlgorithmSignals}, \cite{LiuData-drivenLearning} and magnetic saturation and hysteresis in potential and current transformers cause measurement errors in current and voltage phasors \cite{McCamishAMeasurementsb}. Under these conditions, the state estimations based on the weighted least squares method suffer from masking and smearing effects, thus yielding inaccurate results. The situation exacerbates under heavy penetration of renewable energy sources (RES) and distributed generations (DGs) due to the stochastic dynamics that they introduce in the power grid. \{For a large-scale power system, performing classical Monte Carlo (MC) simulations of the thousands of realizations for uncertainty quantification requires high computational power.
Therefore, developing robust and computationally efficient models and tools that process real measurements to analyze the stochastic dynamics of a power system is of paramount importance. 

In the literature, several stochastic power flow methods have been proposed to carry out sensitivity analysis and uncertainty quantification \cite{Lin2018ComparisonGeneration}. Among them, the most popular methods are the MC simulations and meta-models. 
As discussed earlier, MC methods turn out to be computationally inefficient when thousands of simulation runs are needed to achieve meaningful statistical results in uncertainty quantification.

{
Meta-models, also known as emulators, surrogates, or response surfaces, only statistically represent the deterministic power flow simulator. Those based on Gaussian processes are non-parametric reduced-ordered models in which the model output realizations are assumed to follow a Gaussian distribution \cite{Xu2020ProbabilisticEmulator}.
% \cite{Rocchetta2020AGrids}, \cite{Rocchetta2018ADeficiency}.
Other types of meta-models extensively  developed in the literature are based on polynomial chaos \cite{ren2015probabilistic, ni2016basis,wu2016probabilistic,wang2019data,xu2019probabilistic,laowanitwattana2018probabilistic,laowanitwattana2021probabilistic,ye2022generalized}.  Ni $et$ $al.$ \cite{ni2016basis} developed a sparse polynomial chaos expansion to tackle a large number of random input variables considering correlation among them.  Wu $et$ $al.$ \cite{wu2016probabilistic} proposed generalized polynomial chaos (gPC) with rectangular formulations to preserve the non-linearity of power flow. Wang $et$ $al.$  \cite{wang2019data} extended the gPC to the data-driven gPC to better deal with the dependent correlated uncertainties among input variables.  Xu $et$ $al.$ \cite{xu2019probabilistic} developed hierarchical polynomial chaos analysis of variance (ANOVA) for efficient extension of the gPC to large-scale systems without falling prey to the curse of dimension. A few methods using neural networks include deep neural network models \cite{yang2019fast,xiang2020probabilistic}, graph convolutional network models \cite{wang2020probabilistic}, and graph neural network \cite{wu2021probabilistic} to overcome the computational challenge.} 

Once developed, the meta-model, which  statistically represents the power flow simulator, is run while considering thousands of new input variables to perform sensitivity analysis and uncertainty quantification. \{However, none of them are robust enough to be trained on real-time data, which makes them unsuitable to perform time-series stochastic power flow analysis.

The accuracy of the results obtained from a meta-model is highly dependent on the quality of the training data. A well-trained meta-model requires data points that fill the input design space. As a result, the data points are sampled from an assumed probability distribution in the design space of stochastic input variables. The assumed probability distributions typically are the Gaussian distribution for the load, the Weibull distribution for the wind speed, and the Beta distribution for the solar irradiance, among others. However, in practice, these distributions may not represent the actual data \cite{Ye2016IdentificationData},\cite{Li2020OutlierApplication}, yielding inaccurate uncertainty quantification results. The conventional meta-modeling methods are not designed to handle {the misrepresentations of power curve distribution}, yielding biased results.

Consequently, modern data-driven models are introduced in the literature, which make use of the raw observational data as a consequence of the proliferation of sensing and metering devices. This approach captures the natural stochasticity of the underlying process. For example, Wang ~\emph{et al.} in \cite{wang2020data} developed a data-driven emulator using polynomial chaos that estimates the statistics of the voltage phasors while Xu~\emph{et al.} in \cite{xu2020data} proposed a fully non-parametric approach to avoid assuming a parametric distribution for the Gaussian process meta-model. However, all these methods are relying on raw data without considering outliers. It is well known that wind generation (WG) time series data are frequently contaminated with communication errors, wind turbine outages, and curtailments  \cite{Ye2016IdentificationData} while PV time series data are contaminated with large signal noise, sensor failures, communication equipment failures, maximum power tracking abnormalities, array shutdowns, and power limitations, to name a few \cite{Li2020OutlierApplication}.

{As a consequence, various data preprocessing techniques have been proposed to account for these abnormal tendencies. For instance, Long $et$ $al.$ \cite{long2019image} developed an algorithm based on the mathematical morphology operation of wind power curve image for detecting and cleaning the wind turbine abnormal data.  
In \cite{zheng2014raw}, a method for filtering out the outliers in raw wind data considering the degree of similarity between the individual objects is developed. These data-cleaning algorithms combined with non-robust data-driven models such as \cite{wang2020data, xu2020data} will be time-consuming in real-time stochastic analysis for increasing power system size with limited computational power and databases for newer wind or solar farms. Their suitability in real-time statistical analysis is arguable. Unlike the proposed method, they are not effectively integrated with the estimation process and are not robust against all types of arising outliers in the time frame of the operation.
} 
 
In this paper, we develop a real-time data-driven time-series stochastic power flow analysis based on a robust process model (RPM). The proposed RPM makes use of the Schweppe-type generalized maximum likelihood estimator (SHGM) that can handle up to $25\%$ of outliers in the training data set. Recall that outliers may be either vertical outliers or bad leverage points.
{
The other estimators proposed in robust statistics for linear regression are M-estimators, which are not robust against bad leverage points\cite{rousseeuw2005robust,hampel1986robust}}.
In power systems, leverage points are power flows on relatively short lines or power injection on buses with relatively many incident lines \cite{Mill1996robustStatistics}.
\ {We assess the robustness of the RPM only theoretically by using three statistical concepts, namely, influence function, finite-sample breakdown point, and asymptotic maximum bias curve.
The robustness of the RPM is experimentally demonstrated on the radial IEEE-33 bus power distribution system and a real-world $240$-bus power distribution system located in the Midwest U.S. for which the training data are manually added with vertical outliers and bad leverage points up to $25\%$.} 

This paper is organized as follows. Section II discusses the conventional GPM followed by theoretical background on statistical robustness concepts that are used to access the robustness of the RPM. Section III presents the development of the proposed methodology. Section IV demonstrates the performance of the RPM on the IEEE 33-bus  distribution system and a real-world 240-bus distribution system. Section V concludes the paper and outlines future work. 
\section{BACKGROUND}
\subsection{Formulation of the Time-series Stochastic Power Flow in the Gaussian Process Framework}
The power flow simulator, represented by $f(\cdot)$, is assumed to be a function of active and reactive power injection measurements at all the $p$ buses, $\mathbf{x}_{t_{i}}\in \mathbb{R}^{2p}$ in training interval $\bm{t}=[1,\hdots,n]$. The output variables are $y_{t_{i}} = f(\mathbf{x}_{t_{i}}) + {\epsilon}_{{t}_{i}}$ are voltage magnitude or phase angle measurement at a bus with an independent and identically distributed additive Gaussian measurement noise, ${\epsilon}_{{t}_{i}}\sim\mathcal{N}(0,\sigma_{0}^{2})$.
In the Gaussian process modeling framework, the uncertainty about the power flow simulator output is characterized as a Gaussian process having a specific mean function $m(\cdot)$, and a covariance (kernel) function $k(\cdot, \cdot)$. As a Gaussian process distribution, the simulator output variables $f(\mathbf{x}_{t_1}),f(\mathbf{x}_{t_2}),\hdots,f(\mathbf{x}_{t_n})$ follow a multivariate normal distribution. 
To incorporate our belief about the simulator, the prior distribution is formulated as a Gaussian process with mean function $m_0(\cdot)$ and covariance function $k_0(\cdot,\cdot)$. Formally, we have 
\begin{equation}\label{0004}
    f(\cdot)|\; \bm{\beta},\bm{l},\tau^{2}\sim \textrm{GP}(m_0(\cdot),k_0(\cdot,\cdot)),
\end{equation}
where the mean function $m_0(\cdot)$ takes the form 
\begin{equation}
    m_0(\mathbf{x})=\bm{h}(\mathbf{x})^{T}\bm{\beta}.
\end{equation}
Here,  $\bm{h}(\mathbf{x}_{t_i}):\mathbb{R}^{2p}\rightarrow\mathbb{R}^{q}$ denotes the basis function that can be chosen to model the assumed degree of non-linearity of the power system, that is,  $\bm{h}(\mathbf{x}_{t_{i}})=[1,\mathbf{x}_{t_{i}}, \mathbf{x}_{t_{i}}^2, \mathbf{x}_{t_{i}}^3,\hdots]^{T}$.
For example, the constant, linear, and quadratic basis functions are respectively given by $\bm{h}(\mathbf{x}_{t_{i}})=[1]$, 
$\bm{h}(\mathbf{x}_{t_{i}})=[1,{x}_{t_{i1}},\hdots,
{x}_{t_{i2p}}]^{T}$, and  $\bm{h}(\mathbf{x})=
[{1},{x}_{t_{i1}},\hdots,{x}_{t_{i2p}},{x}_{t_{i1}}^2,
\hdots,{x}_{t_{i2p}}^2]^{T}\in 
\mathbb{R}^q,\;q=4p+1,\; i=1,2,\hdots,n$.
The kernel function $k_0(\mathbf{x}_{t_i},\mathbf{x}_{t_j})$ denotes the covariance between corresponding output points $({y}_{t_i},{y}_{t_j})$.  
A commonly used covariance function is the radial basis function given by 
\begin{equation}
     k_{0}(\mathbf{x}_{{t}_{i}},\mathbf{x}_{{t}_{j}}|\bm{l})=\tau^{2}\textrm{exp}\left(-\sum_{k=1}^{2p}\frac{(\textrm{x}_{{t}_{ik}}-\textrm{x}_{{t}_{jk}})^{2}}{2{lk}^{2}}\right), 
\end{equation}
where $\bm{l}=(l_{1},\hdots,l_{2p})$ denotes the characteristic length-scale, which models the rapidity of the process; $i,j=1,\hdots,n$. Some other covariance functions are listed in Table \ref{tab11}.
\begin{table*}[!ht]
\small
\centering
\caption{Commonly used Kernel Functions}
  \scalebox{1}{\begin{tabular}{|p{3cm}|p{9cm}|}
     \hline 
      Type   & Expression  \\
      \hline
      Exponential $k_{E}(\mathbf{x}_{t_i},\mathbf{x}_{t_j})$  & $\tau^2\textrm{exp}\left(-\sum_{k=1}^{2p}\frac{|\textrm{x}_{t_{ik}}-\textrm{x}_{t_{jk}}|}{{l}_k}\right)$   \\ \hline
      Matern ${3}/{2}$ $k_{M}(\mathbf{x}_{t_i},\mathbf{x}_{t_j})$ & $\tau^2\left( 1+\sum_{k=1}^{2p}\frac{\sqrt{3}(\textrm{x}_{t_{ik}}-\textrm{x}_{t_{jk}})}{l_{k}}  \right)\textrm{exp}\left(-\sum_{k=1}^{2p}\frac{\sqrt{3}|\textrm{x}_{t_{ik}}-\textrm{x}_{t_{jk}}|}{l_k}\right)$ \\ \hline
      Rational Quadratic $k_{RQ}(\mathbf{x}_{t_i},\mathbf{x}_{t_j})$  & $\tau^2\left(1+ \textrm{exp}\left(-\sum_{k=1}^{2p}\frac{(\textrm{x}_{t_{ik}}-\textrm{x}_{t_{jk}})^2}{2{l}_k^2\alpha}\right) \right)^{-\alpha}$ \\
      \hline
    \end{tabular}}
    \label{tab11}
\end{table*}

Let us gather $n$ input and output measurements into the matrix $\mathbf{X}=[\mathbf{x}_{t_1},\hdots,\mathbf{x}_{t_n}]^{T}$ and the vector $\mathbf{y}=[y_{t_1},\hdots,y_{t_n}]^{T}$, respectively. The matrix of of basis functions is then represented as $\mathbf{H}(\mathbf{X})=[\bm{h}(\mathbf{x}_{t_1}),\hdots,\bm{h}(\mathbf{x}_{t_n})]^{T}$. 
The distribution of the output vector $\mathbf{y}$ of the power system according to \eqref{0004} is a multivariate normal random vector having a covariance function diagonally additive with noise elements $\bm{\epsilon}\sim \mathcal{N}(\bm{0},\sigma^{2}_{n} \mathbf{I}_n)$. Formally, we have 
\begin{equation}\label{0001}
    \mathbf{y}|\mathbf{X},\bm{\beta},\bm{l},\tau^{2},\sigma^{2}_{n}\sim\mathcal{N}\left(\mathbf{H}(\mathbf{X})\bm{\beta},\bm{\Sigma}(\mathbf{X})\right),
\end{equation}
where $\bm{\Sigma}(\mathbf{X})=\bm{k}_{0}(\mathbf{X},\mathbf{X})+\sigma^{2}_{n} \mathbf{I}_n$.
The noise elements $\bm{\epsilon}$ with zero mean and variance $\sigma^{2}_{n}$, also called "nugget", account for model uncertainty and numerical stability.
The training data set is constituted by $(\mathbf{y},\mathbf{X})$.

{For the stochastic power flow analysis, let us consider that we draw $K$ samples of the input test variable $\mathbf{x}_{t^{*}}$ at instances $ {\bm{t^{*}}}=[1,\hdots,n^{*}]$ in the prediction interval. In time-series analysis, the assumption of stationarity and ergodicity for a specific range of time is often made. The latter means that the sample average, commonly known as the ensemble average, is equal to the time average. The assumption of ergodicity allows us to model a stochastic time-series power flow using a single real-time measurement per instance. Let us group the K sampled test predictors for instance ${t^{*}_{i}}$ denoted as $\mathbf{x}^{(k)}_{t^{*}_{i}}$ for $k$ in $[1 ,\; K]$ into $\mathbf{X}_{i}^{*}=[\mathbf{x}^{{(1)}}_{t^{*}_{i}},\hdots,\mathbf{x}^{(K)}_{t^{*}_{i}}]^{T}$.
% Let us now consider the output of the power flow simulator $f(\cdot)$ at the next $n^{*}$ time instances, for which the associated test point in the input feature space is denoted by $\mathbf{x}^*$.
Using a hierarchical formulation, the model output variables $\mathbf{y}^{*}$ obtained through the power flow simulator $f(\cdot)$ at the test points $\mathbf{X}_{i}^{*}$ together with the training output variables follow a joint multivariate Gaussian distribution given by
\begin{equation}\label{eq5}
    \begin{bmatrix}
      \mathbf{y}\\
     \mathbf{y}^{*}|\mathbf{X}_{i}^{*}
    \end{bmatrix}\sim \mathcal{N}\left( \begin{bmatrix}
      \bm{m}_{0}(\mathbf{X})\\
      \bm{m}_{0}(\mathbf{X}_{i}^*)\\
    \end{bmatrix},\begin{bmatrix}
    \bm{\Sigma}({\mathbf{X}})& \mathbf{C}(\mathbf{X}_{i}^{*})\\
        \mathbf{C}^{T}(\mathbf{X}_{i}^{*})&\mathbf{V}(\mathbf{X}_{i}^{*})\\
    \end{bmatrix} \right),
\end{equation}
 where $\mathbf{C}(\mathbf{X}_{i}^{*})=\bm{k}_{0}(\mathbf{X},\mathbf{X}_{i}^{*}),\mathbf{C}^{T}(\mathbf{X}_{i}^{*})=\bm{k}_{0}(\mathbf{X}_{i}^{*},\mathbf{X})$ and $\mathbf{V}(\mathbf{X}_{i}^{*})=\bm{k}_{0}(\mathbf{X}_{i}^{*},\mathbf{X}_{i}^{*})$. The covariance matrix, $\mathbf{\Sigma}(\mathbf{X})$, is represented by $\mathbf{\Sigma}$ hereafter. Furthermore, we assume an a priori Gaussian probability distribution for the simulator output at the test points, ${f}(\mathbf{X}_{i}^{*})|\mathbf{X}_{i}^{*}$, that is, 
 \begin{equation}\label{eq8}
    {f}(\mathbf{X}_{i}^{*})|\mathbf{X}_{i}^{*}\sim\textrm{GP}\left( \bm{m}_{0}(\mathbf{X}^*),\mathbf{V}(\mathbf{X}_{i}^{*})\right).
    \end{equation}
Upon conditioning and using the standard techniques in multivariate distributions, we get 
\begin{equation}\label{eq9}
  {f}(\mathbf{X}_{i}^{*})|\mathbf{X}_{i}^{*},\mathbf{y},\mathbf{X},\bm{\beta},\bm{l},\tau^{2},\sigma^{2}_{n}\sim\textrm{GP}\left(\bm{\mu}^{*}(\mathbf{X}),\bm{\Sigma}^{*}(\mathbf{X})\right),
\end{equation}
where the estimated mean function $ \widehat{\bm{\mu}}^{*}(\mathbf{X}_{i}^{*})$ is given by
\begin{equation}\label{eq10}
 \widehat{\bm{\mu}}^{*}(\mathbf{X}_{i}^{*})
  = \widehat{\bm{m}}_{0}(\mathbf{X}_{i}^{*})+\widehat{\mathbf{C}}^{T}(\mathbf{X}_{i}^{*})\widehat{\mathbf{\Sigma}}^{-1}\mathbf{r},
\end{equation} 
and the estimated covariance function $\widehat{\bm{\Sigma}}^{*}(\mathbf{X}^{*})$ is expressed as
\begin{equation}\label{eq11}
    \widehat{\bm{\Sigma}}^{*}(\mathbf{X}_{i}^{*})=\widehat{\mathbf{V}}(\mathbf{X}_{i}^{*})-\widehat{\mathbf{C}}^{T}(\mathbf{X}_{i}^{*})\widehat{\mathbf{\Sigma}}^{-1}\widehat{\mathbf{C}}(\mathbf{X}_{i}^{*}),
\end{equation}
for all the instances in the predictive interval,  $i=1,\hdots,n^{*}$.}
The estimate of the mean function given by \eqref{eq10} acts as a computationally efficient surrogate model that captures the behavior of the power flow while the covariance matrix estimate given by \eqref{eq11} quantifies the associated uncertainty.

Let us apply a weak prior for $(\bm{\beta},\tau^{2})$, $p(\bm{\beta},\tau^{2})\propto \frac{1}{\tau^{2}}$, combining with \eqref{0001} and using Bayes' theorem yields a posterior distribution for $(\bm{\beta},\tau^{2})$, which is normal inverse-gamma distribution given by 
\begin{equation}
    \bm{\beta}|\mathbf{y},\mathbf{X},\bm{l},\tau^{2},\sigma^{2}_{n}\sim \mathcal{N}(\hat{\bm{\beta}},\tau^{2}(\mathbf{H}^{T}\mathbf{\Sigma}^{-1}\mathbf{H})^{-1}),
\end{equation}
where $\hat{\bm{\beta}}$ is the weighted least squares estimate given by  $\hat{\bm{\beta}}=(\mathbf{H}^{T}\mathbf{\Sigma}^{-1}\mathbf{H})^{-1}\mathbf{H}^{T}\mathbf{\Sigma}^{-1}\mathbf{y}$, and
\begin{equation}
    \tau^{2}|\mathbf{y},\mathbf{X},\bm{l},\sigma^{2}_{n}\sim \textrm{InvGamma}\left(\frac{n-q}{2},\frac{(n-q-2)\hat{\tau}^{2}}{2} \right), 
\end{equation}
where $\hat{\tau}^{2}=\frac{\mathbf{y}^{T}(\mathbf{\Sigma}^{-1}-\mathbf{\Sigma}^{-1}\mathbf{H}(\mathbf{H}^{T}\mathbf{\Sigma}^{-1}\mathbf{H})^{-1}\mathbf{H}^{T}\mathbf{\Sigma}^{-1})\mathbf{y}}{(n-q-2)}$.

\subsection{Smearing and Masking Effects in Conventional GPM Design}
% In the design of conventional GPM, the hyperparameter estimation technique is based on the weighted least squares are prone to smearing and masking effects. 
% Let us consider a linear regression model of GPM given by 
% \begin{equation}\label{1*}
%     \mathbf{y}=\mathbf{H}(\mathbf{X})\bm{\beta}+\bm{e},
% \end{equation}
% where $\bm{e}\sim\mathcal{N}(\bm{0},\mathbf{\Sigma})$. 
% The generalized least-squares estimate of the weight vector $\bm{\beta}$ yields  
% \begin{equation}
%     \widehat{\bm{\beta}}=(\mathbf{H}^T\mathbf{R}^{-1}\mathbf{H})^{-1}\mathbf{H}^T\mathbf{R}^{-1}\mathbf{y}.
% \end{equation}
The residual vector is defined as the difference between the observation vector,  $\mathbf{y}$, and the estimated vector, $\widehat{\mathbf{y}}$. Formally, we have
\begin{equation}\label{3*}
    \bm{r}=\mathbf{y}-\widehat{\mathbf{y}},\\
\end{equation}
where 
\begin{equation}\label{4*}
  \hat{\mathbf{y}}=\mathbf{S}\mathbf{y},  
\end{equation}
and where $\mathbf{S}$ is the hat matrix given by 
$\mathbf{S} = \mathbf{H}(\mathbf{H}^T\mathbf{\Sigma}^{-1}\mathbf{H})^{-1}\mathbf{H}^T\mathbf{\Sigma}^{-1}.$
Substituting \eqref{4*} into \eqref{3*} yields\
\begin{align}\label{2*}
  \bm{r}&=\mathbf{y}-\mathbf{S}\mathbf{y},\\
  &= (\mathbf{I}-\mathbf{S})\mathbf{y},\\
  &=\mathbf{W}\mathbf{y} \label{02*}.
\end{align}
% \begin{equation*}
%     \bm{r}=\mathbf{W}(\mathbf{F}\bm{\beta}+\bm{e})
% \end{equation*}
Substituting the expression $\mathbf{y}=\bm{m}(\mathbf{X})+\bm{e};\; \bm{e}\sim \mathcal{N}(\bm{0},\mathbf{\Sigma})$ into \eqref{02*} yields
\begin{equation}\label{03*}
    \bm{r}=\mathbf{W}\bm{e}.
\end{equation}
Here, $\mathbf{W}$ is called the residual sensitivity matrix as it expresses the sensitivity of the residuals to the errors. For the WLS estimator, the outlier detection and identification statistical tests suffer from the smearing and masking effect of the outliers on the residuals as shown next.
\begin{itemize}
    \item { Smearing effect}\\
    Let us assume that ${e}_{1}\neq0$ and ${e}_{i}={0}$ for $i=2,3,\hdots,n$. From \eqref{2*}, we have
\begin{align}
        {r}_{1}&={W}_{11}{e}_{1}\neq0,\\
        {r}_{2}&={W}_{21}{e}_{1}\neq0,\\
        {r}_{n}&={W}_{n1}{e}_{1}\neq0.
\end{align}
This is known as the smearing effect of one outlier on the residuals, which makes the identification of that outlier using the residual statistical test difficult to achieve.
     \item { Masking effect}\\
     Let us assume that the ${e}_{1}\neq0$ and ${e}_{2}\neq0$ while ${e}_{i}=0$ for $i=3,4,\hdots,n$.
     The associated residuals are expressed as
     \begin{align}
         {r}_{1}&={W}_{11}{e}_{1}+{W}_{12}{e}_{2},\\
         {r}_{2}&={W}_{21}{e}_{1}+{W}_{22}{e}_{2}.
     \end{align}
     Therefore, there exist $e_{1}$ and $e_{2}$ so that the residuals ${r}_{1}\approx0$ and  ${r}_{2}\approx0$. This is known as the masking effect of the outliers on the residuals, which results in the failure of the outlier residual identification test. 
    \end{itemize}
\subsection{Robustness Concepts}
In this subsection, we briefly review the definition of the statistical efficiency of an estimator and the robustness concepts developed in robust statistics, namely, the asymptotic influence function, the breakdown point, and the asymptotic maximum bias curve. 

\subsubsection{Asymptotic Influence Function}
For an $\epsilon$-contaminated model $G(r)=(1-\epsilon)\Phi(r)+\epsilon F(r)$, where $\Phi$ is the cumulative Gaussian distribution function and $F$ is the unknown distribution function of residuals. The influence function is based on the Gateaux derivative that quantifies the local sensitivity of an estimator $T(G)$ to an arbitrary infinitesimal contamination $H=\Delta_{\bm{r}}$. It is expressed as
\begin{equation}\label{eq28}
    \textrm{IF}({r}_i,\bm{h}_{i};\Phi)=\underset{\epsilon \to 0}{\textrm{lim}  }\;\frac{T((1-\epsilon)\Phi+\epsilon \Delta_{r})-T(\Phi)}{\epsilon}.
\end{equation}
The total influence function, IF$(r;\Phi)$, of an M-estimator for a linear regression model is equal to the product of the scalar-valued influence of residuals, IR$({r}_{i};\Phi)$, and the vector-valued influence of position, IP$(\bm{h}_{i};\Phi)$. Formally, we have
\begin{equation}
  \textrm{IF}(r;\Phi)=\textrm{IR}(r_{i};\Phi)\textrm{IP}(\bm{h}_{i};\Phi). 
\end{equation} 
They are given by
IR$(r_{i};\Phi)=\frac{\psi(\frac{r_{i}}{s})}{E\left[\psi^{'}\left(\frac{r_{i}}{s}\right)\right]}$ and IP$(\bm{h}_{i};\Phi)=(\mathbf{H}^{T}\mathbf{H})^{-1}\bm{h}_{i}$.
For an M-estimator, the IR$(\cdot)$ is bounded if the $\psi(\cdot)$ is bounded while the IP$(\cdot)$ is always unbounded, revealing its non-robustness to bad leverage points.
\subsubsection{Finite-Sample Breakdown Point}
The maximum value of $\epsilon$, denoted as $\epsilon^{*}$, for which the maximum bias of an estimator is finite is called the finite-sample breakdown point of that estimator. Formally we have $\epsilon^{*}=max \{\epsilon; b_{max}(\epsilon)<\infty\}$. \cite{donoho1992breakdown} and \cite{huber1992robust} showed that the  maximum finite-sample breakdown point of any regression equivariant estimator under the assumption of general position is given by $\left \lfloor\frac{n-q}{2}\right \rfloor /n$ \cite{maronna1995behavior}. 

\subsubsection{ Asymptotic maximum bias curve}
The asymptotic maximum bias curve is the curve of the upper bound of a bias of an estimator for an increasing level of contamination $ 0\leq\epsilon< \epsilon^{*}$. The asymptotic maximum bias curve of any Fisher consistent estimator, $\hat{\bm{\theta}}$, in its functional form, $\bm{T}$, at any $\epsilon$ contaminated model is defined as  $b_{max}(\epsilon)=\underset{H}{\textrm{sup}}|{\bm{T}}(G)-\bm{\theta}|$. 
In the location case, Huber \cite{huber1992robust} showed that the sample median has the smallest possible asymptotic maximum  bias curve among all location equivariant estimators. In linear regression, the estimator that has the minimum asymptotic bias curve is unknown.

\subsubsection{Statistical Efficiency}
The minimum possible variance that any estimator of location,  $\hat{\bm{\theta}}_{n}$, is able to attend at an assumed probability distribution, $F$, is given by the Cramer-Rao lower bound, which is defined as the inverse of the Fisher information, $I_f$. Formally, we have 
\begin{equation}
    Var(\sqrt{n} \hat{\bm{\theta}}_{n}; F)\geq \frac{1}{I_f}; \; \forall n.
\end{equation}
The ratio of the Cramer-Rao lower bound and the variance of an estimator is called the efficiency of that estimator. An asymptotically efficient estimator is one whose variance attains the Cramer-Rao lower bound for $n$ tending to infinity. 

\section{Robust Data-driven Process Emulator}
% \subsection{Robust Estimation of the Mean Function Hyperparameter} 

In this section, we discuss the development of the proposed RPM model.
We rewrite \eqref{0001} in the form of a regression problem in terms of the mean function hyperparameter given by
\begin{equation}
    \mathbf{y}(\mathbf{X})=\mathbf{H}(\mathbf{X})\bm{\beta}+\bm{e},
\end{equation}
The hyperparameters of the mean and covariance function incorporating the maximum likelihood estimation method are obtained by solving
\begin{equation}\label{eq76}
 {\widehat{\bm{\theta}}}=\underset{({{\bm{\beta}}},\bm{l},\tau^{2},\sigma_n^2)\in\mathbb{R}^{q}\mathbb{R}^{2p}\mathbb{R}^{+*}\mathbb{R}^{+*}}{\mathrm{arg\, max}} \, \textrm{log}\,  \mathcal{L}\left(\mathbf{y}|\mathbf{X},{{\bm{\beta}}},\bm{l},\tau^{2},\sigma_n^2\right),
\end{equation}
where $\bm{\theta}$ represents a vector of hyperparameters $({{\bm{\beta}}},\bm{l},\tau^{2},\sigma_n^2)$. 

\subsection{Schweppe-type Generalized Maximum Likelihood Estimator}
We propose to estimate $\bm{\beta}$ in a robust manner using the SHGM estimator. The SHGM estimator minimizes a weighted loss function of residuals $r_i$ given by
\begin{equation}\label{eq14}
    J(\bm{\beta})=\underset{\hat{\bm{\beta}}}{\mathrm{min}}\sum_{i=1}^{n} w_i^2\rho\bigg(\frac{r_i}{w_i s}\bigg),
\end{equation} 
where $\rho(\cdot)$ is a non-linear loss function of the standardized residuals, $r_{Si}=\frac{r_i}{w_i s}$. The residual scale $s$ is robustly estimated by 
$s^{*}=1.4826$  $b_{m}\;\textrm{median}|\bm{r}|$ when there is a little to none knowledge about the error covariance.
\begin{equation}
    s=1.4826 1+ \frac{5}{n-q}\;\textrm{median}|\bm{r}|.
\end{equation}
We choose the Huber loss function because of its convexity and  its quadratic characteristic at its center. It is defined as
\begin{equation}\label{13}
    \rho(r_{i})=\begin{cases}
    \frac{r_{i}^2}{2}     &\text{for } r_{i}< c,\\
    c|r_{i}|-\frac{c^2}{2}  &\text{for } r_{i}\geq c.\\
\end{cases}
\end{equation}
The threshold parameter $c$ is typically chosen to be equal to $1.5$, which offers a good compromise between a high statistical efficiency at the Gaussian distribution and good robustness against outliers.

\subsection{Weights Based on Projection Statistics}
The weights are calculated using the projection statistics, which are a robust version of the Mahalanobis distances. Formally, they are defined as the maximum of the standardized projection distances obtained by projecting the point cloud in the directions that originate from the coordinate-wise median and that pass through each of the data points \cite{Mill1996robustStatistics}). Let $\bm{h}^{T}(\mathbf{x}_{i})$ be represented by $\bm{h}_{i}^{T}$. Formally, we have 
 \begin{equation}\label{eq36}
     \textrm{PS}_{i}=\underset{||\bm{v}||=1}{\textrm{max}}\; \frac{\bm{h}_{i}^{T}\bm{v}-\underset{j}{\textrm{med}}(\bm{h}_{j}^{T}\bm{v})}{1.4826\;\underset{k}{\textrm{med}}\;|\bm{h}_{k}^{T}\bm{v}-\underset{j}{\textrm{med}}(\bm{h}_{j}^{T}\bm{v})|},
 \end{equation}
where $\bm{v}_j=\frac{\bm{u}_{j}}{||\bm{u}_{j}||}$;  $\bm{u}_j=\bm{h}_{i}-\mathbf{M}$; $j=1,\hdots,n$. Here, $\mathbf{M}$ denotes the coordinatewise median given by 
 \begin{equation*}
     \mathbf{M}=\{\underset{j=1,\hdots,n}{\textrm{med}}\;\bm{h}_{j1},\hdots,\underset{j=1,\hdots,n}{\textrm{med}}\;\bm{h}_{jq}\}.
 \end{equation*}
The weights are calculated as
\begin{equation}\label{15}
\small
   w(\bm{h}_i)=\begin{cases}
    1,&  \textrm{PS}_{i}^2\leq b;\\
    \frac{b}{\textrm{PS}_i^2}, & \textrm{otherwise}.
\end{cases}
\end{equation}
The data point $\bm{h}_{i}$ is considered as a leverage point when the associated PS$_i^2$ is greater than $b$. The weights downweight the bad leverage point and vertical outliers while retaining the good leverage points.
\subsection{Robust Estimation of the Mean Function Hyperparameter}
We estimate the hyperparameter of the mean by setting the gradient of the objective function $J(\bm{\beta})$ with respect to $\bm{\beta}$ to zero, which is given by 
\begin{equation}\label{0005}
    \sum_{i=1}^{n}{w}_{i}\bm{h}_{i}\frac{\partial{\rho}({r}_{Si})}{\partial{r}_{Si}}=0.
\end{equation} 
Let us define the psi-function as $\psi({r}_{Si})=\frac{\partial{\rho}({r}_{Si})}{\partial{r}_{Si}}$. 
\eqref{0005} now becomes
\begin{equation}\label{0006}
     \sum_{i=1}^{n}{w}_{i}\bm{h}_{i}\psi({r}_{Si}).
\end{equation}
Dividing \eqref{0006} by the standardized residuals $r_{S_{i}}$, we get 
\begin{equation}\label{2}
\sum_{i=1}^{m}{q}\bigg(\frac{{r}_i}{{w}_i s}\bigg)\bm{h}_i{r}_i=\bm{0},
\end{equation}
where ${q}({{r}_{Si}})=\frac{\psi({r}_{Si})}{{r}_{Si}}$ is called a weight function. For the case of Huber $\rho$-function, it is defined as
\begin{equation}\label{1}
   {\bm{q}}({r}_{S_i})=\begin{cases}
    1,&  {r}_i\leq c\\
    \frac{b\; \textrm{sign}({r}_{S_i})}{{r}_{Si}}, & \textrm{otherwise}\\
\end{cases}.
\end{equation}
Substituting the expression of the residuals, $r_{i}$, and rewriting \eqref{1} in matrix form yields
\begin{align}
     &\mathbf{H}^T\mathbf{Q}(\mathbf{y}-\mathbf{H}\bm{\beta})=\mathbf{0}\\
      &\bm{\beta}^=(\mathbf{H}^T\mathbf{Q} \mathbf{\Sigma}^{-1}\mathbf{H})^{-1}\mathbf{H}^T\mathbf{Q}^{(k)} \mathbf{\Sigma}^{-1} \mathbf{y},
\end{align}
where $\mathbf{Q}=\textrm{diag}({q}({r}_{S_i}))$. 
Since $\bm{\beta}$ is a function of $\mathbf{Q}$, we solve for $\bm{\beta}$ in an iterative manner by incorporating iterative re-weighted least squares (IRLS) algorithm. Formally, we have
\begin{equation}\label{eq19}
    \bm{\beta}^{(k+1)}=(\mathbf{H}^T\mathbf{Q}^{(k)} \mathbf{\Sigma}^{-1}\mathbf{H})^{-1}\mathbf{H}^T\mathbf{Q}^{(k)} \mathbf{\Sigma}^{-1} \mathbf{y}.
\end{equation}

\subsection{Iterative Procedure for the Hyperparameter Estimation}
In this subsection, the robust estimation of the hyperparameters ($\bm{l},\tau^{2},\sigma^{2}_{n}$) of the RPM associated with the covariance function is discussed.
With the observation set available from the MC simulation of the code, $(\mathbf{y},\mathbf{X})$, we estimate the
hyperparameter $\widehat{\bm{\beta}}$ using the algorithm given by \eqref{eq19}.
The maximum likelihood estimate of the remaining  hyperparameters, namely,  ($\bm{l},\tau^{2},\sigma^{2}_{n}$), is formulated as
\begin{equation}\label{eq77}
  (\widehat{\bm{l}},\widehat{\tau}^{2},\widehat{\sigma}_{n}^2)=\underset{\bm{l},\tau^{2},\sigma^{2}_{n}}{\mathrm{arg\, max}} \, \textrm{log}\,  L\left(\mathbf{Y}|\mathbf{X},\widehat{\bm{\beta}},\bm{l},\tau^{2},\sigma_{n}^2\right).
\end{equation}
% Here, $\widehat{\bm{\beta}}$ represents the converged estimated mean function hyperparameter given by \eqref{eq19}. Simplifying $\textrm{log}\,L$ further, we get
% \begin{multline}\label{eq77}
%     \textrm{log}\,L\left(\mathbf{Y}|\mathbf{X},\widehat{\bm{\beta}},\bm{l},\tau,\sigma_{n}^2\right)\\=\
%     -\frac{1}{2}(\mathbf{Y}-\mathbf{H}\widehat{\bm{\beta}})^T \left[\bm{k}(\mathbf{X},\mathbf{X}|\bm{l},\tau)+\sigma_{n}^2\mathbf{I}_n \right]^{-1}(\mathbf{Y}-\mathbf{H}\widehat{\bm{\beta}})\\-\frac{n}{2}\textrm{log}\, (2\pi)-\frac{1}{2}\textrm{log}\, |[\bm{k}(\mathbf{X},\mathbf{X}|\bm{l},\tau)+\sigma_{n}^2\mathbf{I}_n|.\\
%  \end{multline}
Let us define the resulting $ \textrm{log}\,L$ function by
\begin{align}
    {\Gamma}(\bm{l},\tau^{2},\sigma_{n}^2)= \textrm{log}\;|\bm{k}(\mathbf{X},\mathbf{X}|\bm{l},\tau^{2})+\sigma_{n}^2\mathbf{I}_n|.
\end{align}
Consequently, the maximum likelihood estimate of $(\bm{l},\tau^{2},\sigma_{n}^2)$ reduces to
\begin{equation}
    (\widehat{\bm{l}},\widehat{\tau}^{2},\widehat{\sigma}_{n}^2)= \underset{\bm{l},\tau^{2},\sigma_{n}^2}{\mathrm{arg\, min}} \,\Gamma(\bm{l},\tau^{2},\sigma_{n}^2).
\end{equation}
The hyperparameters,  $(\widehat{\bm{l}},\widehat{\tau}^{2},\widehat{\sigma}_{n}^2)$, are estimated by utilizing a gradient-based optimizer as described in \cite{gpmlbook}. We can then update $\widehat{\bm{\beta}}$ as $\widehat{\widehat{\bm{\beta}}}=\widehat{\bm{\beta}}(\widehat{\bm{l}},\widehat{\tau}^{2},\widehat{\sigma}_{n}^2)$. 
\begin{algorithm}[!htbp]
\caption{Algorithm for Estimating the RDP Hyperparameters}
\begin{algorithmic} [1]
\State Develop the power-system simulator in the case where the measurements are unavailable;  
\State Run the simulator at required input power measurements to obtain the voltage magnitude and the voltage phase angle at each load bus, which  constitutes the training data-set $(\mathbf{X},\mathbf{y})$;
\State Construct $\mathbf{H}$ using a suitable basis function;
\State Calculate the projection statistics of the row vectors of $\mathbf{H}$ given by \eqref{eq36};
\State Calculate the weights $\bm{w}$ based on the PS given by \eqref{15}; 
\State Initialize $\bm{\beta}$ using the weighted least squares solution as $\bm{\beta}_{0}=(\mathbf{H}^{T}\mathbf{\Sigma}^{-1}\mathbf{H})^{-1}\mathbf{H}^{T}\mathbf{\Sigma}^{-1}\mathbf{y}$; 
\State Update $\bm{\beta}$ by executing the IRLS algorithm given by \eqref{eq19} until convergence while setting the hyperparameters $(\bm{l}, \tau^{2}, \sigma_{n}^{2})$ at their initial values to obtain ${\widehat{\bm{\beta}}}$;
\State Update $(\bm{l}, \tau^{2},\sigma_{n}^{2})$ while setting $\bm{\beta}={\widehat{\bm{\beta}}}$;
\State Iterate Steps $7$ and $8$ until convergence, e.g. $||\bm{r}||\leq 0.001$, to obtain the final hyperparameter estimates,  $(\widehat{\widehat{\bm{\beta}}},\widehat{\bm{l}},\widehat{\tau}^{2}\widehat{\sigma}_n^2)$.
\end{algorithmic}\label{tab0}
\end{algorithm}
The algorithm used for estimating the hyperparameters is summarized in Algorithm \ref{tab0}. Once all the hyperparameters of the RPM,  $({\bm{\beta}},\bm{l},{\tau}^{2},\sigma_{n}^{2})$, are estimated, we use \eqref{eq10} as a robust computationally efficient surrogate and \eqref{eq11} to quantify its variance. 

\subsection{Robustness of the RPM}
The influence function of the SHGM estimator is given by 
\begin{equation}\label{eq30}
    \textrm{IF}({r}_{Si},\bm{h}_{i};\Phi)=\frac{\psi(r_{Si})}{E_{\Phi}[\psi^{'}(r_{Si})]}(\mathbf{H}^{T}\mathbf{H})^{-1}\bm{h}_{i}{w}_{i},
\end{equation}
For the SHGM estimator, one can notice that the influence of  position, IP$(\bm{h}_{i},\Phi)=(\mathbf{H}^{T}\mathbf{H})^{-1}\bm{h}_{i}{w}_{i}$, is bounded thanks to the weights calculated using the projection statistics (see Section III), whose breakdown point attains the maximum given by $\frac{[(n-q-1)/2]}{n}$ \cite{maronna1995behavior}. 
Note that the SHGM estimator reduces to an $\ell_2$-norm estimator for small standardized residuals and to the $\ell_1$-norm estimator for larger ones. Therefore, it has a high statistical efficiency at the Gaussian distribution while being robust to outliers.

\section{Case Studies}
{In this section, we compare  the performance of the proposed model RPM to that of the Gaussian process model (GPM) when applied to a standard IEEE 33-bus system (Case A) and to a real-world 240-bus distribution system located in the Midwest U.S. with high penetration of RESs and DGs (Case B). 
We add vertical outliers, i.e., outliers in $\mathbf{y}$, bad leverage points, i.e., outliers in $\mathbf{X}$, and good leverage points, i.e., outliers in both $(\mathbf{X},\mathbf{y})$ up to $25\%$ in the training data. To demonstrate the good performance of the RPM for non-Gaussian distribution noises, we assume that the noise follows the Student's t distribution with $10$  degrees of freedom. This distribution is chosen because it has heavier tails for low degrees of freedom, producing sampling values that may fall far from its median. We compare the performances of the GPM and the RPM using mean absolute error and the root mean square index for each of the cases.} 
\subsection{IEEE 33-Bus System}
\begin{figure*}%
    \centering
    \subfloat[\centering ]{{\includegraphics[height=2.5cm,width=8.5cm]{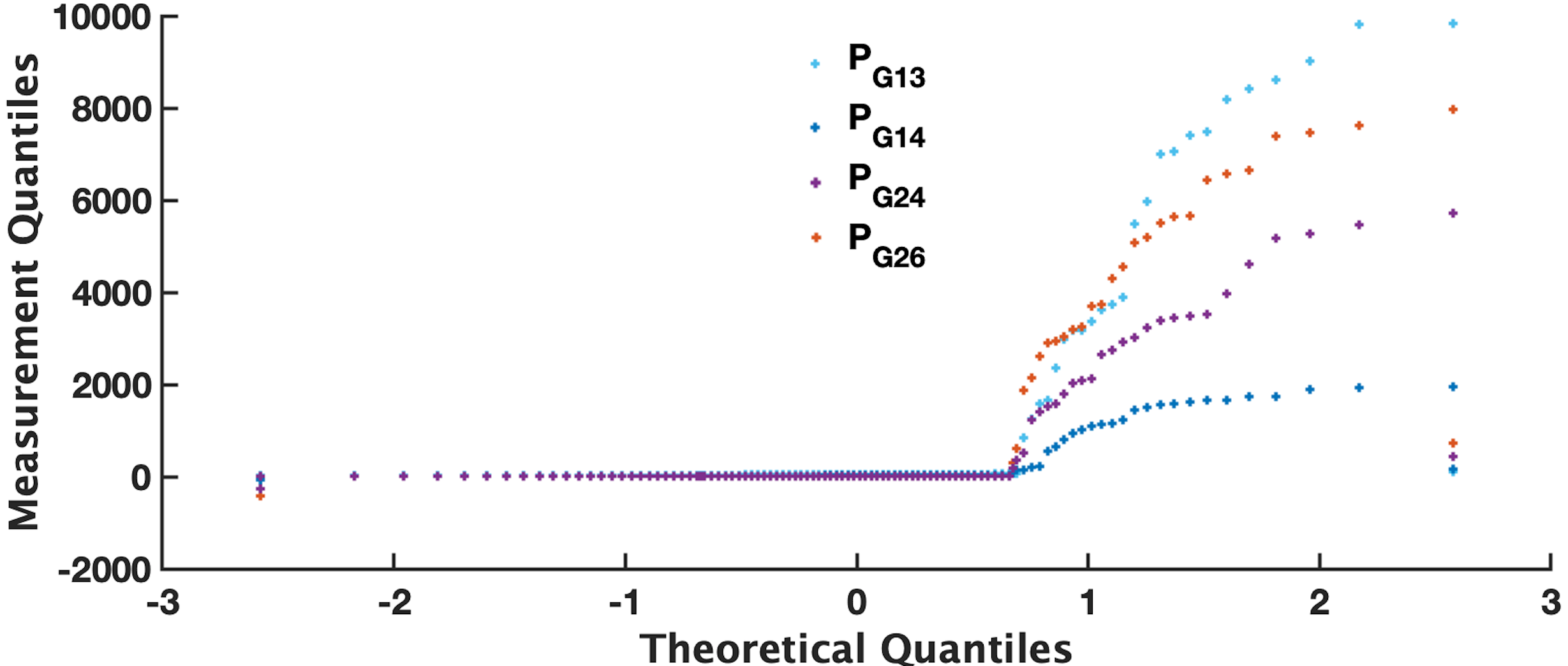} }}%
    \qquad
    \subfloat[\centering  ]{{\includegraphics[height=2.5cm,width=8.5cm]{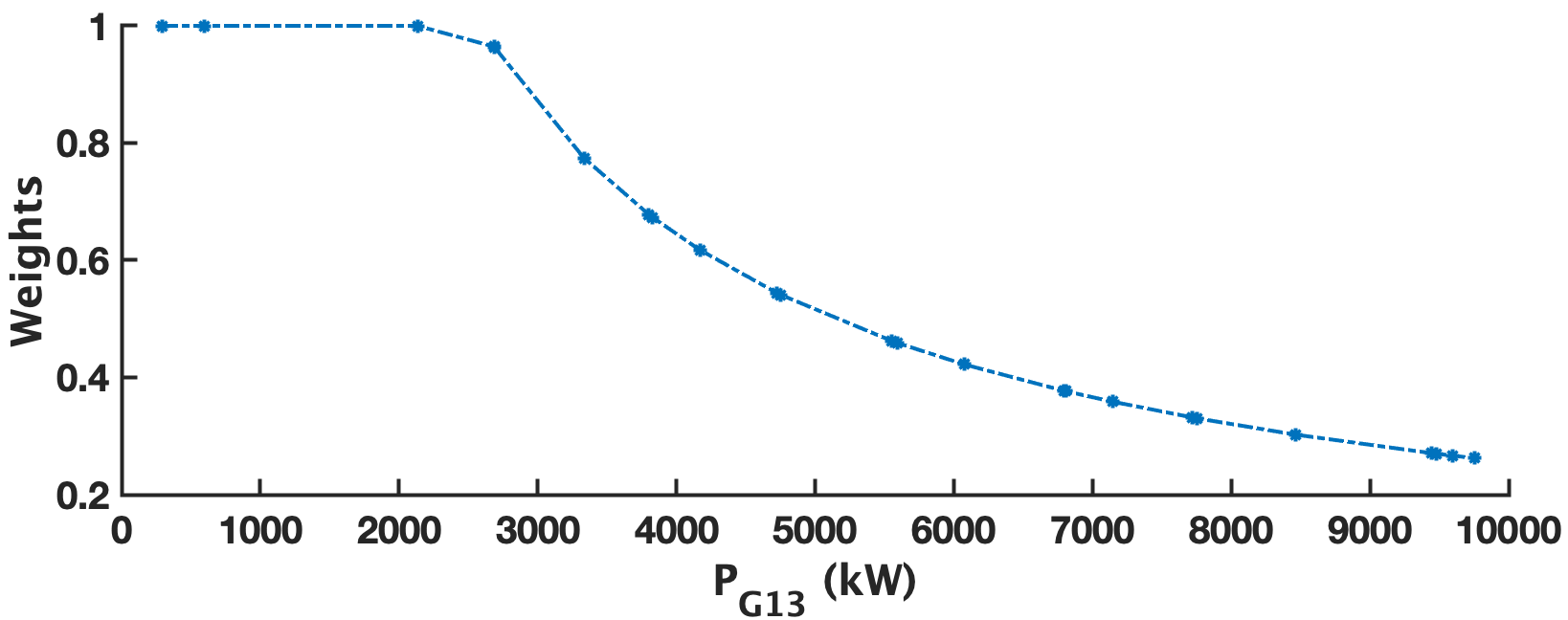}}}%
    \caption{Outliers corrupting the training data set; (a) QQ-plot of the measurements corrupted with $25\%$ of outliers; (b) plot of the weights using the PSs vs. the outlier magnitudes.}%
    \label{DATA}
\end{figure*}

\begin{figure}%
    \centering
    \subfloat[\centering ]{{\includegraphics[height=4cm,width=9.1cm]{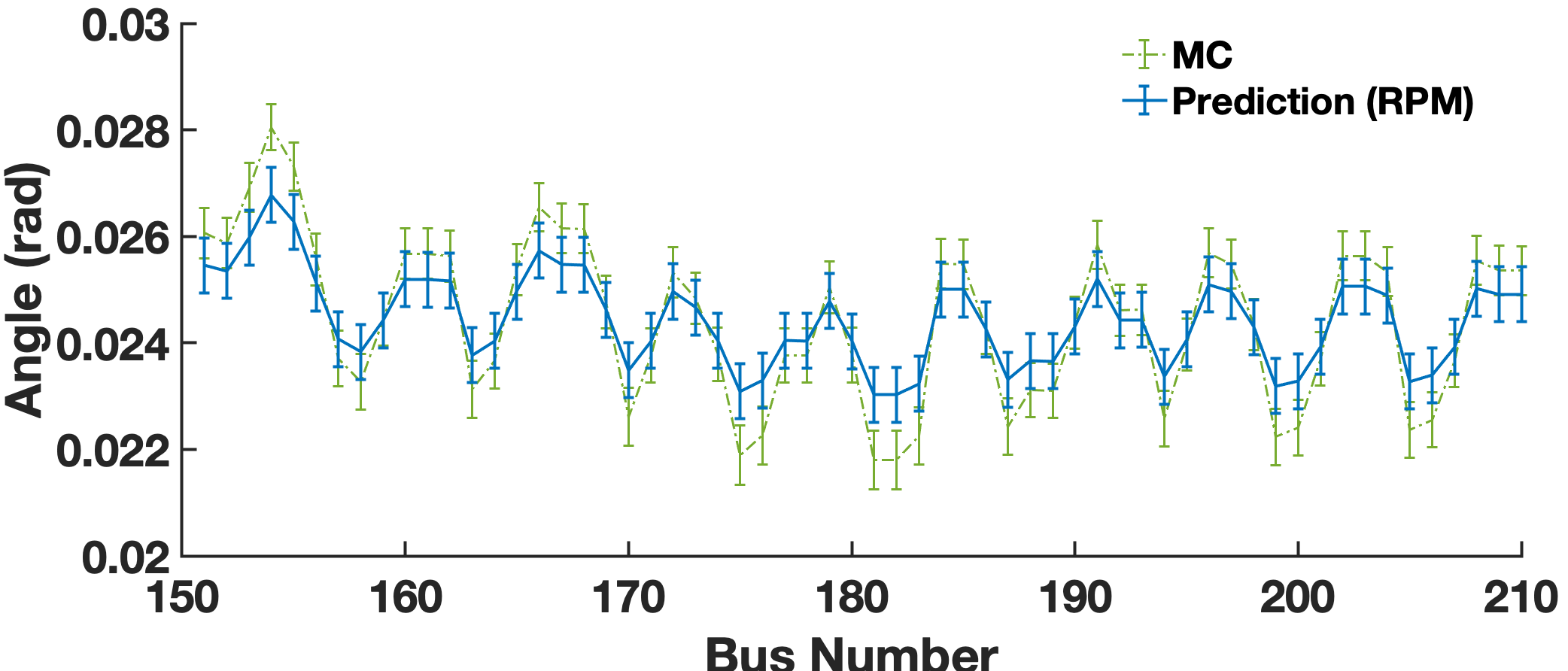}}}%
 \qquad
    \subfloat[\centering  ]{{\includegraphics[height=4.5cm,width=4.5cm]{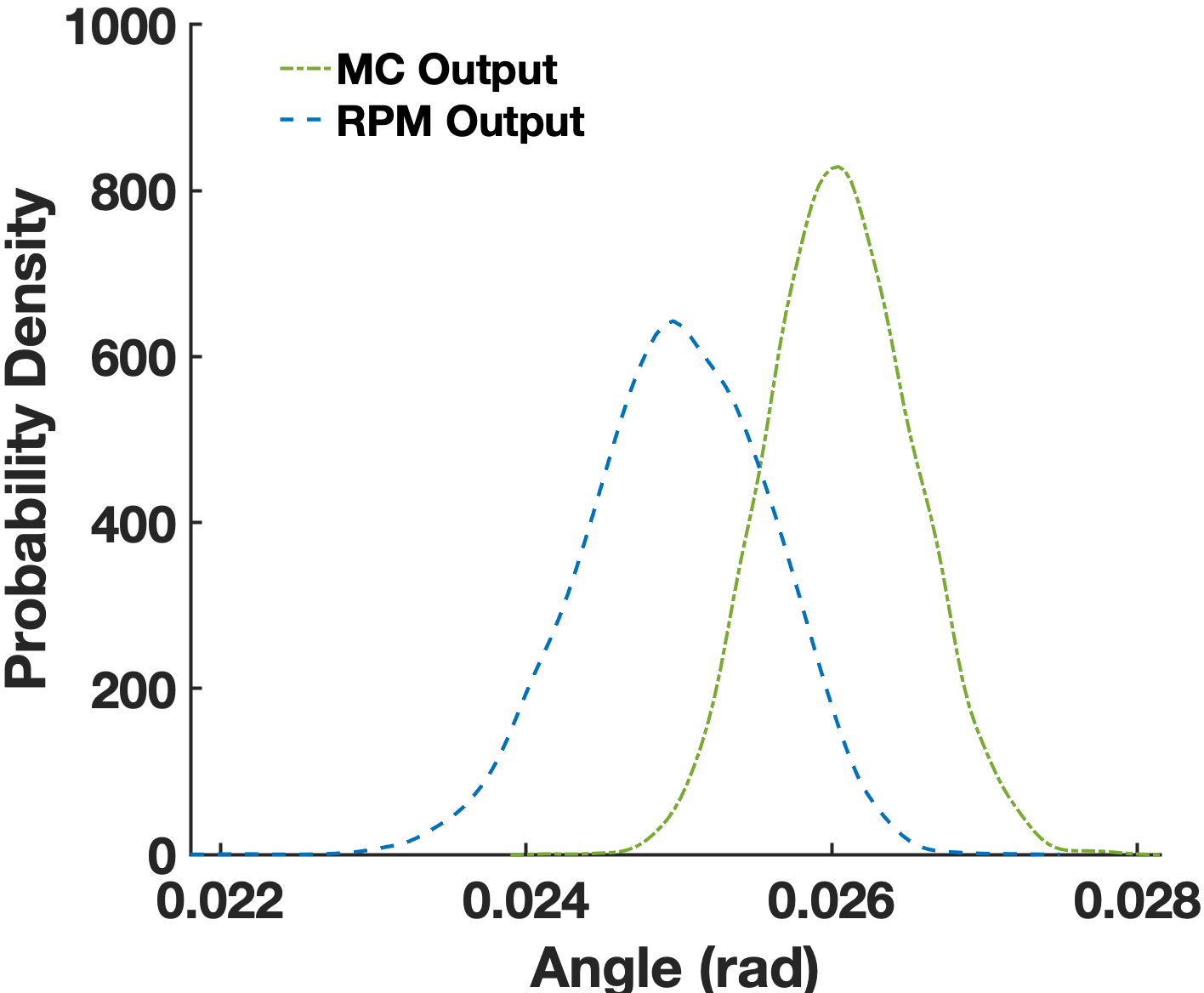}}}%
    \subfloat[\centering ]{{\includegraphics[height=4.5cm,width=4.5cm]{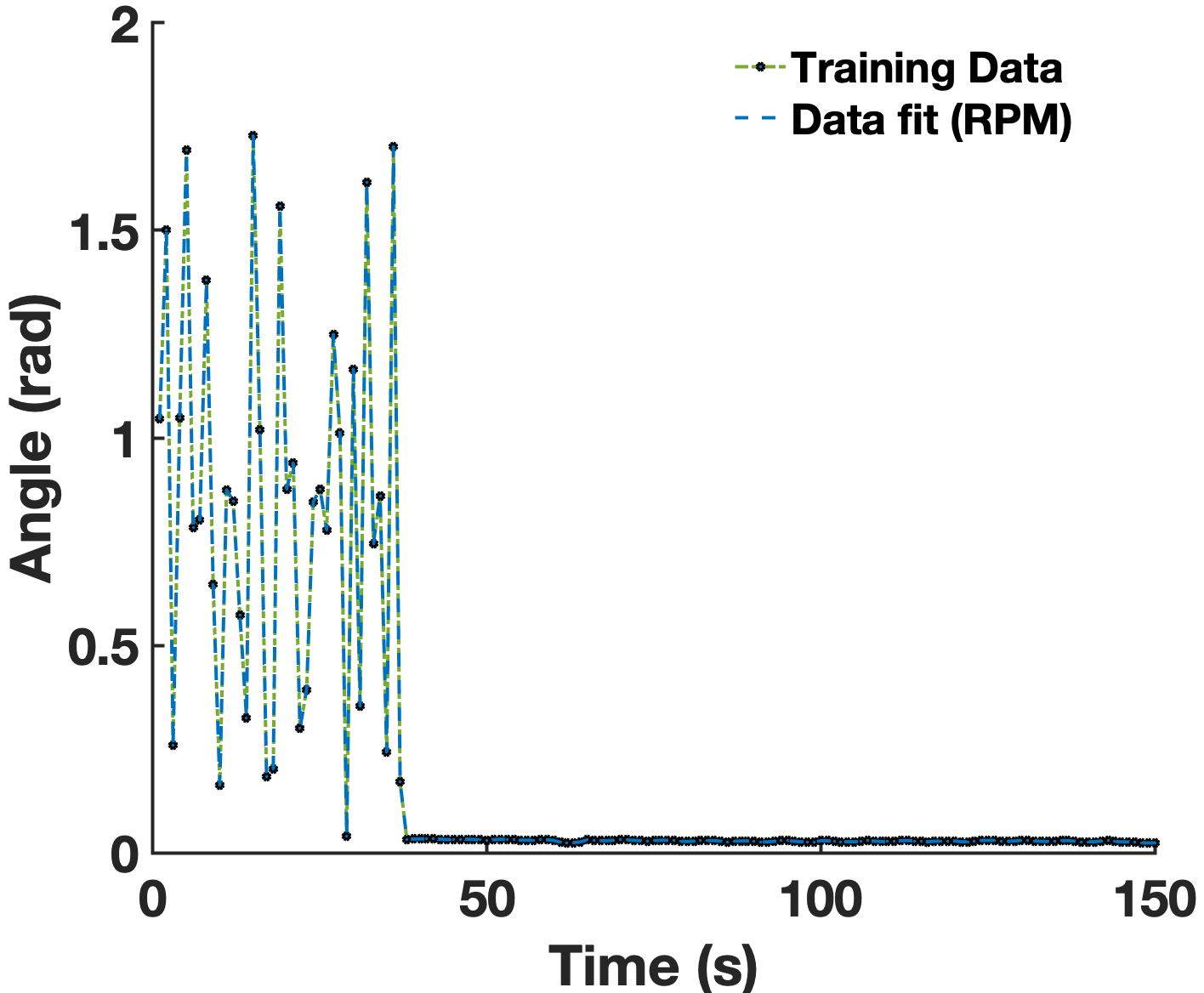} }}%
    \caption{RPM results for the voltage phase angle at Bus 19: (a) prediction at the test points; (b) probability density at the test points; and (c) fitted values over the training data points.}%
    \label{VoltageAngle19}
\end{figure}

\begin{figure}[t]%
    \centering
  \subfloat{\includegraphics[height=4cm,width=9cm]{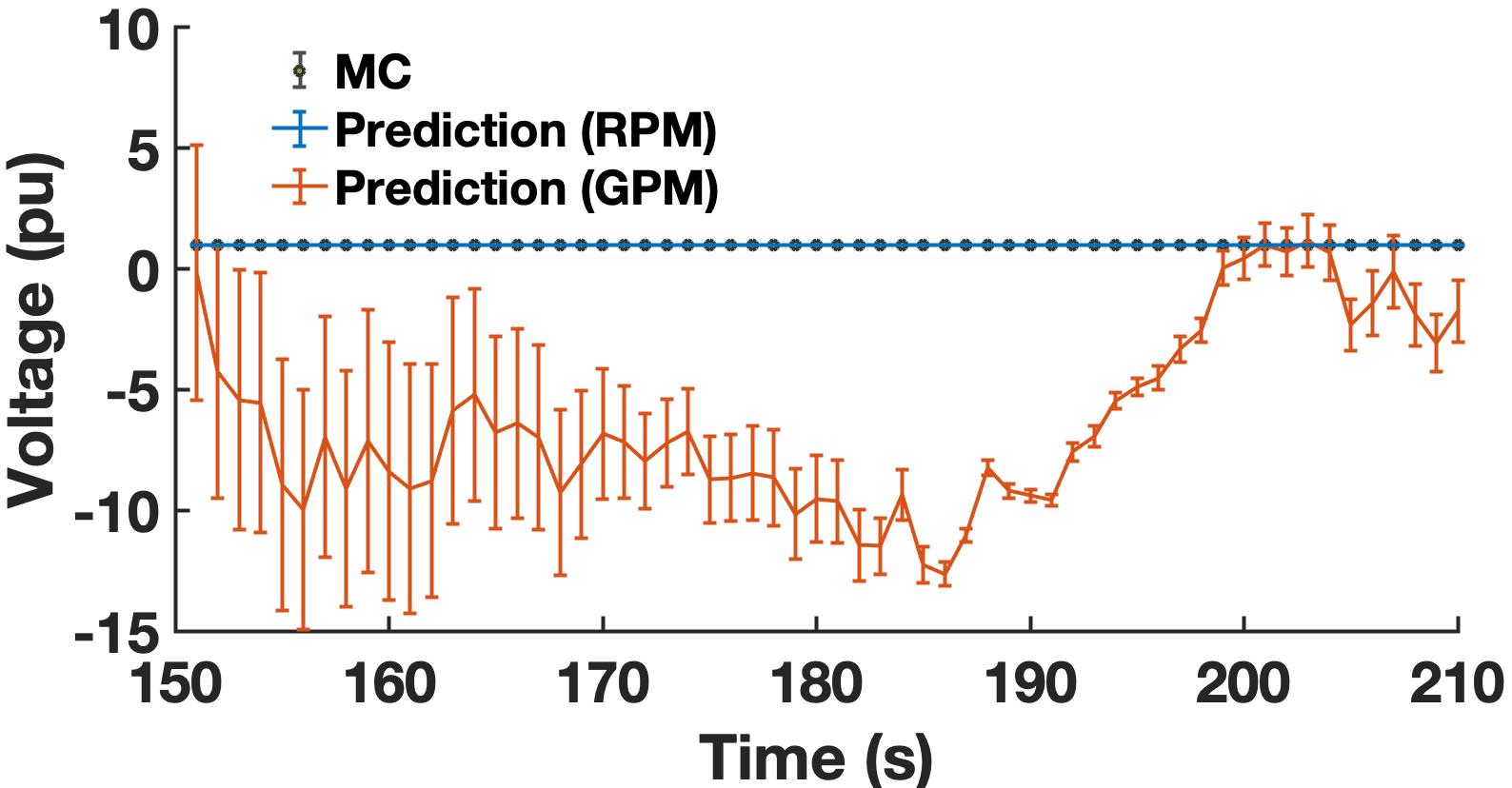}}%
  \qquad 
   \qquad
    \subfloat{\includegraphics[height=3.5cm,width=8.9cm]{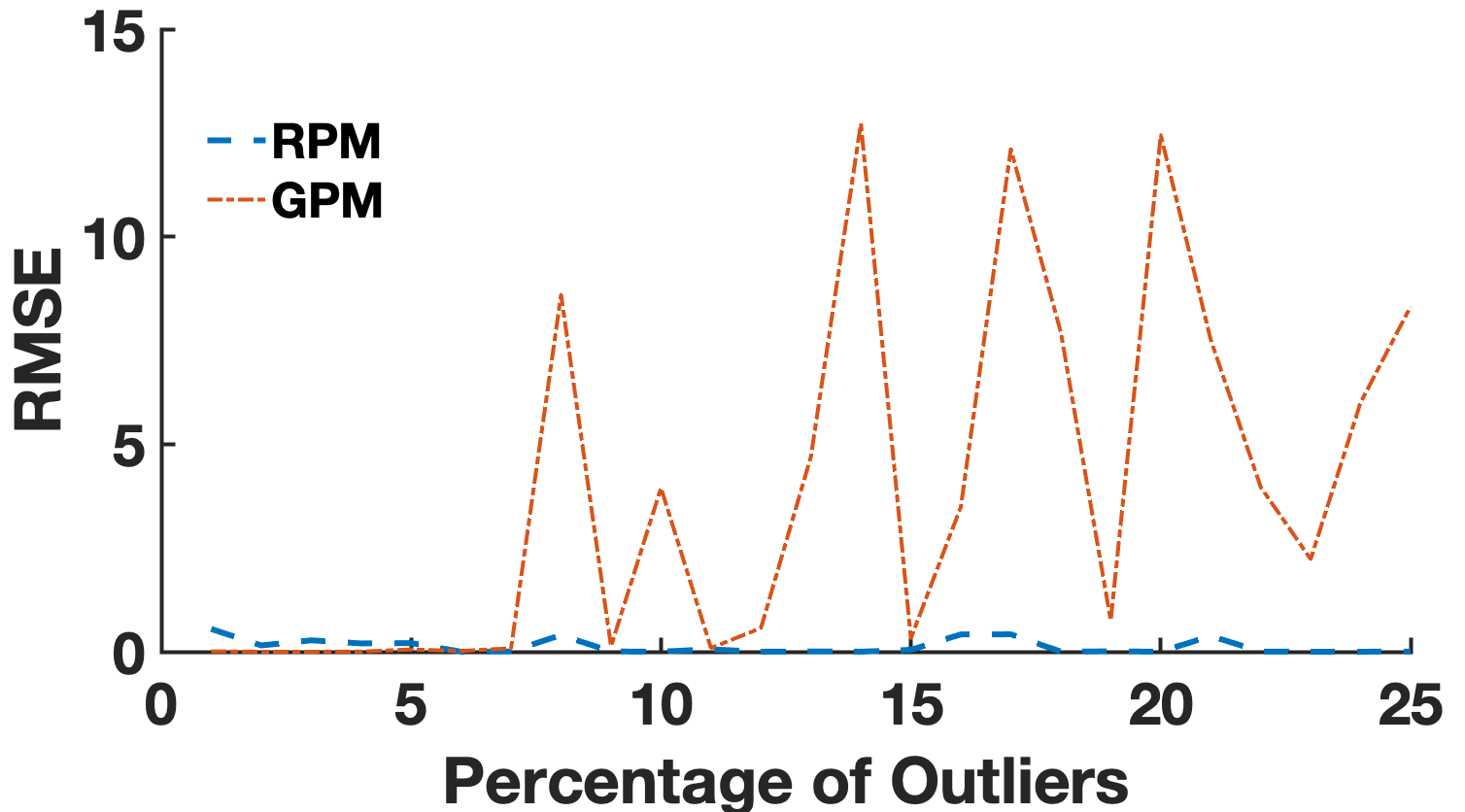}}%
    \caption{Comparison between the performance of the  RPM and the GPM: (a) voltage magnitude at Bus 19; (b) RMSE values.}
    \label{Comparison}%
\end{figure}

\begin{figure}%
    \centering
  \subfloat[\centering ]{{\includegraphics[height=3cm,width=9cm]{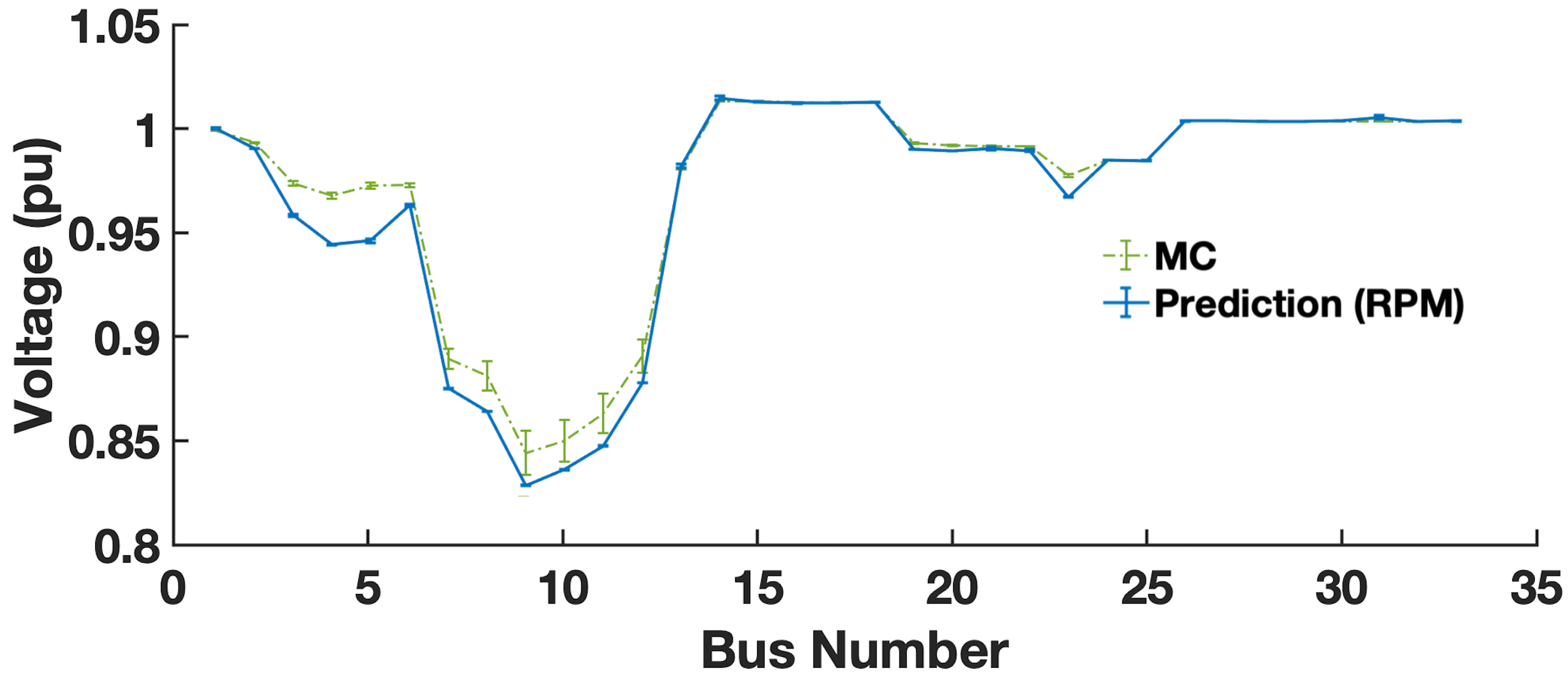}}}%
    \qquad
    \subfloat[\centering ]{{\includegraphics[height=3cm,width=9cm]{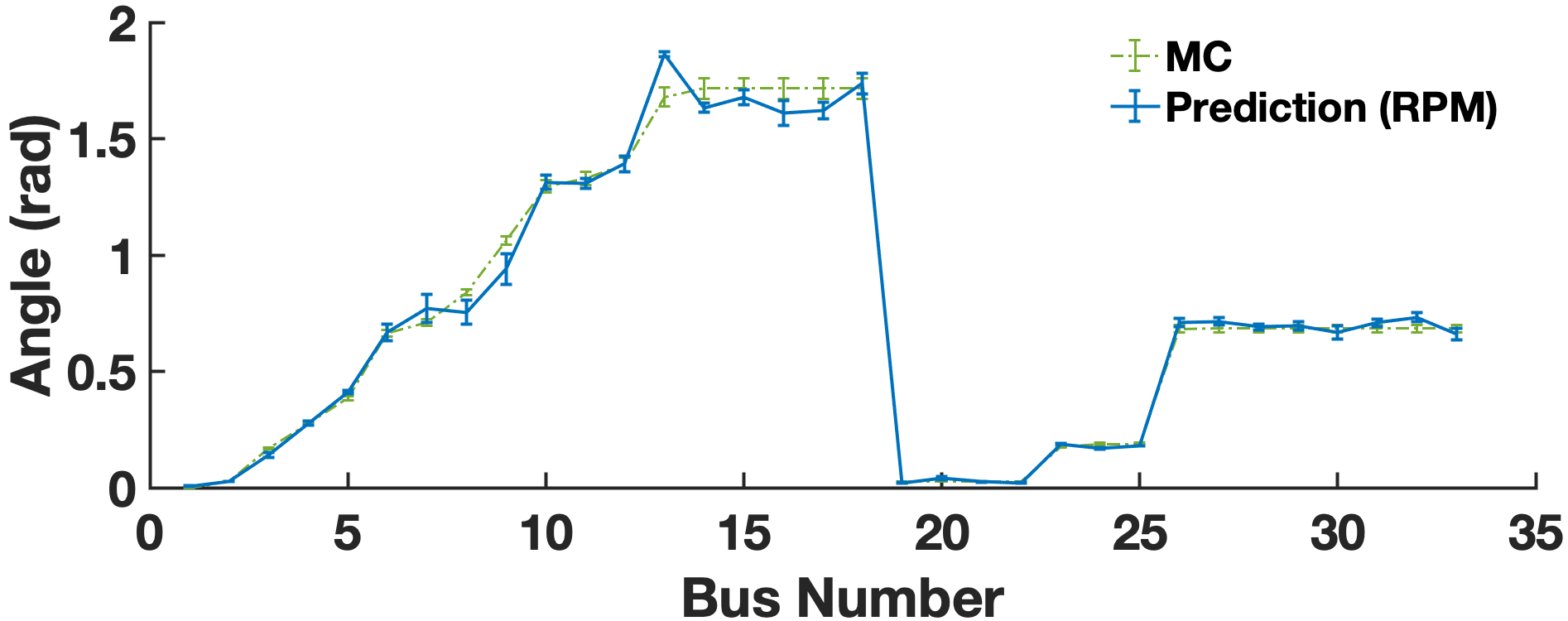}}}%
    \caption{RPM predictions for the IEEE 33-bus system with the training data set added with $25\%$ of outliers: (a) voltage magnitudes; (b) voltage phase angles.}%
    \label{outlierL}
\end{figure}

The RPM is applied to a standard IEEE $33$-bus system, to which are attached four RES, namely, a PV ($P_{G24}$) to Bus 24, and three WGs ($P_{G13}, P_{G14}, P_{G26}$) to Buses 13, 14, and 26 of capacity 1 kW, 50 kW, 10kW, 10kW, respectively. The time-series data considered for the RES power outputs and loads are the real measurements with a resolution of $1s$. We run the power flow simulator at $n=150$ input data points $\mathbf{X}=[\mathbf{x}_{t_1},\hdots,\mathbf{x}_{t_{150}}]^{T}$ to obtain the corresponding voltage magnitude and angle values $\mathbf{y}=[y_{t_1},\hdots,y_{t_{150}}]^{T}$ that constitutes the training data. 
Trained on $(\mathbf{y},\mathbf{X})$, the RPM and the conventional GPM are used to make predictions for the next $n^*=60$ data points constituted as the validation data set at instances $\bm{t}^{*}=[t_{151},\hdots,t_{210}]$. {To perform stochastic analysis, Latin hypercube sampling is employed to generate $7000$ samples of the input variables at each instance in the validation data set following the Weibull distribution for WGs $(P_{t^{*}_{i}, G13}, P_{t^{*}_{i}, G14}, P_{t^{*}_{i}, G24} \sim \textrm{Weibull}(2.06,7.1))$ and the Beta distribution for the PV $(P_{t^{*}_{i}, G26} \sim \textrm{Beta}(2.06,2.5))$; $i=151,\hdots,210$.} The results obtained from the Monte Carlo (MC) simulations performed at these samples stand as reference values for comparing the results obtained from the RPM and the GPM.  
The robustness of the RPM is demonstrated by the addition of $25\%$ outliers as shown in Fig. \ref{DATA} (a) in the training data set.
{To be precise, we impose a worst-case scenario by adding bad leverage points to the input data points $[\mathbf{x}_{t_1},\hdots,\mathbf{x}_{t_{37}}]$,  i.e., to the measurements $(\mathbf{P}_{G13},\mathbf{P}_{G14},\mathbf{P}_{G24},\mathbf{P}_{G26})$ and to the load consumption of the load buses, $\{\mathbf{P}_{L1},\mathbf{P}_{L2},\hdots,\mathbf{P}_{L33}\}$. Similarly,  vertical outliers are added to the output data points $[y_{t_1},\hdots,y_{t_{37}}]$, i.e., to the measurements of voltage phasors.}
We observe from the weights displayed in Fig. \ref{DATA} (b) that the SHGM estimator downweights the bad leverage points and vertical outliers. The prediction results of the voltage magnitude and angle for Bus 19 with the percentage of outliers up to $25$ in the training data constitute a benchmark for this study. {The mean and standard deviation values (indicated as error bars) of the prediction results for the voltage angle of Bus 19 are displayed in Fig. \ref{VoltageAngle19} (a), where the error bars represent standard deviation values. Fig. \ref{VoltageAngle19} (c) depicts the data fit for the training duration $[t_{1}-t_{150}]$ obtained for the RPM. Fig. \ref{VoltageAngle19} (b) compares the probability density of the voltage angle of Bus 19 calculated from the $7000$ realizations at the next instance $t^{*}_{151}$ obtained from the RPM to the MC simulation output.}  Fig. \ref{Comparison} (a) displays  the predicted values of the voltage magnitude at Bus 19 obtained from the RPM and from the GPM. We observe that the predicted values from the GPM deviate largely from the true values. This is due to the fact that the estimate of the mean function hyperparameter of the conventional GPM is centered at the basic weighted least squares estimate. Therefore, it fails to represent the simulator in presence of outliers while the RPM succeeds. Also, the prediction accuracy is displayed in Fig. \ref{Comparison} (b) using root mean square error (RMSE) values when the training data set is added with an increasing percentage of outliers up to $25\%$. We notice that the RPM consistently exhibits low RMSE values. {The RMSE and mean absolute error (MAE) values for the forecast of the voltage phasors at Bus 19 are listed in Table \ref{tab:ieee33} for the cases of training data added with and without the outliers for both the linear and quadratic basis function. We observe that the prediction results are more accurate for the quadratic basis function in the case of added outliers. Therefore, by the principle of parsimony, we choose a quadratic basis to obtain the results for the voltage phasors for all the $33$-buses in the network plotted in Fig. \ref{outlierL}. }
% \begin{figure}%
%     \centering
%     \subfloat[\centering ]{{\includegraphics[height=3.3cm,width=9cm]{Voltage16.png} }}%
%     \qquad
%     \subfloat[\centering  ]{{\includegraphics[height=4.6cm,width=4.5cm]{DensityV.png} }}%
%     \subfloat[\centering ]{{\includegraphics[height=4.6cm,width=4.5cm]{fitV.png} }}%
%     \caption{RPM results for the voltage magnitude at Bus 19: (a) prediction results at the test points; (b) probability density of the voltage magnitude at the test points; (c) fitted values over the training data points.}
%     \label{Voltagebus19}
% \end{figure}%

\begin{table*}[]
\setlength{\tabcolsep}{2pt}
\centering
\caption{{The RMSE and MAE for the Bus 19 of the IEEE $33$-bus system}}
  \begin{tabular}{|c|cc|cc|cc|cc|cc|cc|cc|cc|}
  \hline
  {} & \multicolumn{8}{c|}{Quadratic Basis} & \multicolumn{8}{c|}{Linear Basis}\\
\hline
 {} & \multicolumn{4}{c|}{With $25\%$ outliers} & \multicolumn{4}{c|}{Without outliers} & \multicolumn{4}{c|}{With $25\%$ outliers} & \multicolumn{4}{c|}{Without outliers}\\
 \hline
  {} & \multicolumn{2}{c}{RPM} & \multicolumn{2}{|c|}{GPM}& \multicolumn{2}{c}{RPM} & \multicolumn{2}{|c|}{GPM}& \multicolumn{2}{c}{RPM} & \multicolumn{2}{|c|}{GPM}& \multicolumn{2}{c}{RPM} & \multicolumn{2}{|c|}{GPM} \\
\hline
 Measure & $V$ & $A$ &  $V$ & $A$&$V$ & $A$ &  $V$ & $A$  & $V$ & $A$ &  $V$ & $A$ & $V$ & $A$ &  $V$ & $A$\\
\hline
RMSE & $0.0034$& $9.7810e^{-4}$&$1.0274$&$0.1527$&$0.0931$&$0.0058$&$0.0657$&$0.1468$&$0.0264$&$0.01264$&$0.7995$&$0.0206$&$0.01005$&$0.01747$&$0.00838$&$0.01645$\\
MAE &$0.003$&$8.2815e^{-4}$&$7.3087$&$0.1287$&$0.0672$&$2.5516e^{-4}$&$0.0334$&$0.1669$&$0.0047$&$9.7962e^{-4}$&$4.2355$&$0.0029$&$0.00058$&$0.00176$&$0.00042$&$0.0018$\\
\hline
\end{tabular}
\label{tab:ieee33}
\end{table*}

\subsection{Real-World 240-Bus System}
\begin{figure*}
\centering
 {\includegraphics[height=12cm,width=19.5cm]{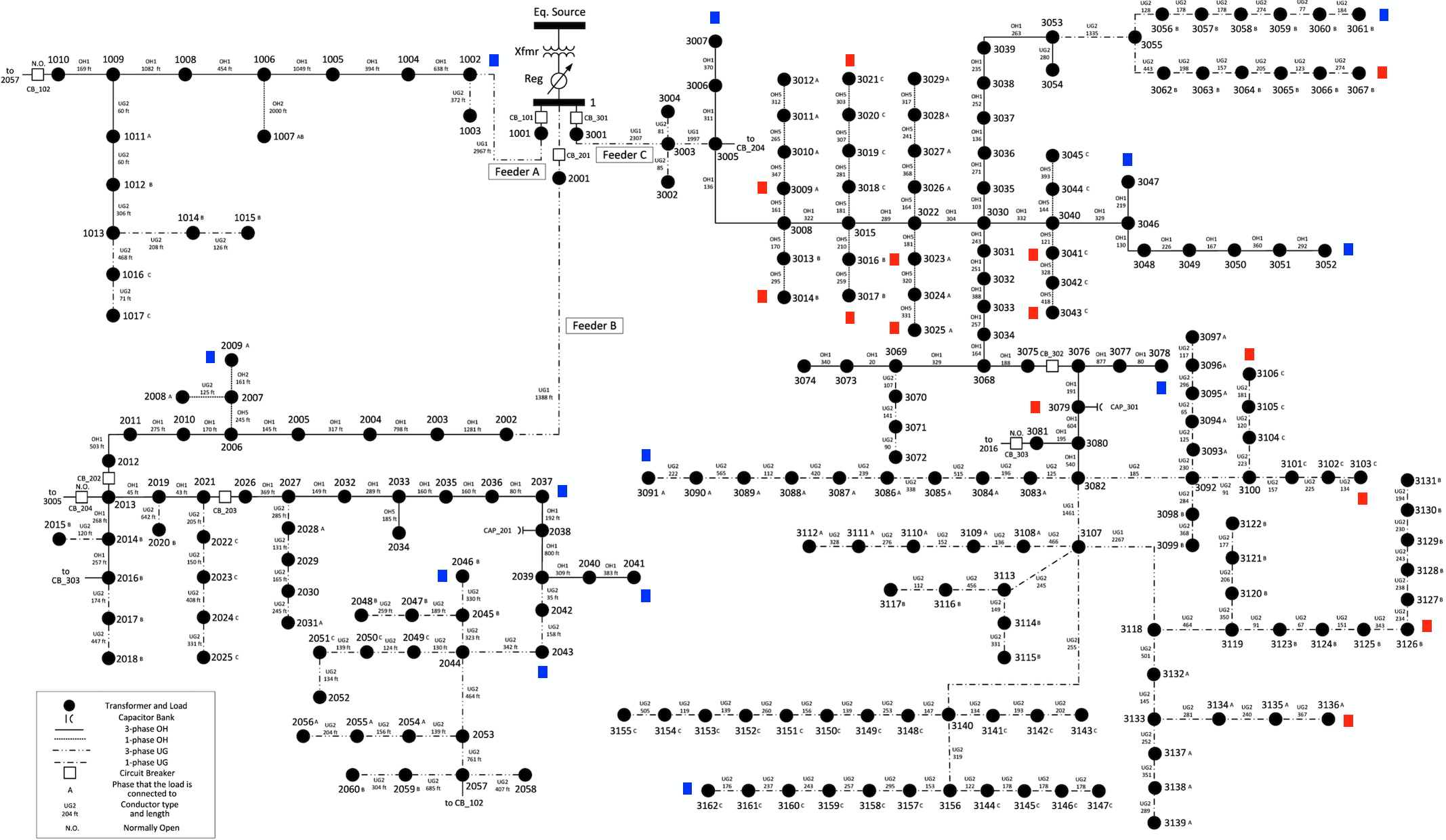}}
    \caption{The online diagram of the 240 bus system integrated with RES. Blue and red squares indicate the PVs and WGs, respectively}
    \label{network}
\end{figure*}

We integrate the $240$-bus radial distribution system \cite{bu2019time} with RES, namely $35$ PVs and $35$ WGs distributed across the network. Their locations in the network are displayed in Fig. \ref{network}. Please note that the RES are connected to each phase of the indicated buses in the network.
\begin{figure}%
    \centering
  \subfloat[\centering ]{{\includegraphics[height=4cm,width=4.5cm]{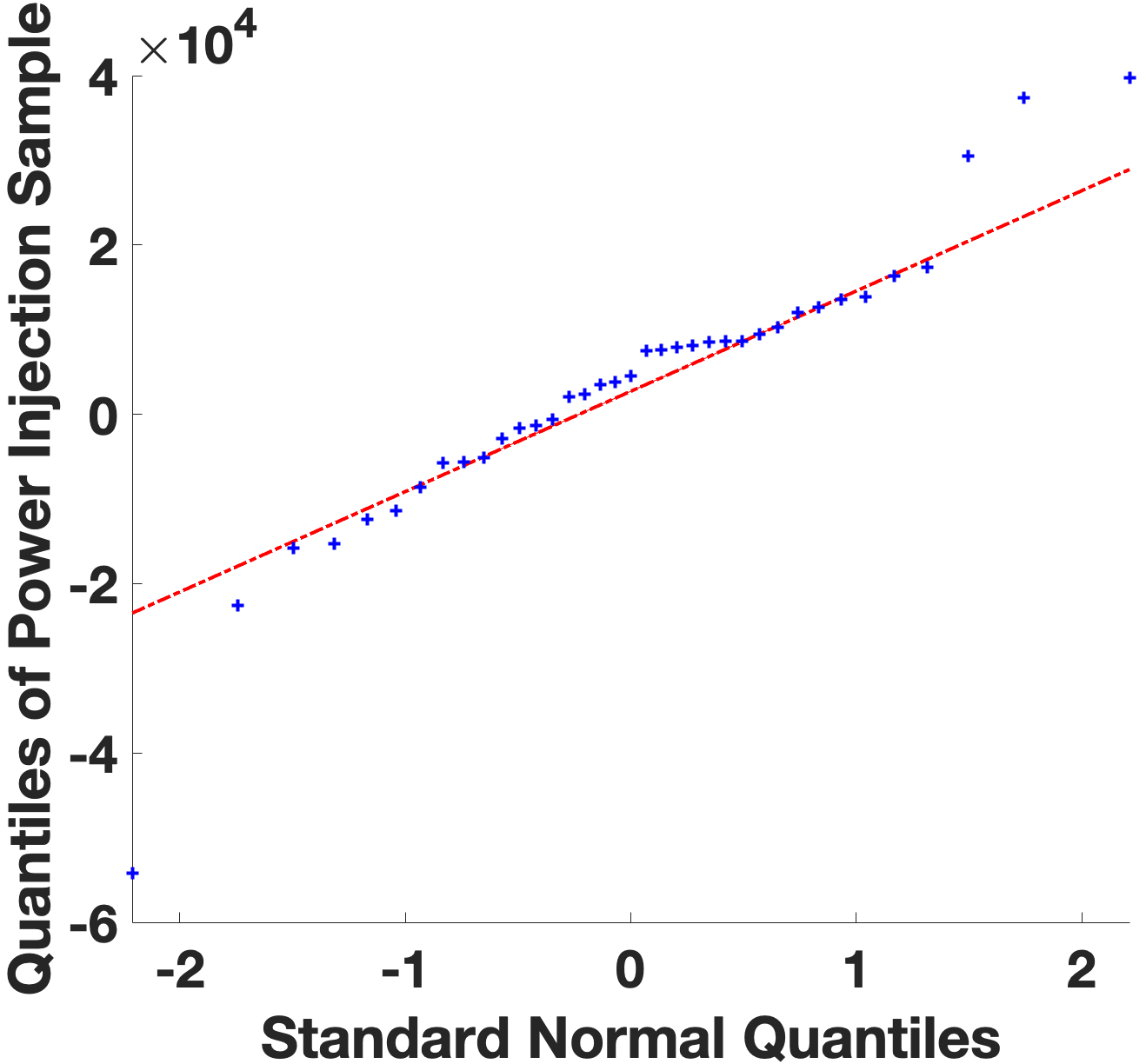}}}%
    \subfloat[\centering ]{{\includegraphics[height=4cm,width=4.5cm]{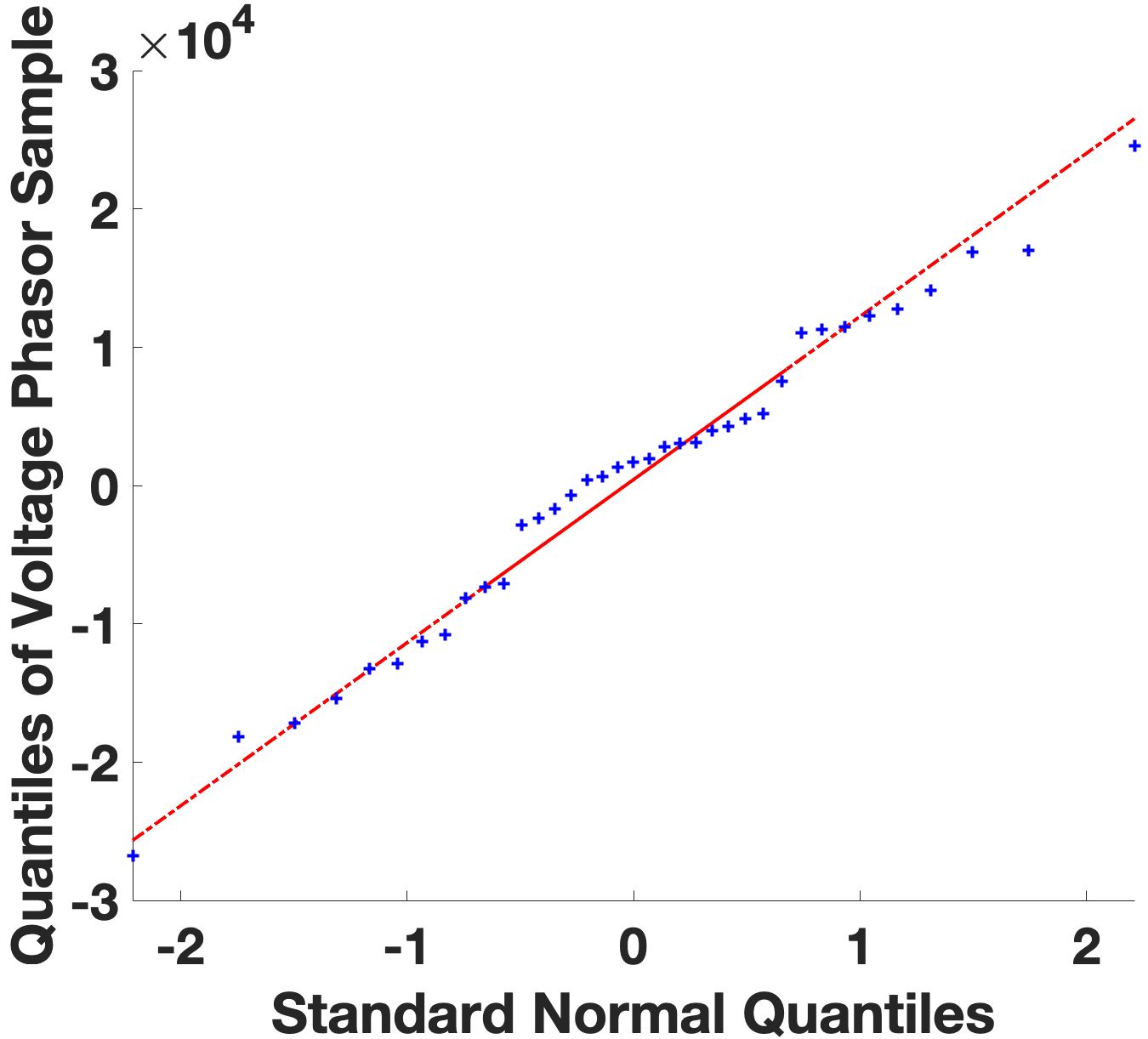}}}%
    \caption{QQ plot of $25\%$ outliers added in (a) power injection measurements; (b) voltage magnitude measurements.}%
    \label{Q}
\end{figure}

\begin{figure*}%
    \centering
  \subfloat[\centering ]{{\includegraphics[height=3cm,width=8.5cm]{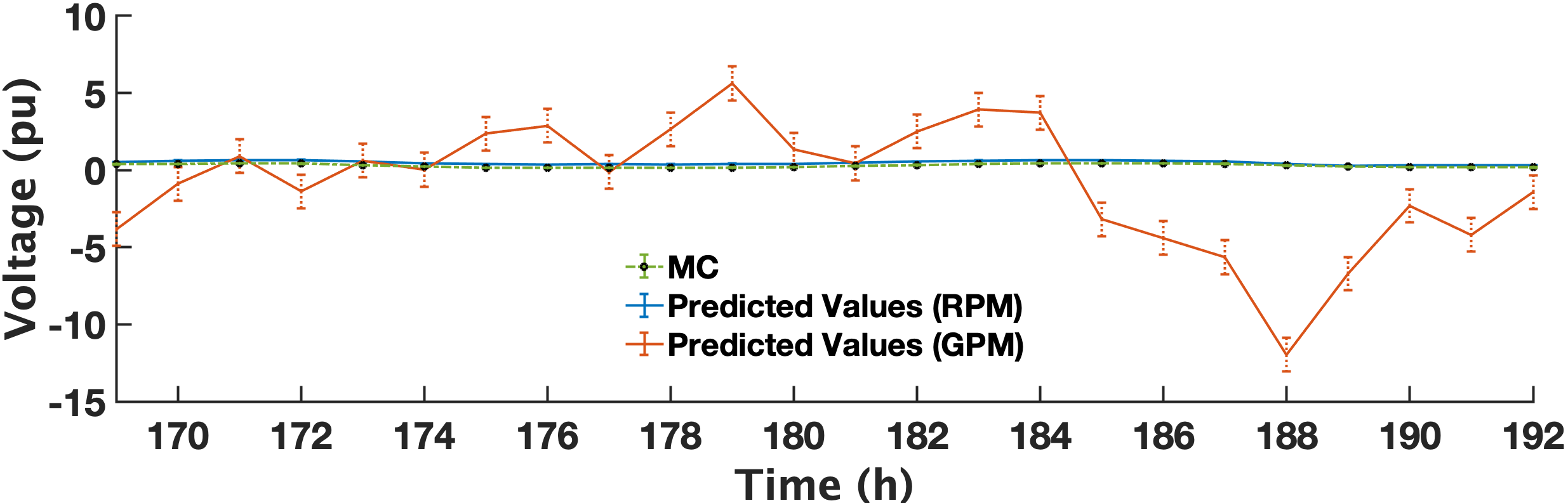}}}%
    \qquad
    \subfloat[\centering ]{{\includegraphics[height=3cm,width=8.5cm]{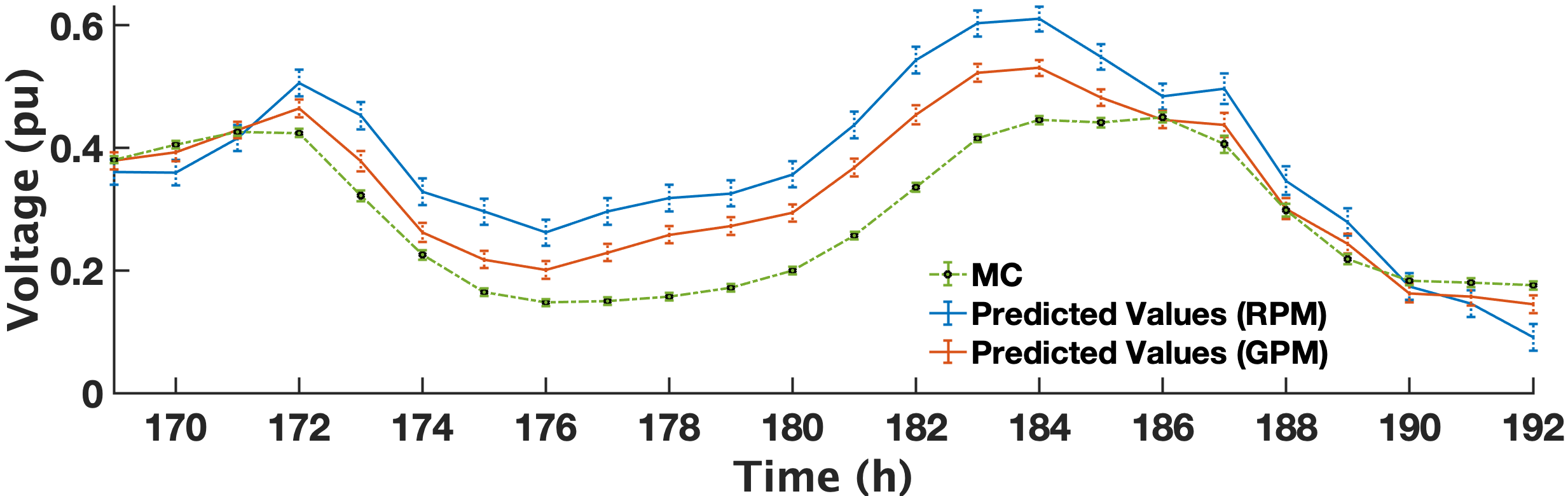}}}%
    \qquad
  \subfloat[\centering ]{{\includegraphics[height=3cm,width=8.5cm]{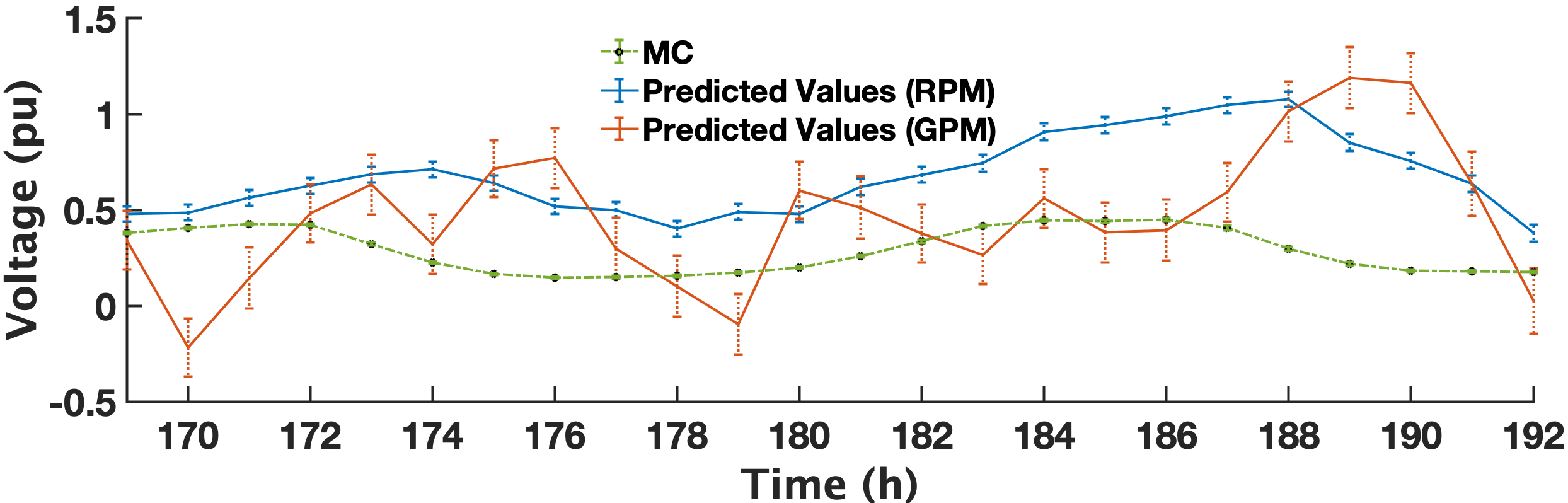}}}%
    \qquad
    \subfloat[\centering ]{{\includegraphics[height=3cm,width=8.5cm]{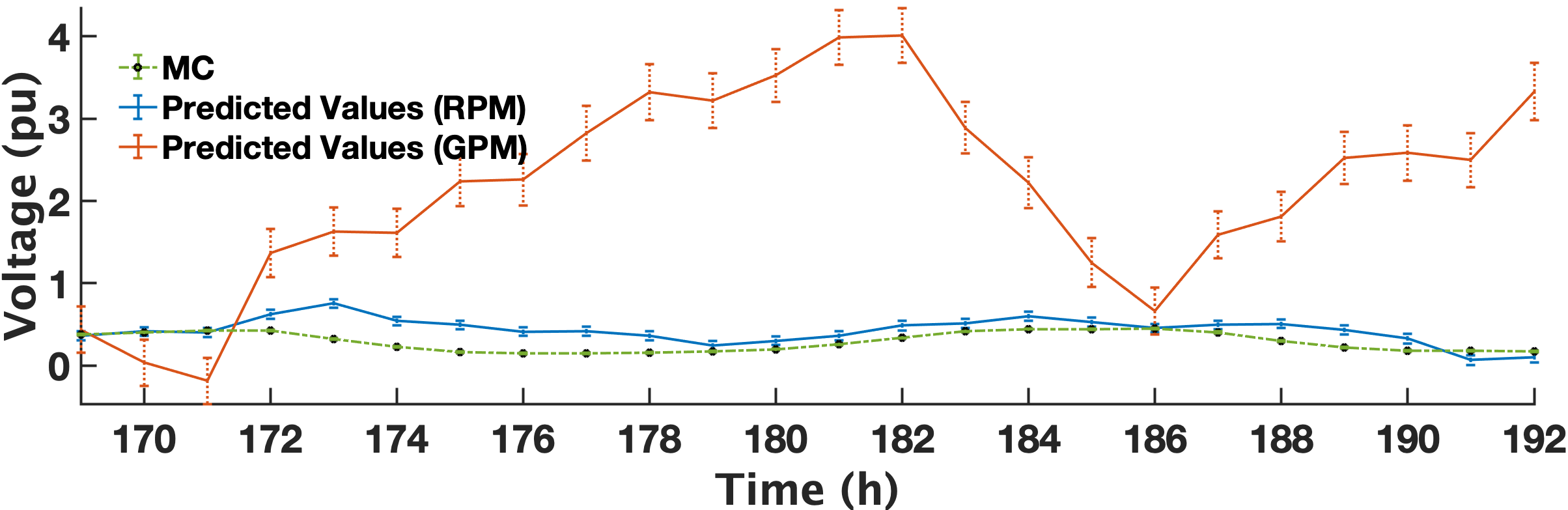}}}%
    \caption{ {Comparison between the GPM and the RPM forecast results for the voltage magnitude of Bus $2003.2$ in the $240-$bus network when (a) the training data set is added with $25\%$ of outliers ; (b) training data set is not added with outliers for linear basis; (c) the training data set is added with $25\%$ of outliers ; (d) training data set is not added with outliers for quadratic basis.} }
    \label{V}
\end{figure*}

\begin{figure*}%
    \centering
  \subfloat[\centering ]{{\includegraphics[height=3cm,width=8.5cm]{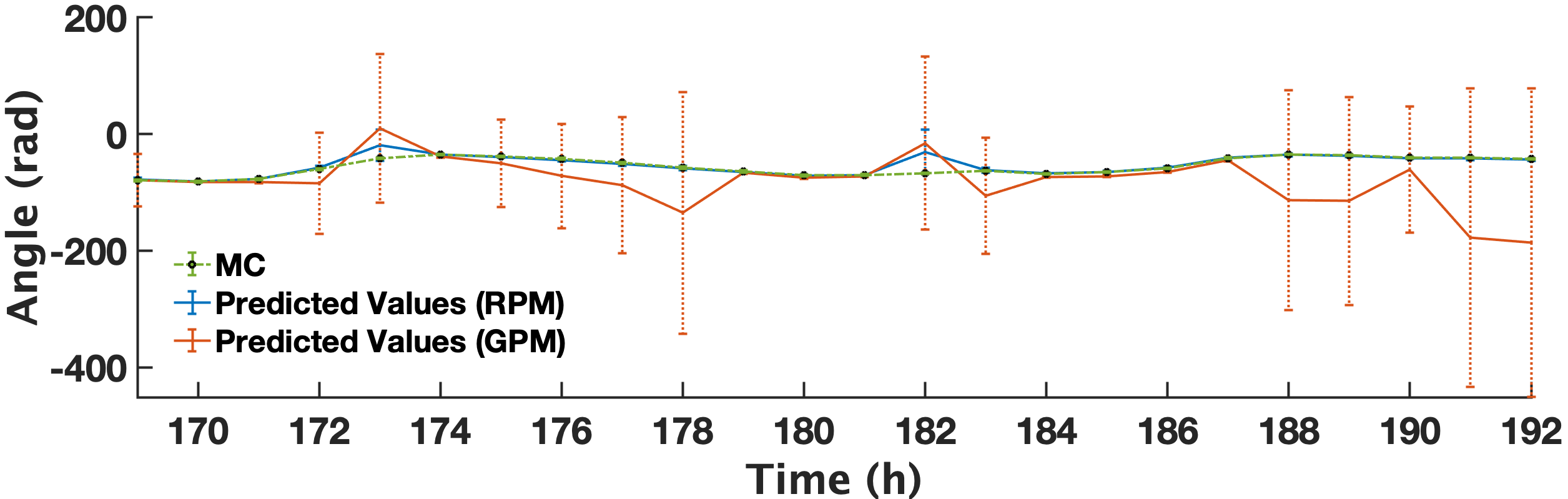}}}%
    \qquad
    \subfloat[\centering ]{{\includegraphics[height=3cm,width=8.5cm]{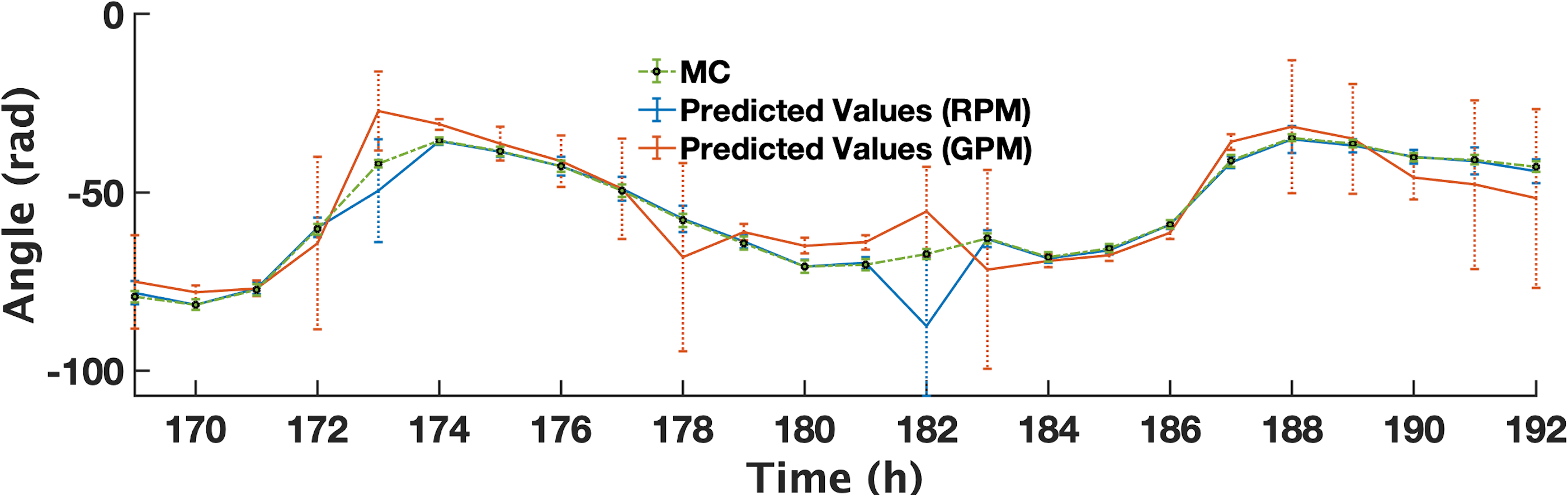}}}%
    \qquad
  \subfloat[\centering ]{{\includegraphics[height=3cm,width=8.5cm]{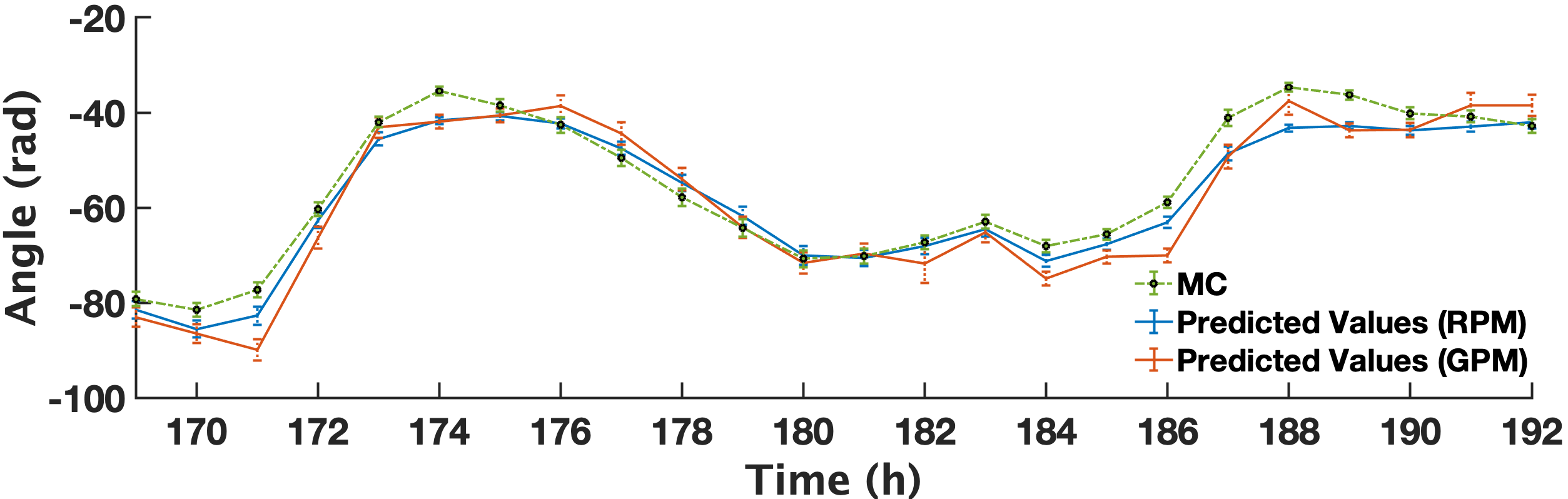}}}%
  \qquad
    \subfloat[\centering ]{{\includegraphics[height=3cm,width=8.5cm]{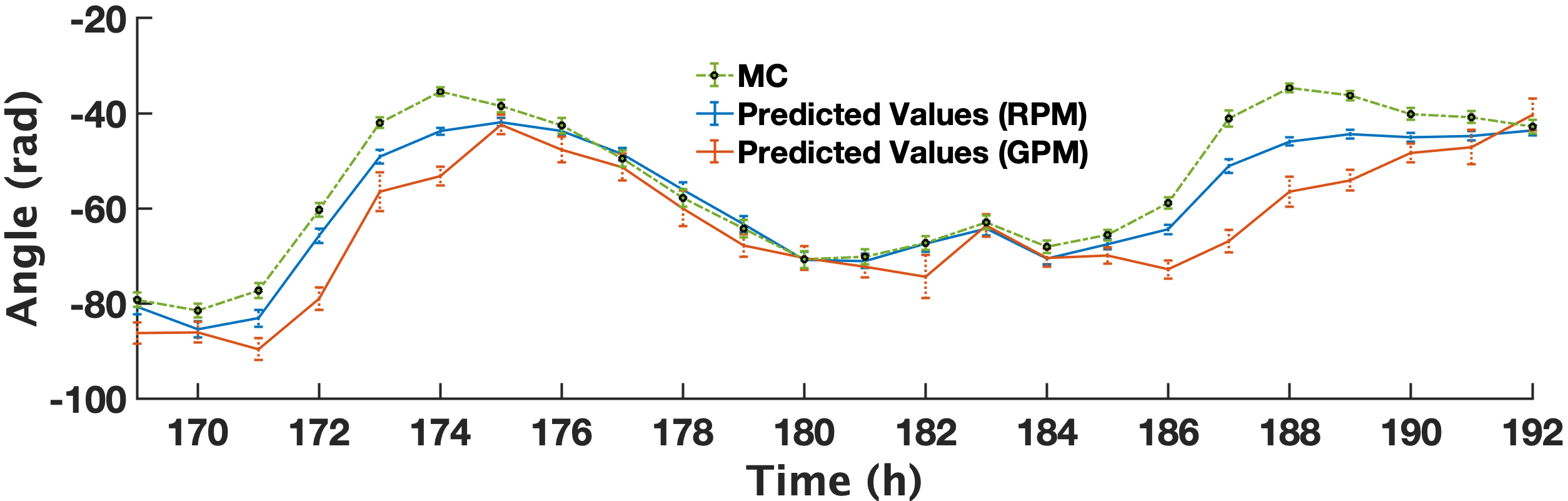}}}%
    \caption{ {Comparison between the GPM and the RPM forecast results for the voltage angle of Bus $2003.2$ in the $240-$bus network when (a) the training data set is added with $25\%$ of outliers ; (b) training data set is not added with outliers for linear basis; (c) the training data set is added with $25\%$ of outliers ; (d) training data set is not added with outliers for quadratic basis.}}
    \label{A}
\end{figure*}

The training data set $(\mathbf{X},\mathbf{y})$ is obtained by running the three-phase power flow simulator for the hourly spaced load and active and reactive power injection measurements for $7$ days i.e. a total of
$n=168$ data points constitute a training data set for the design of the GPM and the RPM. Both the GPM and RPM are analyzed using the prediction results obtained for the $n^{*}=24$ test data points, which are the validation data points. For the probabilistic analysis, $7000$ samples are drawn using Latin hypercube sampling from input variables of load following Gaussian distribution $P_{{t}_{i}^{*},L} \sim \mathcal{N}(P_{{t}^{*}_{i},L},0.05P_{{t}^{*}_{i},L})$, ${i}=169, \hdots,192$, the WGs' output following the Weibull distribution, and the PVs' output following the Beta distribution with the shape and scale parameters same as mentioned in Section 4.A. Voltage phasors predictions at Bus $2003.2$ for a day ahead forecast constitute as a benchmark for this study. {We display the mean of the prediction results obtained from the GPM and the RPM of the voltage magnitude at Bus $2003.2$ using linear and quadratic basis functions in Figs. \ref{V} (b) and (d), respectively, where the error bars represent the standard deviations as showing their probability distributions individually will be inconvenient. Similarly, the prediction results of the voltage angle at Bus $2003.2$ are plotted in Figs. \ref{A} (b) and (d)}.

\begin{figure}%
    \centering
  \subfloat[\centering ]{{\includegraphics[height=4cm,width=4cm]{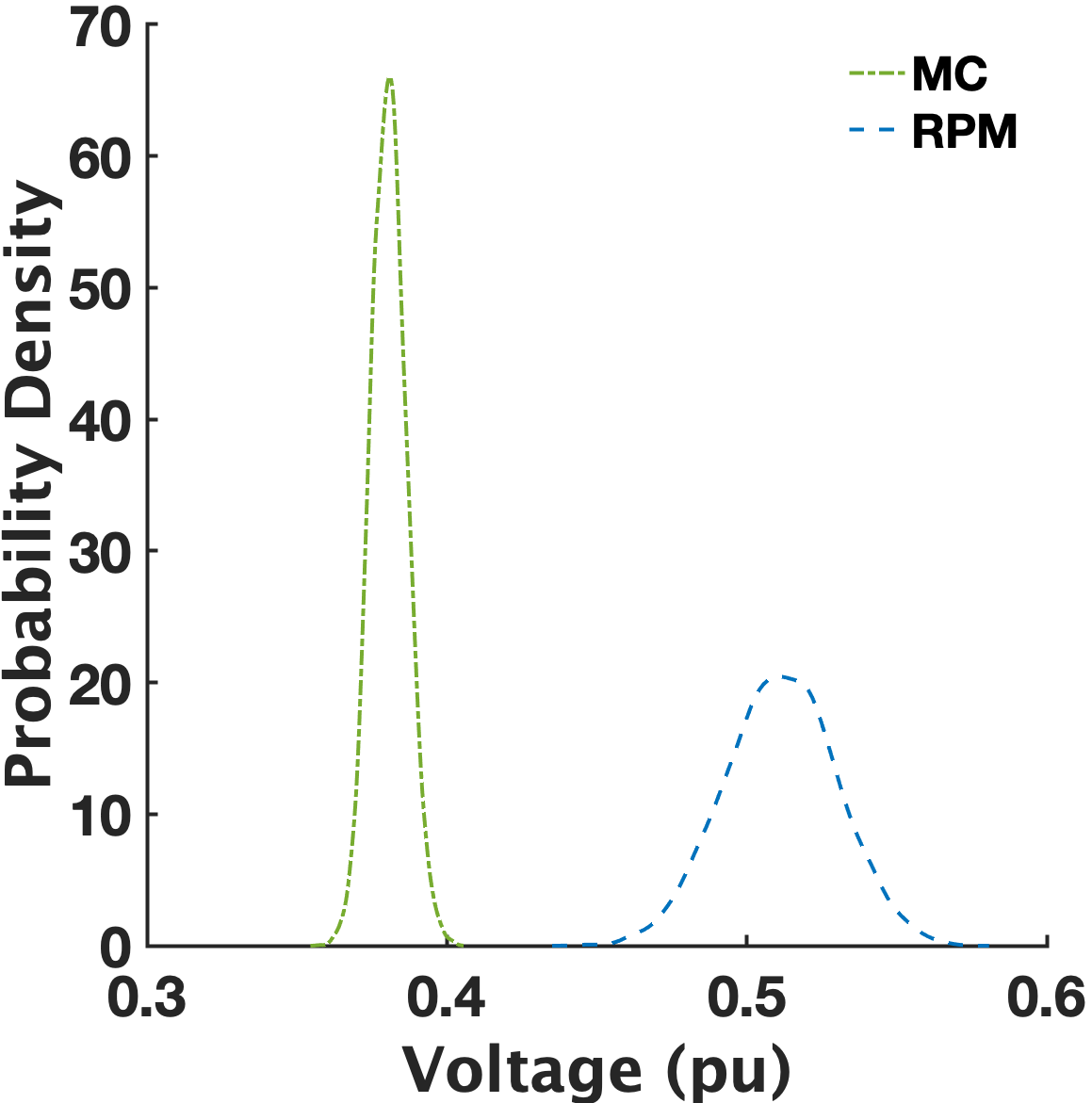}}}%
    \qquad
    \subfloat[\centering ]{{\includegraphics[height=4cm,width=4cm]{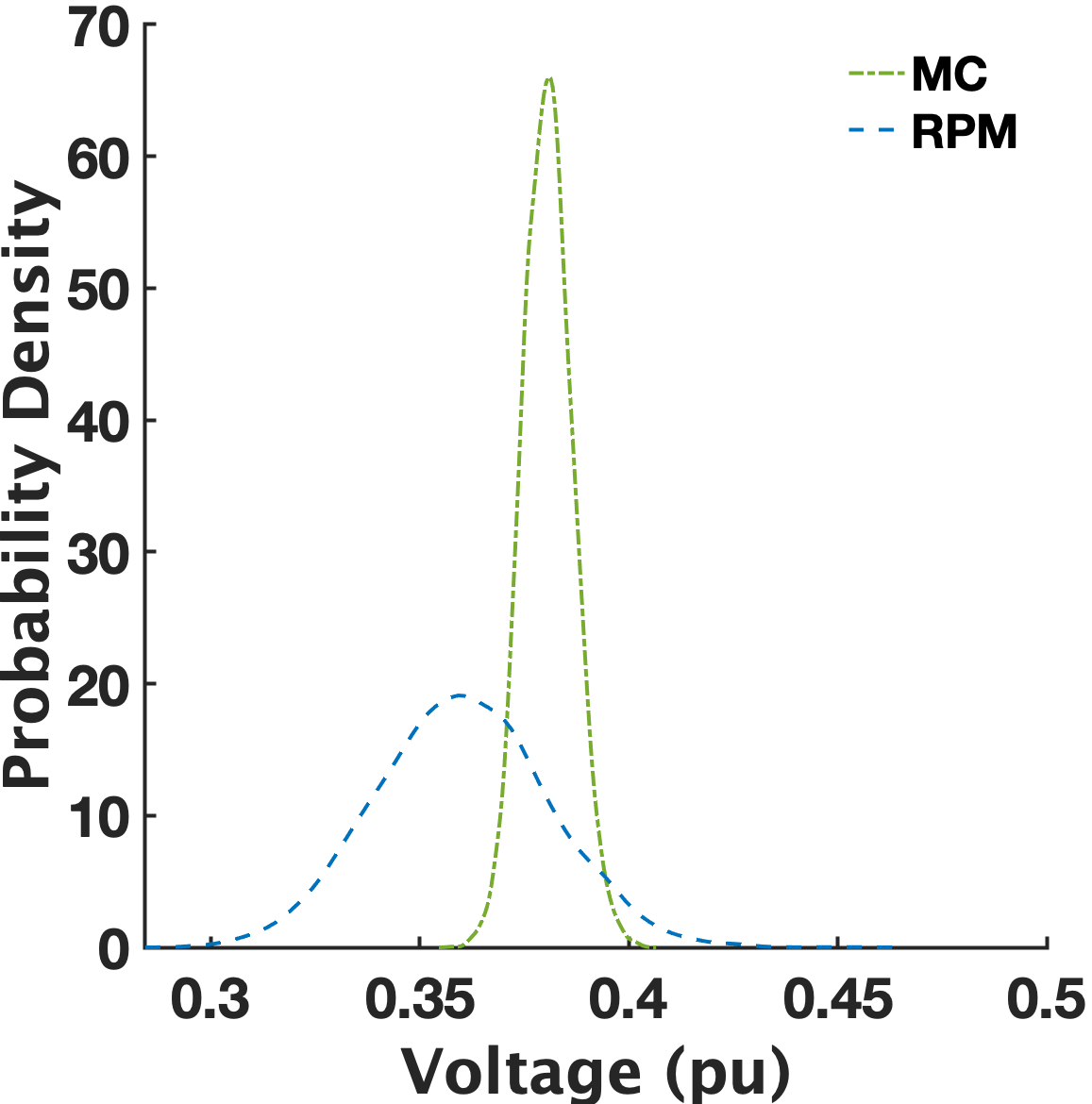}}}%
    \qquad
  \subfloat[\centering ]{{\includegraphics[height=4cm,width=4cm]{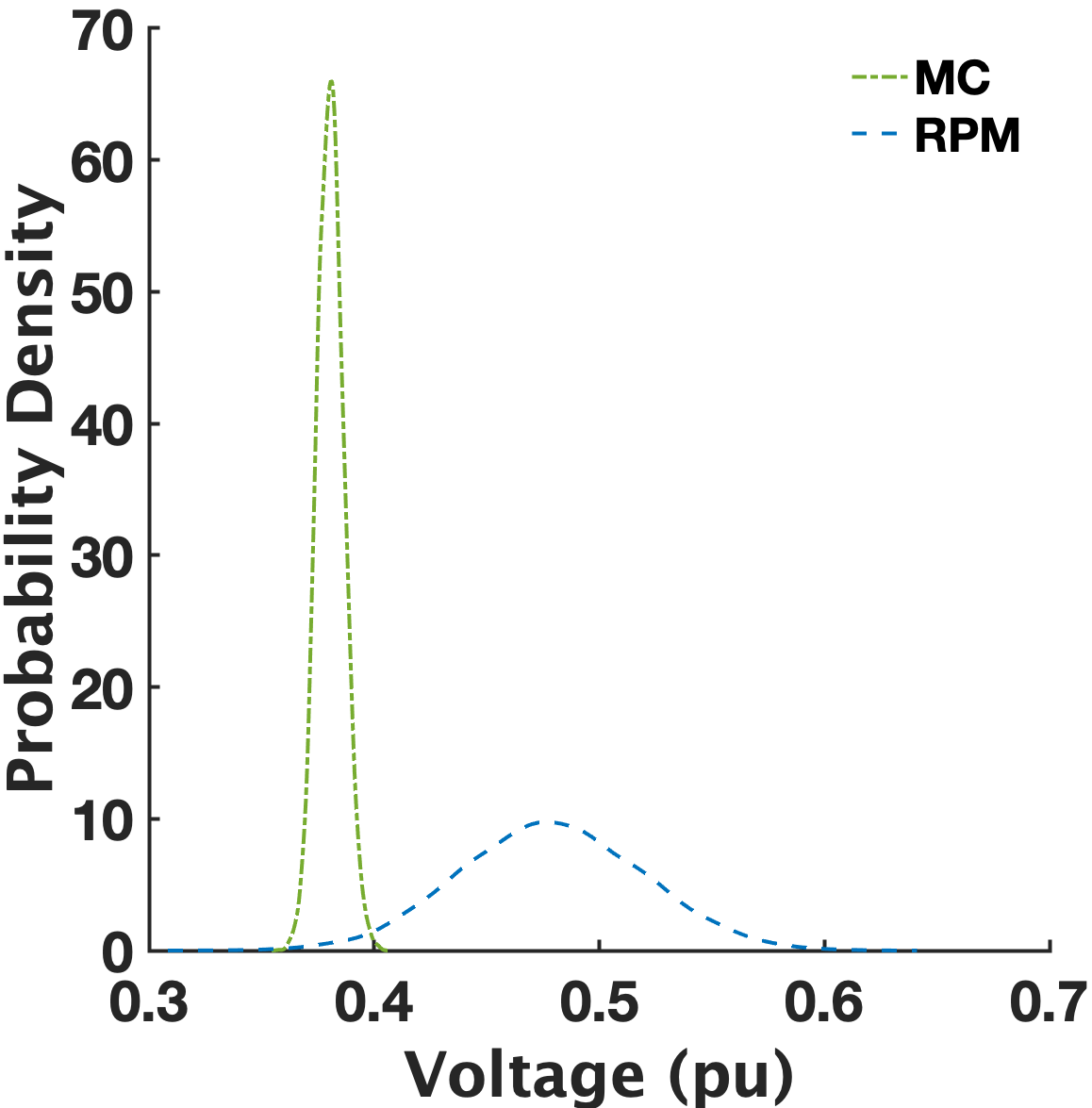}}}%
    \qquad
    \subfloat[\centering ]{{\includegraphics[height=4cm,width=4cm]{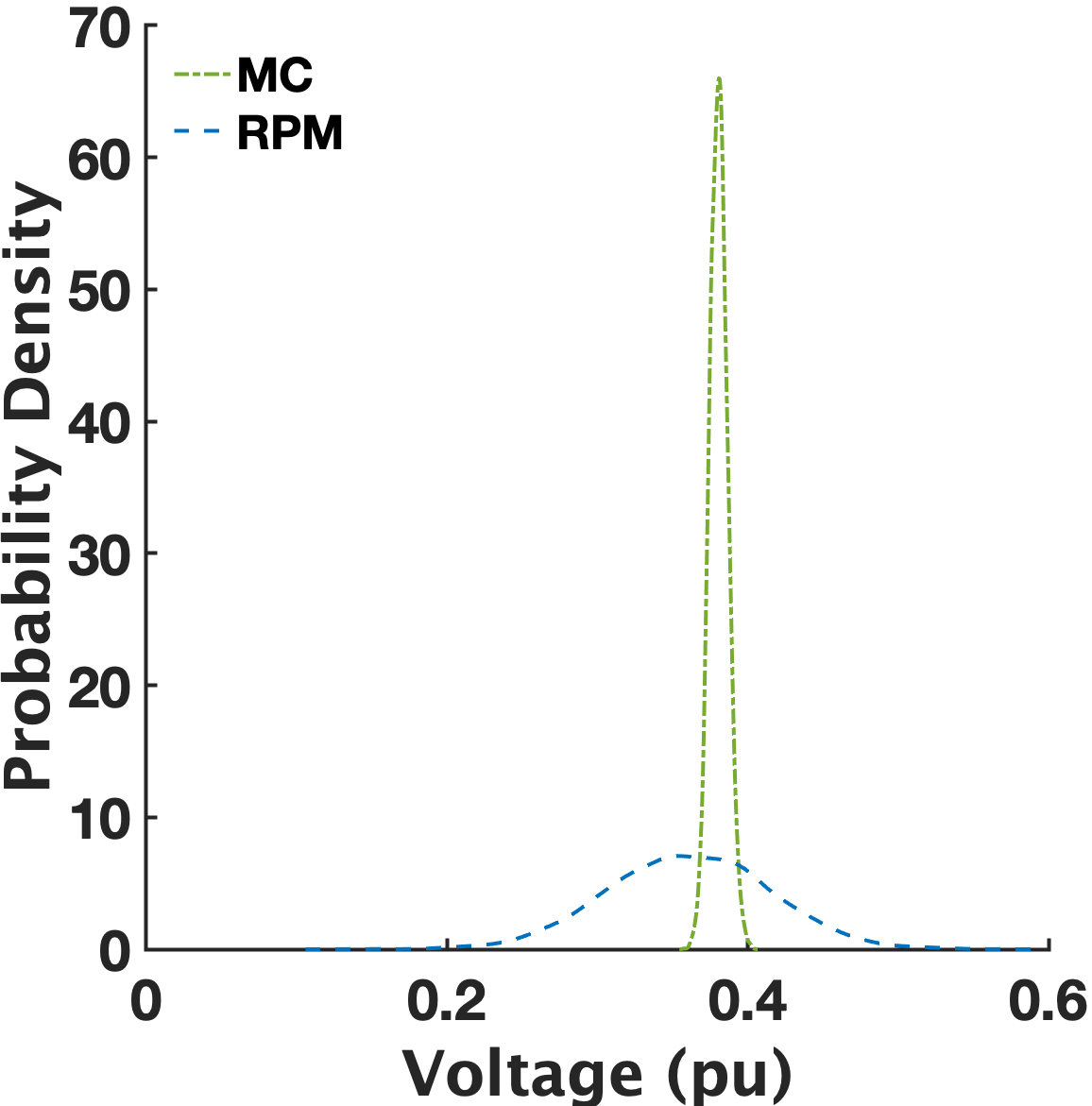}}}%
    \caption{ {Comparison between the GPM and the RPM probability density results for the voltage magnitude of Bus $2003.2$ in the $240-$bus network when (a) the training data set is added with $25\%$ of outliers ; (b) training data set is not added with outliers for linear basis; (c) the training data set is added with $25\%$ of outliers ; (d) training data set is not added with outliers for quadratic basis.} }
    \label{r_V_density}
\end{figure}

\begin{figure*}%
    \centering
  \subfloat[\centering ]{{\includegraphics[height=4cm,width=8.5cm]{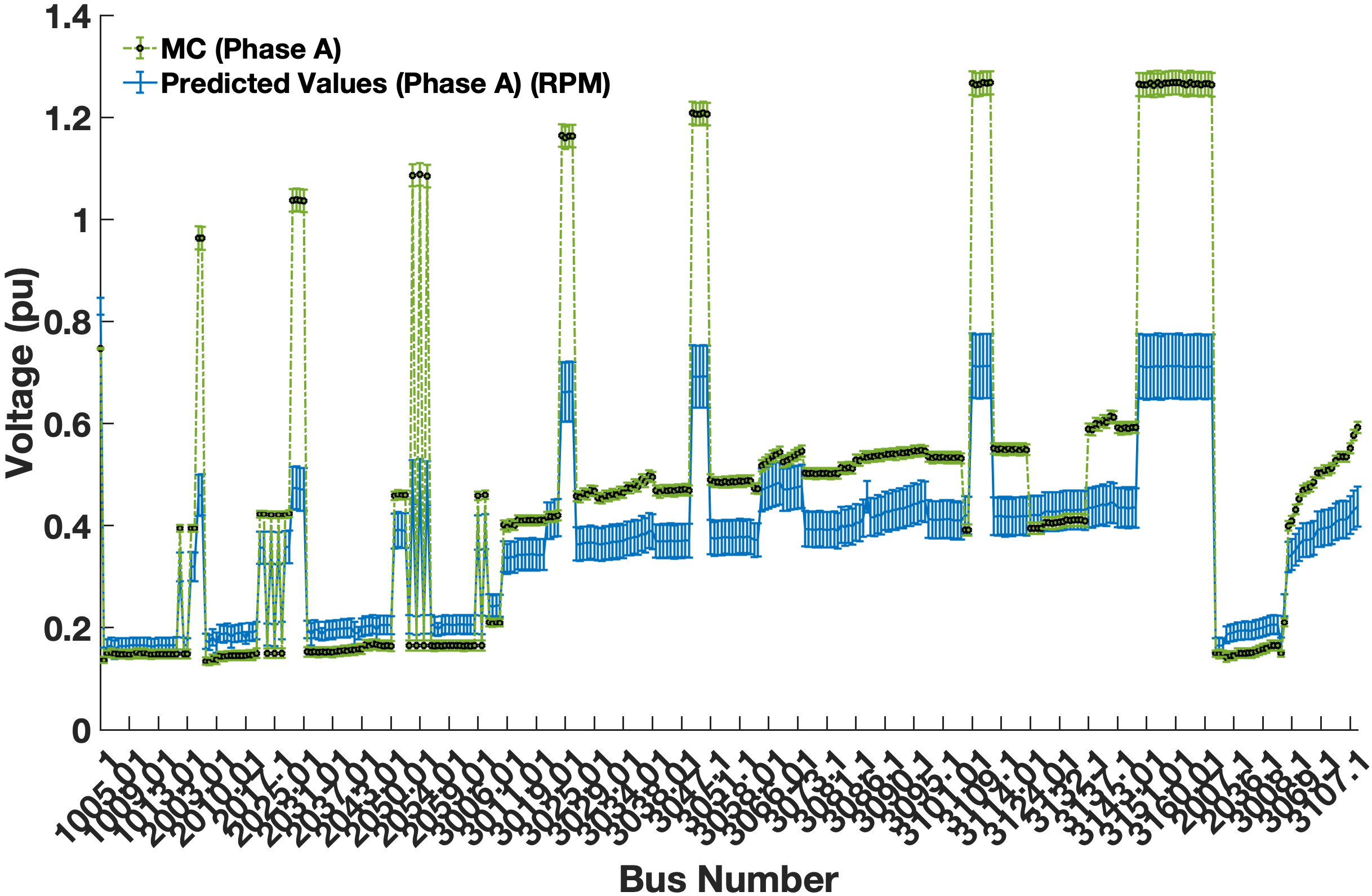}}}%
    \qquad
    \subfloat[\centering ]{{\includegraphics[height=4cm,width=8.5cm]{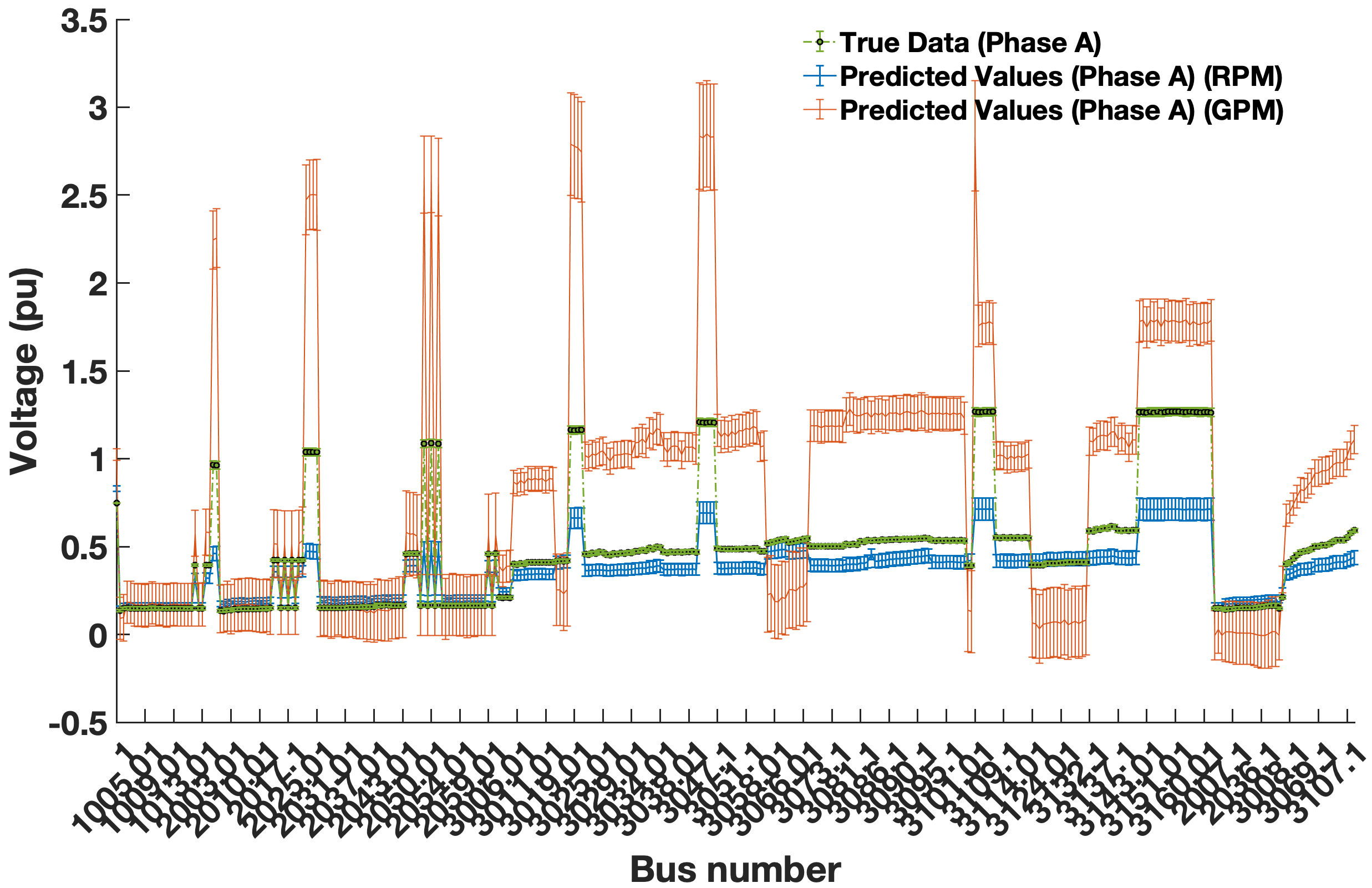}}}%
    \qquad
  \subfloat[\centering ]{{\includegraphics[height=4cm,width=8.5cm]{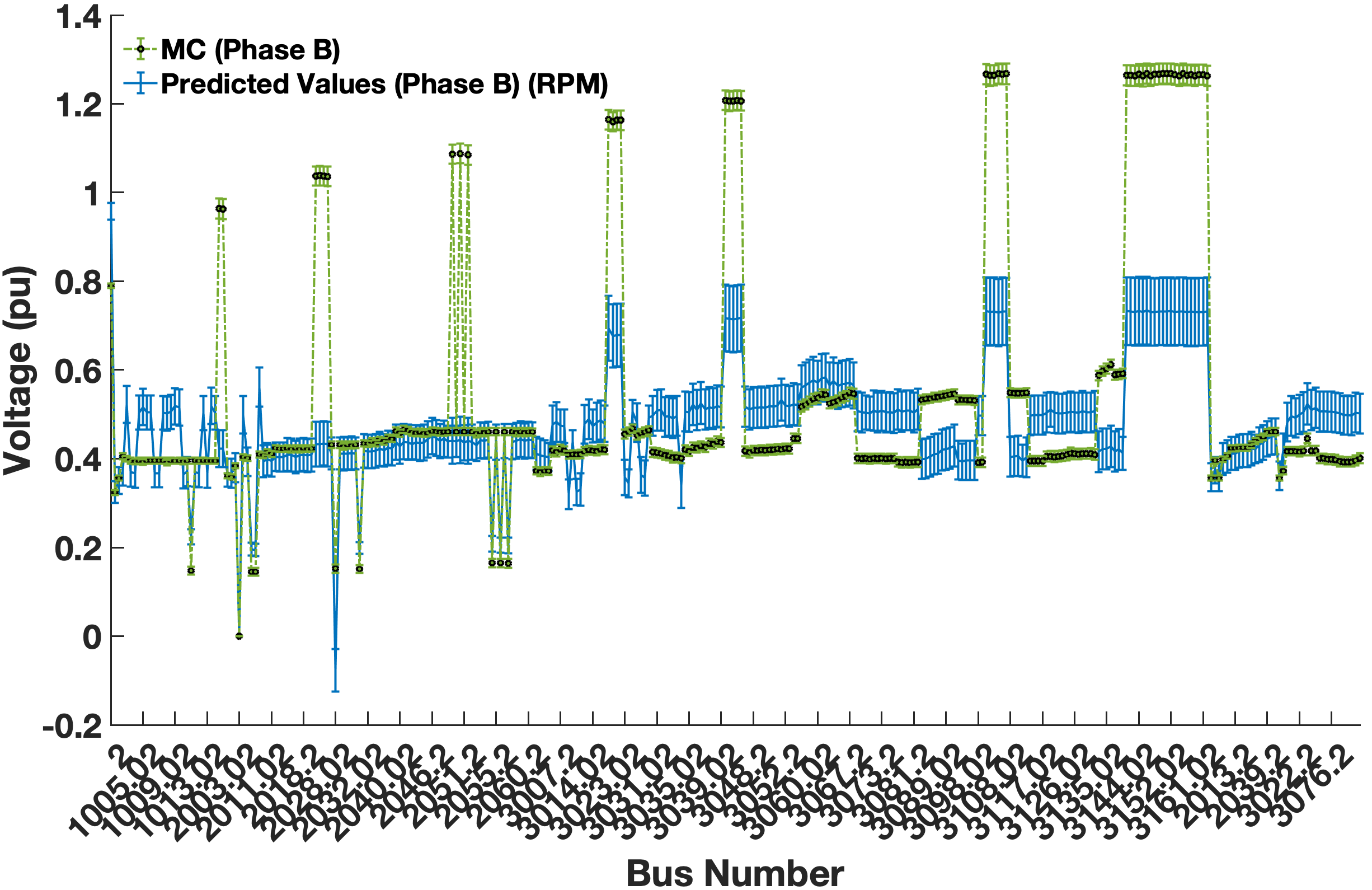}}}%
    \qquad
    \subfloat[\centering ]{{\includegraphics[height=4cm,width=8.5cm]{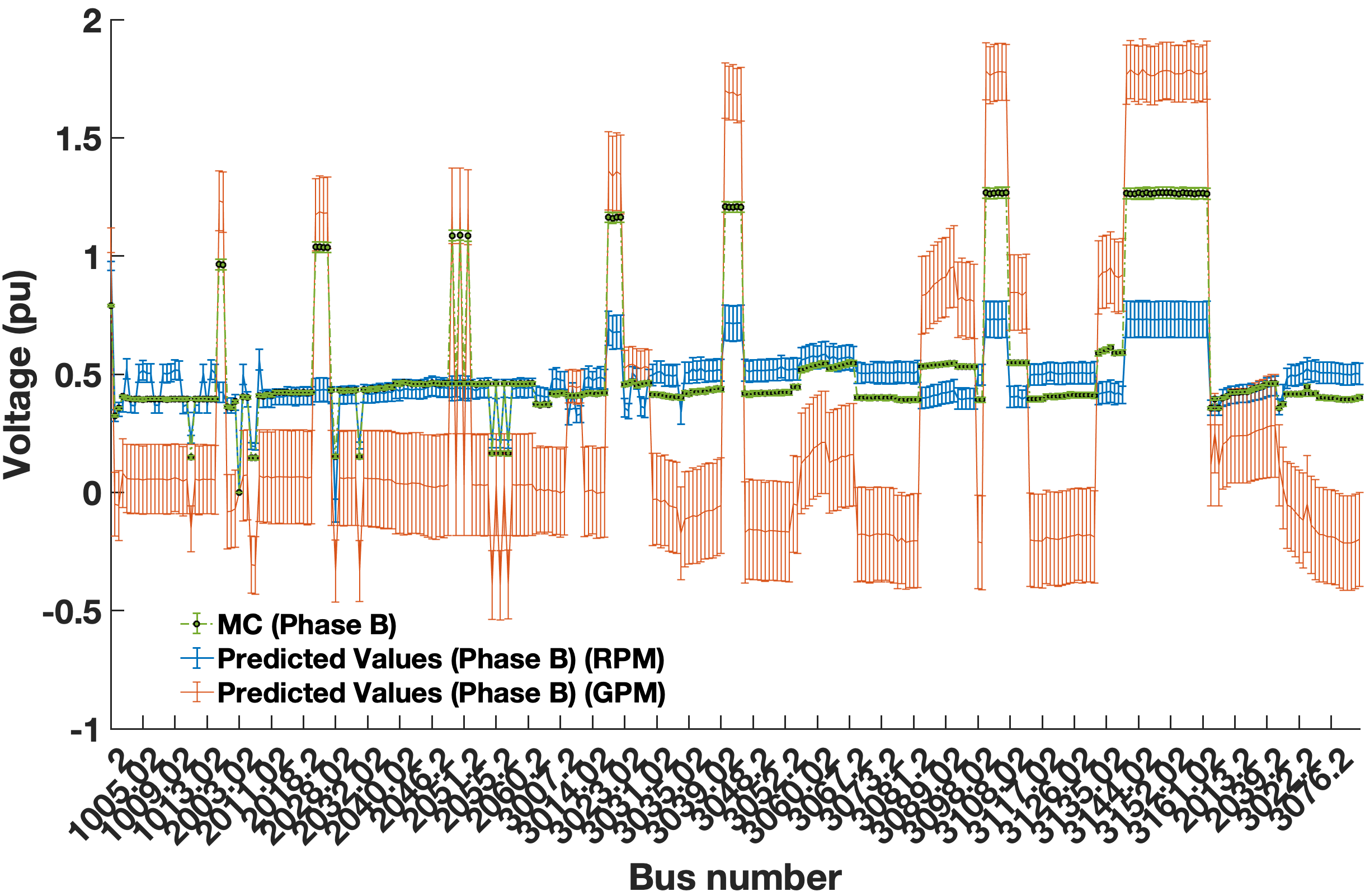}}}%
     \qquad
  \subfloat[\centering ]{{\includegraphics[height=4cm,width=8.5cm]{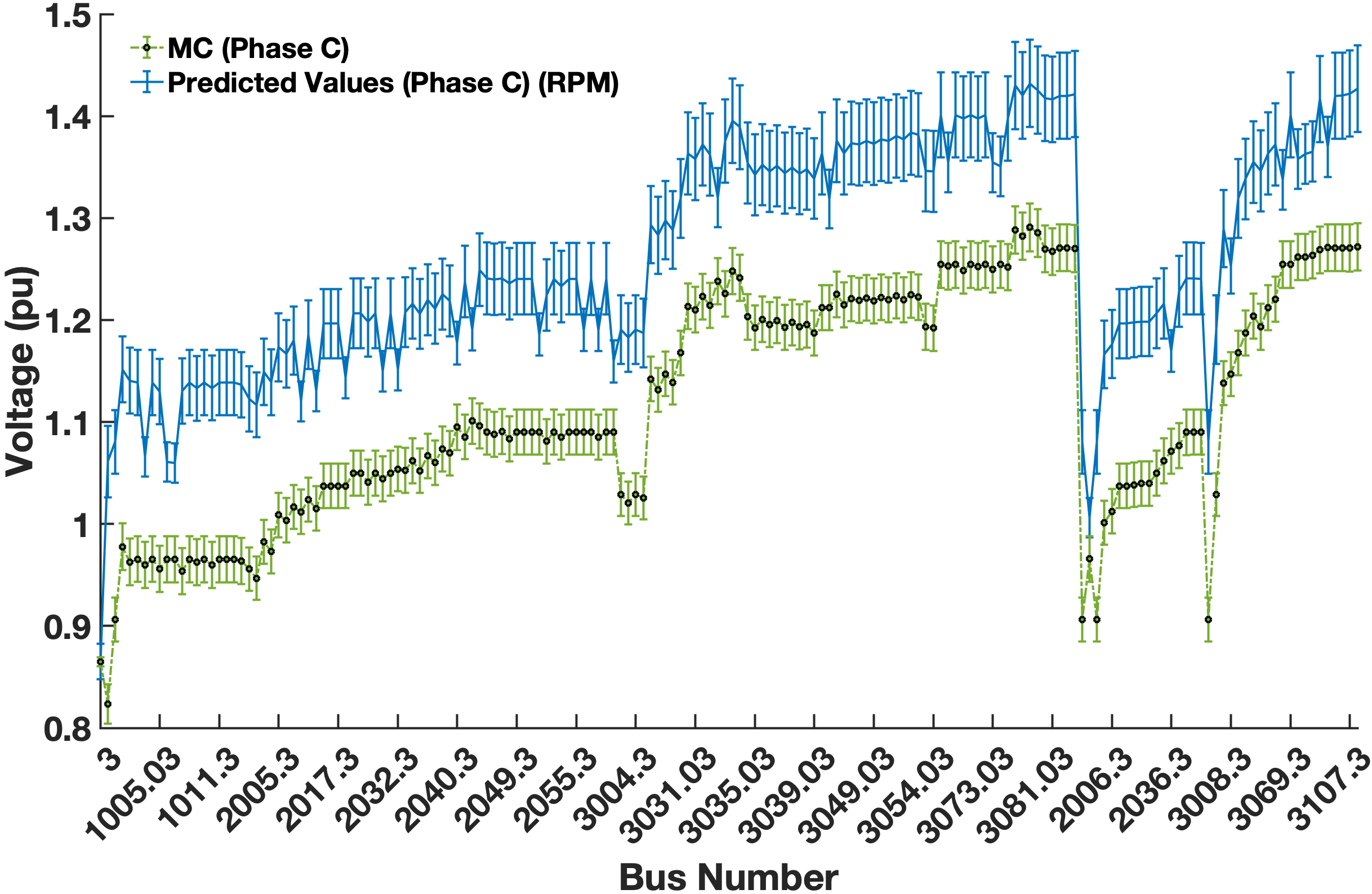}}}%
   \qquad
    \subfloat[\centering ]{{\includegraphics[height=4cm,width=8.5cm]{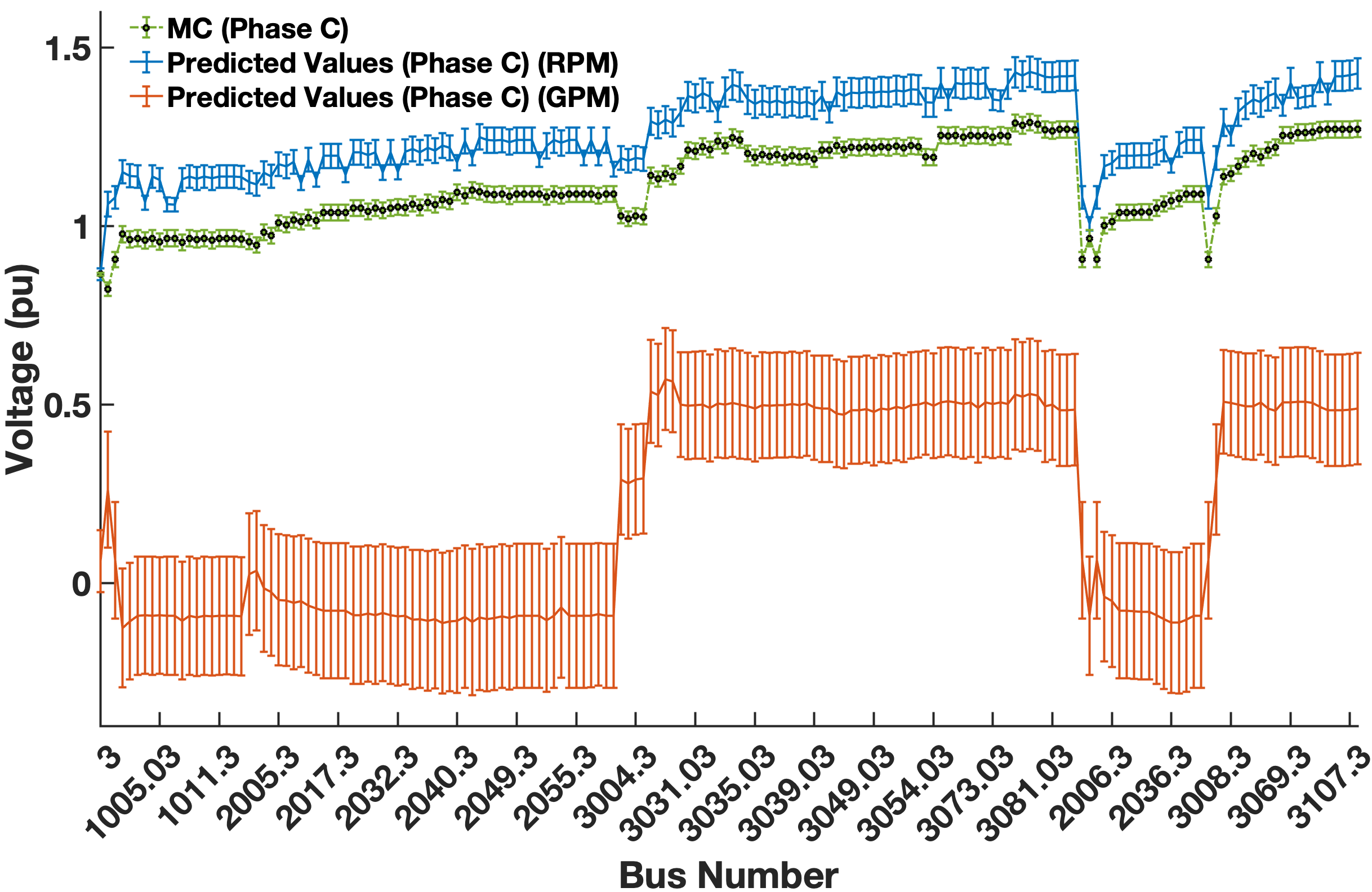}}}%
    \caption{The obtained prediction results of voltage magnitude from the RPM compared with those obtained from the GPM of $240-$bus system with $25\%$ of outliers added in training data. (a) The results obtained from the RPM for phase a; (b) comparison between the GPM and the RPM for phase a;       
    (c) The results obtained from the RPM for phase b; (d) comparison between the GPM and the RPM for phase b; 
    (e) The results obtained from the RPM for phase c; (f) comparison between the GPM and the RPM for phase c.}%
    \label{3_phase}
\end{figure*}

\begin{figure*}[!hbtp]%
    \centering
    \vspace{-0.2cm}
  \subfloat[\centering ]{{\includegraphics[height=4cm,width=8.5cm]{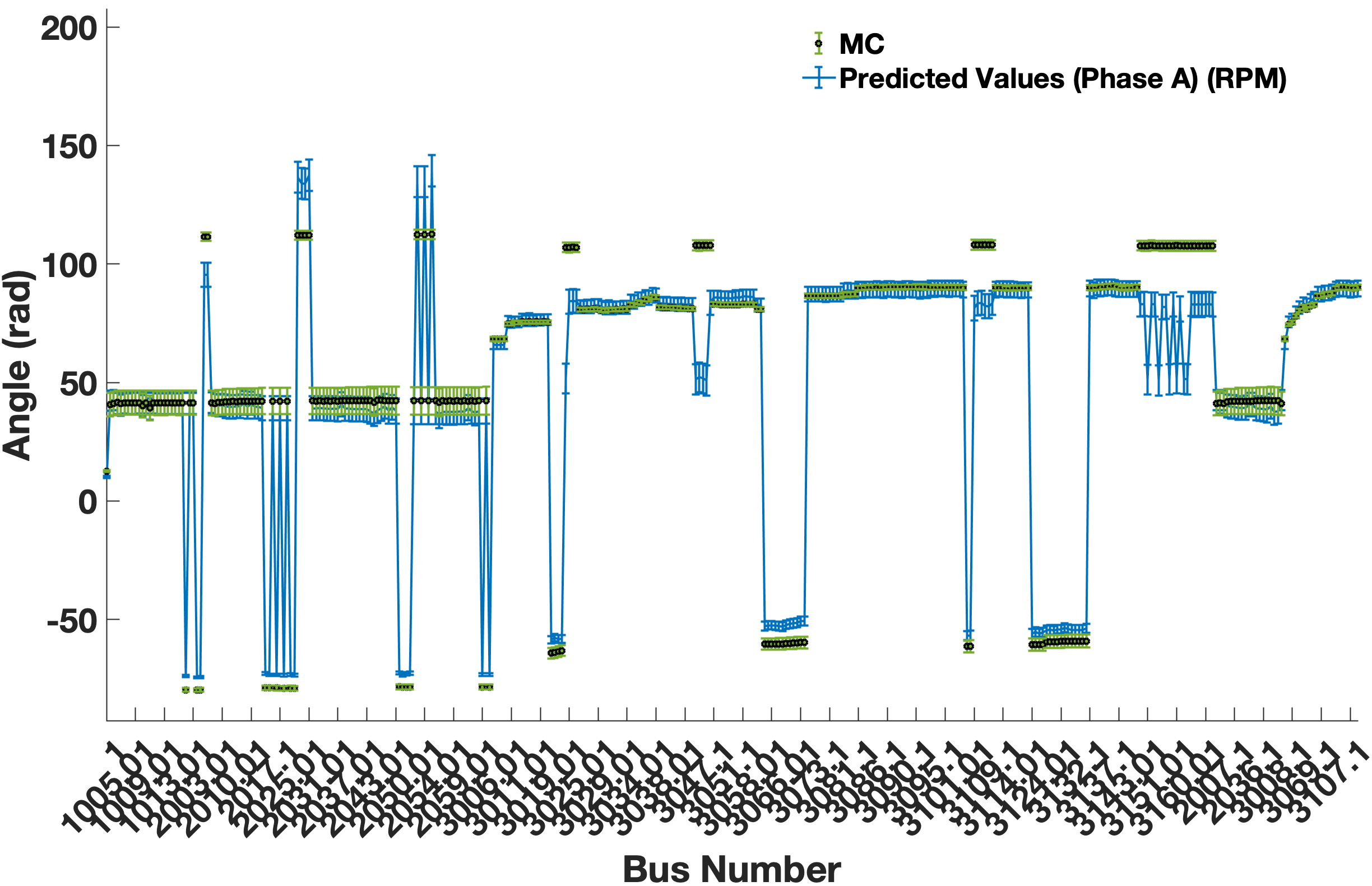}}}%
    \qquad
    \subfloat[\centering ]{{\includegraphics[height=4cm,width=8.5cm]{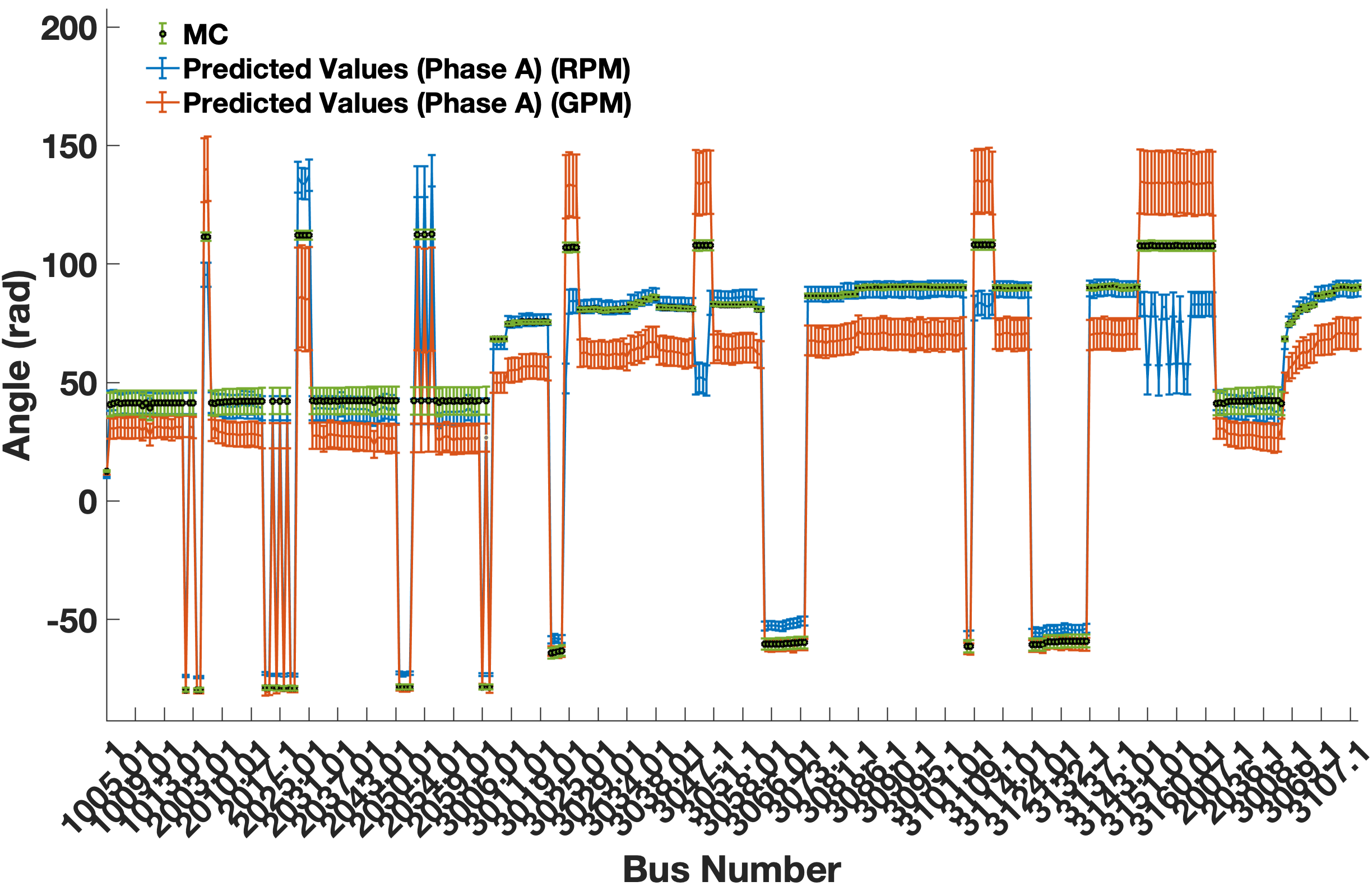}}}%
    \qquad
  \subfloat[\centering ]{{\includegraphics[height=4cm,width=8.5cm]{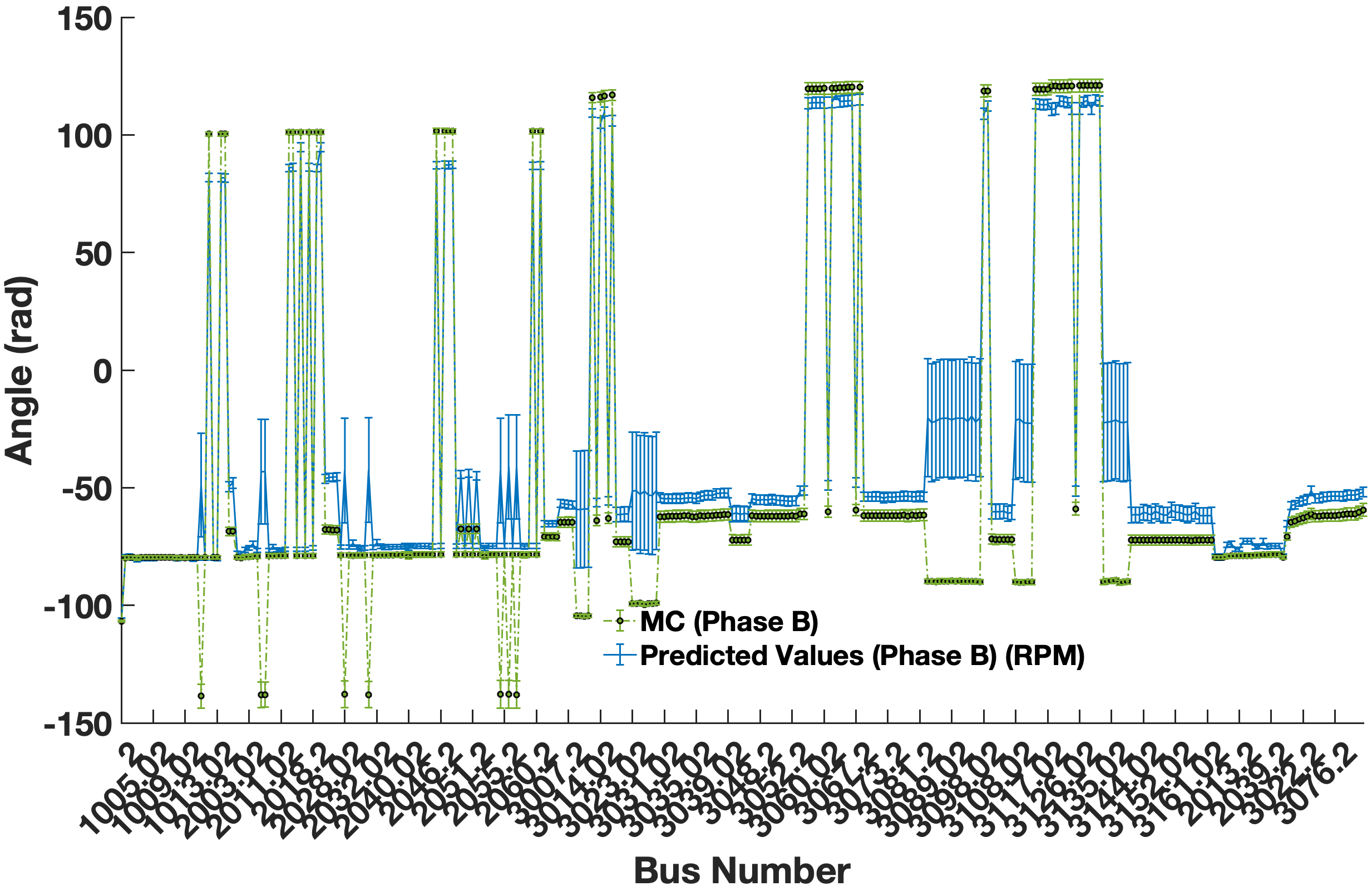}}}%
    \qquad
    \subfloat[\centering ]{{\includegraphics[height=4cm,width=8.5cm]{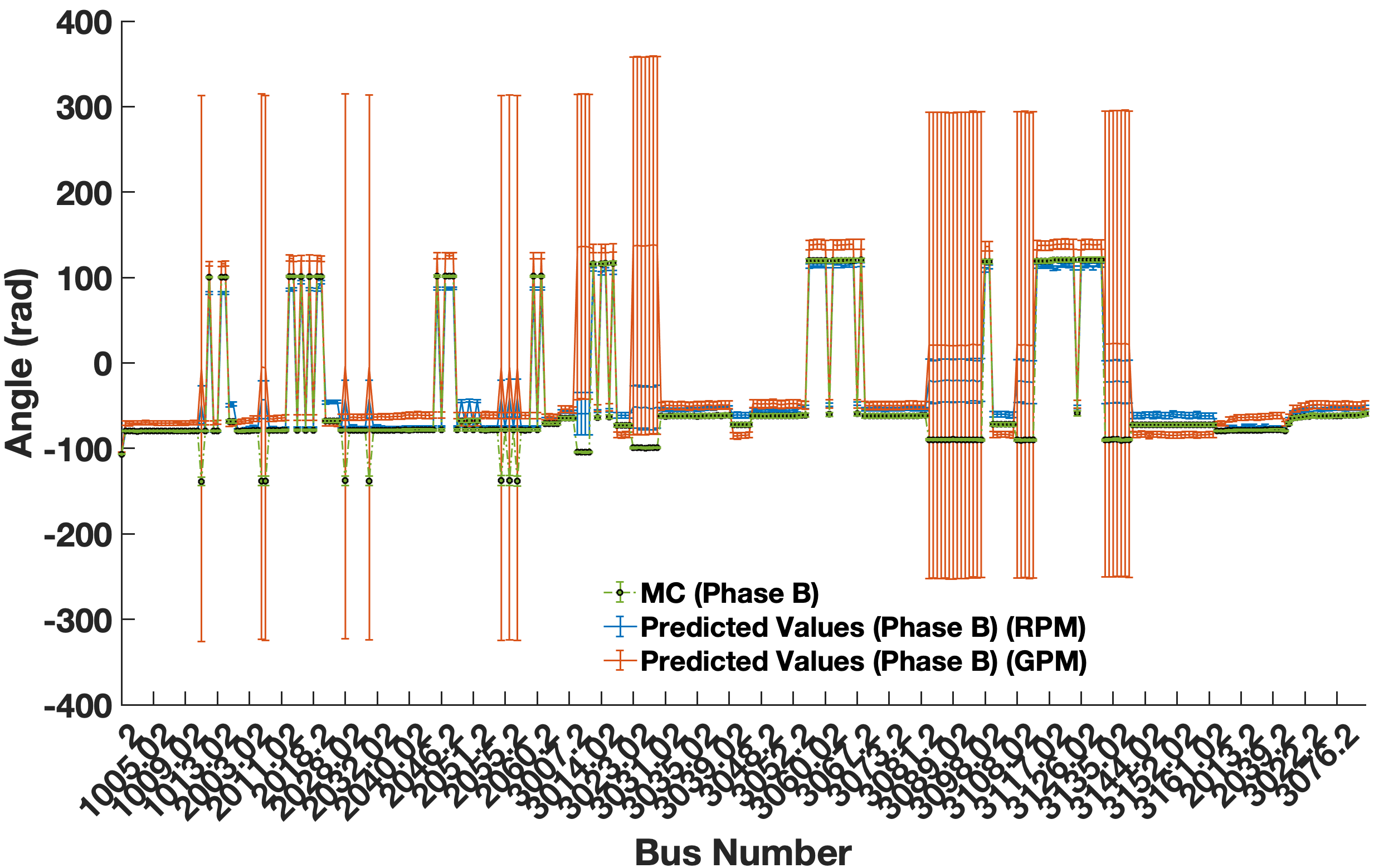}}}%
    \qquad
  \subfloat[\centering ]{{\includegraphics[height=4cm,width=8.5cm]{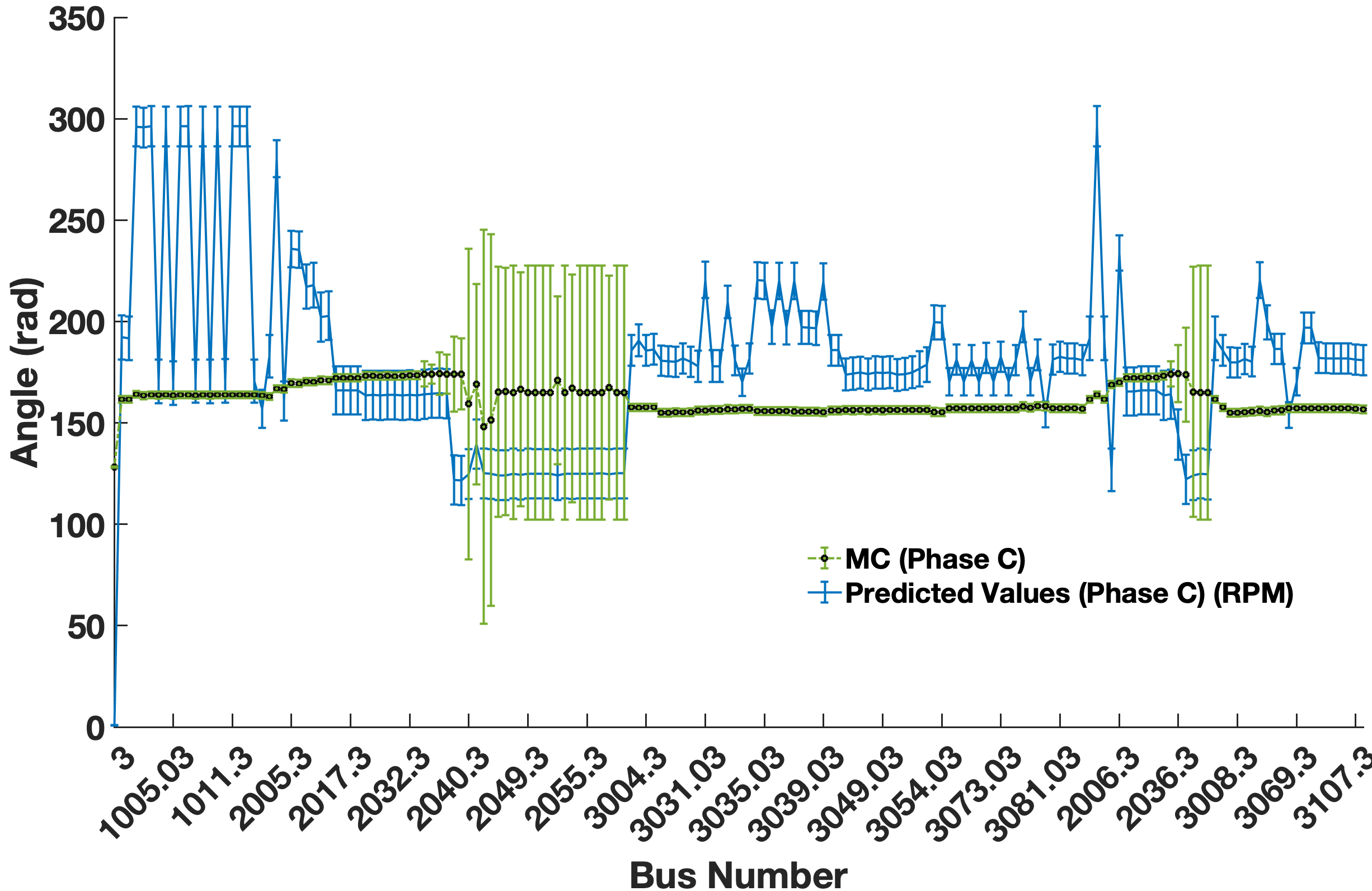}}}%
    \qquad
    \subfloat[\centering ]{{\includegraphics[height=3cm,width=8.5cm]{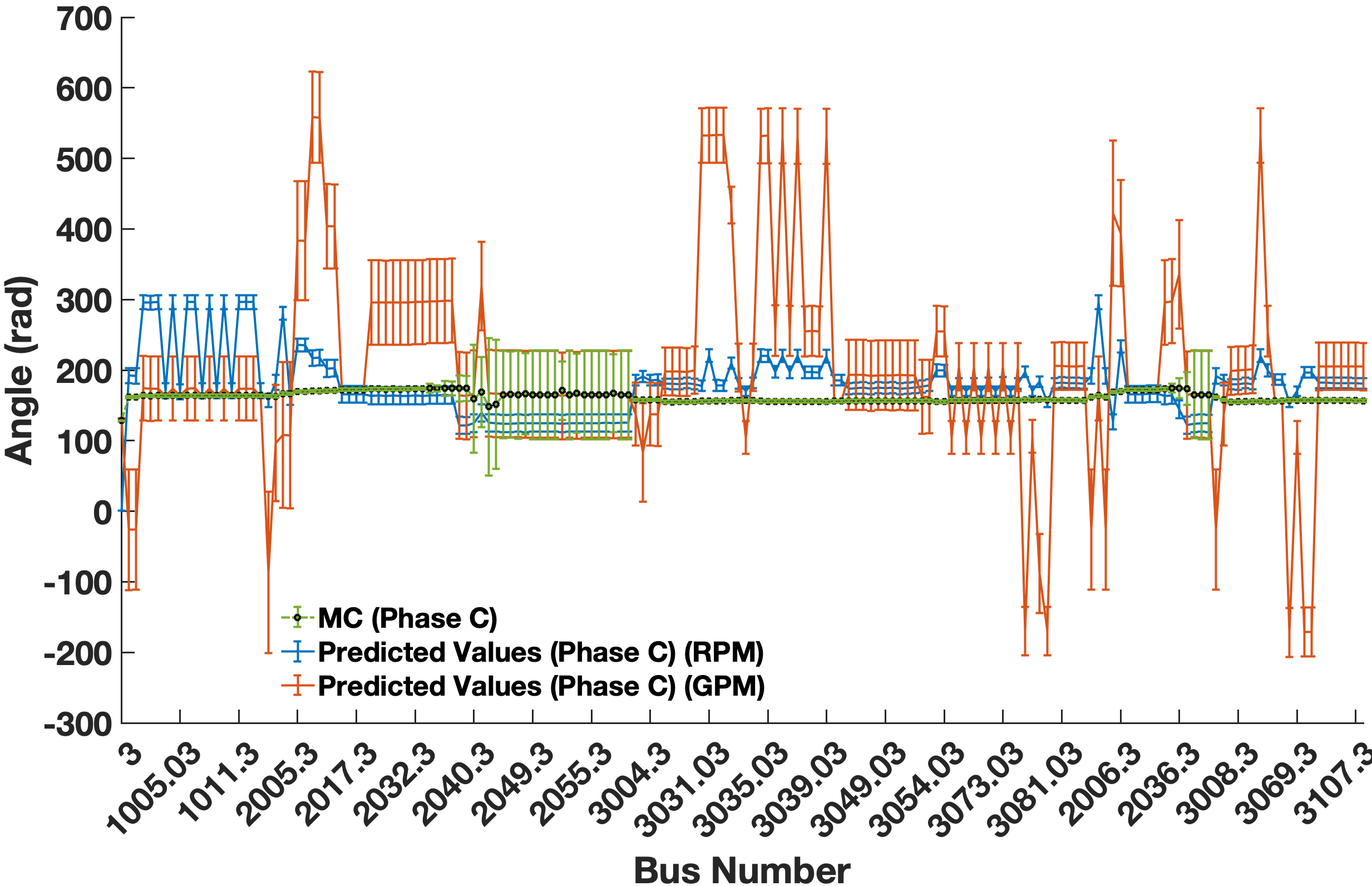}}}%
    \caption{The obtained prediction results of voltage angle from the RPM compared with those obtained from the conventional GPM of $240-$bus system with $25\%$ of outliers added in training data. (a) The results obtained from the RPM for phase a; (b) comparison between the GPM and the RPM for phase a;       
    (c) The results obtained from the RPM for phase b; (d) comparison between the GPM and the RPM for phase b; 
    (e) The results obtained from the RPM for phase c; (f) comparison between the GPM and the RPM for phase c.}%
    \label{3_phase_C}
\end{figure*}

{We now add up to $25\%$ of outliers in the power injection measurements and voltage magnitudes and phase angles, i.e., to the first $42$ input and output data points in the training dataset. To best demonstrate the robustness of the RPM, we choose the outlier distribution to be Student's t with $10$ degrees of freedom because of heavier tails as displayed in Fig. \ref{Q}. Similarly, the same percentage of outliers is included in the measurements of active and reactive load power.} {We display in Figs. \ref{r_V_density} (b) and (d) the probability density function  of the voltage magnitude at Bus $2003.2$ for the test input at instance ${t}^{*}_{169}$ obtained using linear and quadratic basis function. Similarly, the probability density results of the voltage angle at Bus $2003.2$ are plotted in Figs. \ref{r_V_density} (a) and (b). We compare the prediction results of the voltage magnitude at Bus $2003.2$ obtained from the GPM and the RPM with basis function as linear and quadratic in Figs. \ref{V} (a) and (c), respectively. The voltage angle results are displayed in Figs. \ref{A} (a) and (c).} { The RMSE and MAE values of the prediction results for the cases of addition of outliers in the training data set and without the addition of outliers, both with the linear and basis function, are listed in Table \ref{tab:240}. We observe that the RMSE and the MAE values obtained from the GPM and the RPM for the voltage magnitudes' predictions are lesser with linear basis functions than the ones with the quadratic basis functions for both cases. As for the voltage angles, both models' performance is better with the quadratic basis function for the case with outliers in the training data set. The linear basis function yields better performance for the case without outliers. Therefore, we choose the linear basis to plot the voltage magnitudes and the quadratic basis function for the voltage angles to obtain further predictions at all the buses in the system for the case with outliers in the training data set.} {The mean values (indicated as dots) and standard deviations (indicated as error bars) of both the models' prediction results for the voltage magnitudes (see, Fig. \ref{3_phase}) and phase angles (see, Fig. \ref{3_phase_C}) at the buses of the $240-$bus system are compared with the MC simulation results. We observe that the results obtained from the comparable GPM deviate significantly from the true values in both the mean and standard deviation. The performance of the RPM is comparably accurate on account of the trade-off between the accuracy and robustness of the SHGM estimator. Conventional GPM is strongly biased towards outliers, thus the prediction results deviate further away from the  MC results. The bias is particularly significant in phase C results because of the high variance of voltage phasors due to the large power flow in lines. For a large magnitude of outliers, the resulting bias is the worst-case scenario that can be imposed on power system measurements. The proposed RPM keeps this bias finite as long as the added outliers are added without exceeding the breakdown point, whereas the bias is unbounded for the results of conventional GPM.}

\begin{figure*}%
    \centering
  \subfloat[\centering ]{{\includegraphics[height=4cm,width=4cm]{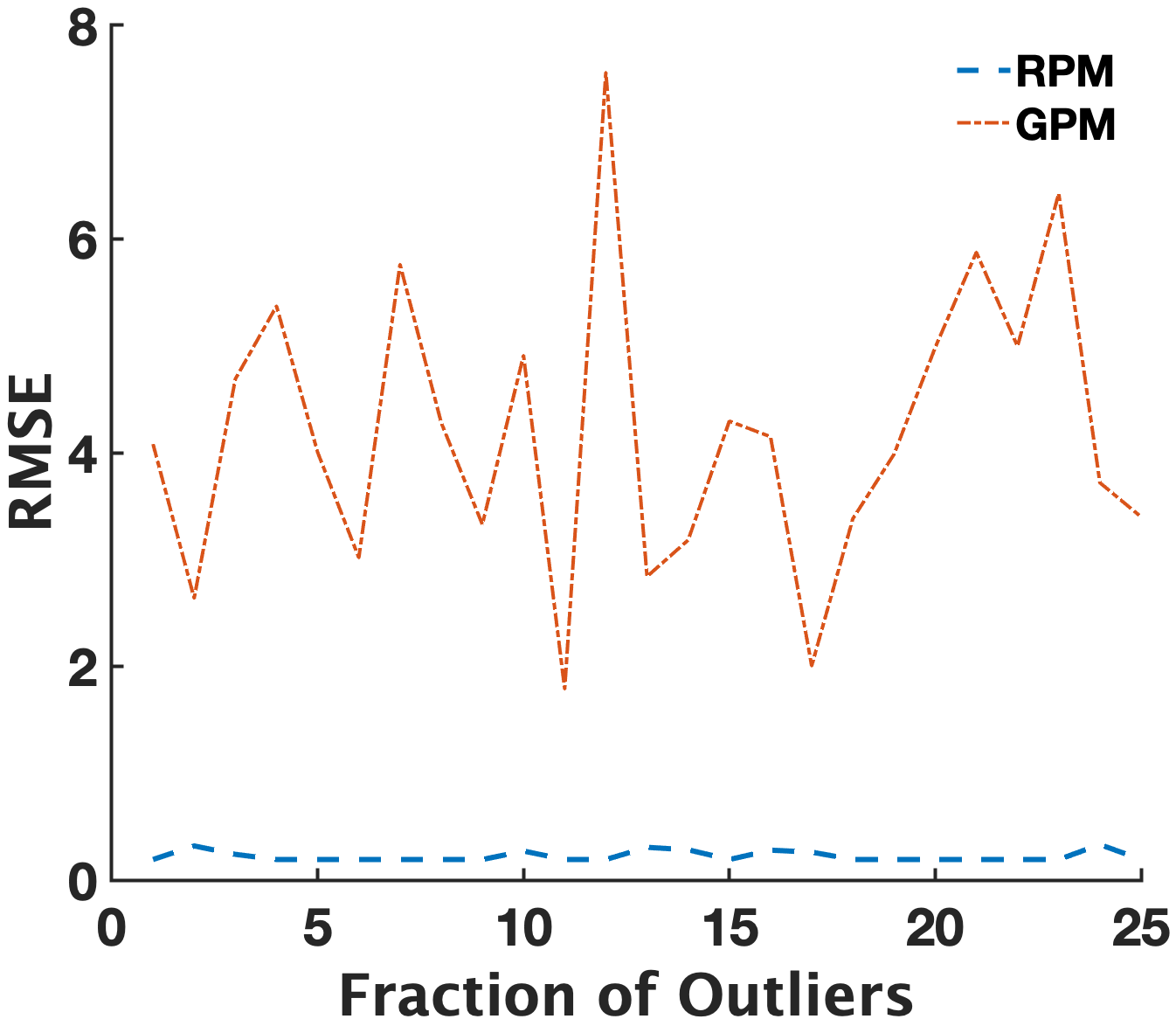}}}%
 \qquad 
    \subfloat[\centering ]{{\includegraphics[height=4cm,width=4.1cm]{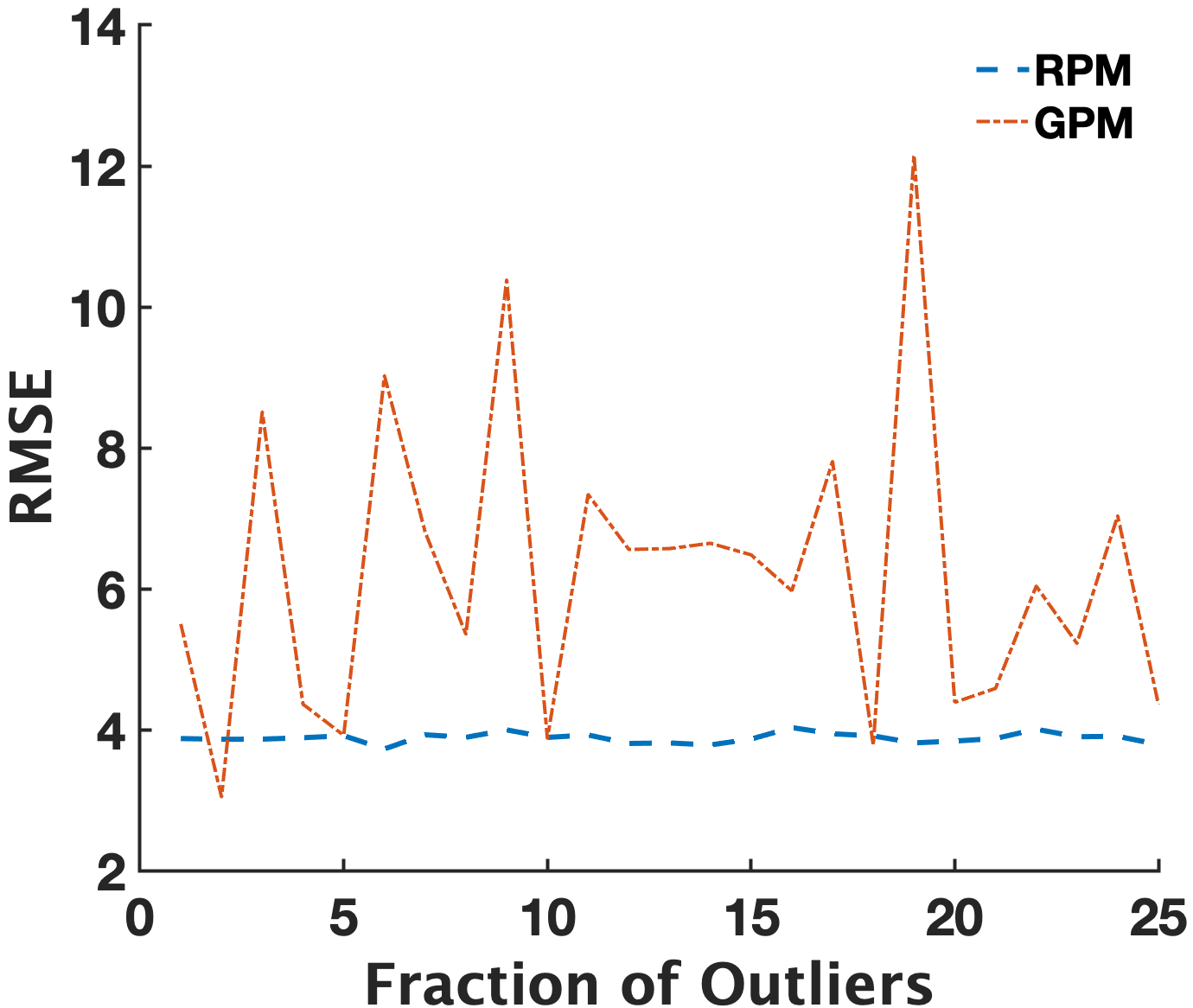}}}%
    \caption{{ RMSE vs the percentage of outliers added in training data for the prediction results at Bus $2003.2$ (a) voltage magnitude; (b) voltage angle.}}%
    \label{r1}
\end{figure*}

The RMSE of the predicted values for the voltage magnitude and the angle at Bus $2003.2$ with an increasing percentage of outliers added in the training data are plotted in Fig.\ref{r1} (a) and (b), respectively. We observe that the RMSE values obtained from the GPM are higher than the ones obtained from the RPM which provides consistently low RMSE results. 
\begin{table*}[]
\small
\setlength{\tabcolsep}{2pt}
\centering
\caption{{The RMSE and MAE for the Bus $2003.2$ of the $240$-bus system}}
  \begin{tabular}{|c|cc|cc|cc|cc|cc|cc|cc|cc|}
  \hline
  {} & \multicolumn{8}{c|}{Quadratic Basis} & \multicolumn{8}{c|}{Linear Basis}\\
\hline
 {} & \multicolumn{4}{c|}{With $25\%$ outliers} & \multicolumn{4}{c|}{Without outliers} & \multicolumn{4}{c|}{With $25\%$ outliers} & \multicolumn{4}{c|}{Without outliers} \\
 \hline
  {} & \multicolumn{2}{c}{RPM} & \multicolumn{2}{|c|}{GPM}& \multicolumn{2}{c}{RPM} & \multicolumn{2}{|c|}{GPM}& \multicolumn{2}{c}{RPM} & \multicolumn{2}{|c|}{GPM}& \multicolumn{2}{c}{RPM} & \multicolumn{2}{|c|}{GPM}\\
\hline
Measure &$V$ & $A$ &  $V$ & $A$&$V$ & $A$ &  $V$ & $A$&$V$ & $A$ &  $V$ & $A$&$V$ & $A$ &  $V$ & $A$\\
\hline
RMSE &0.4222&3.8734&0.4242&5.5007&0.1888&4.9835&2.2246&11.1942&0.1921&8.8619&4.0814&53.4306&0.1183&4.4287&0.0637&6.1481\\
MAE &0.3845&3.1791&0.3164&4.5781&0.1544&3.8252&1.9426&8.5223&0.1859&3.3976&3.0779&34.3391&0.1026&1.5122&0.0513&4.9164\\
\hline
\end{tabular}
\label{tab:240}
\end{table*}

\begin{figure}%
    \centering
  \subfloat[\centering ]{{\includegraphics[height=4cm,width=4cm]{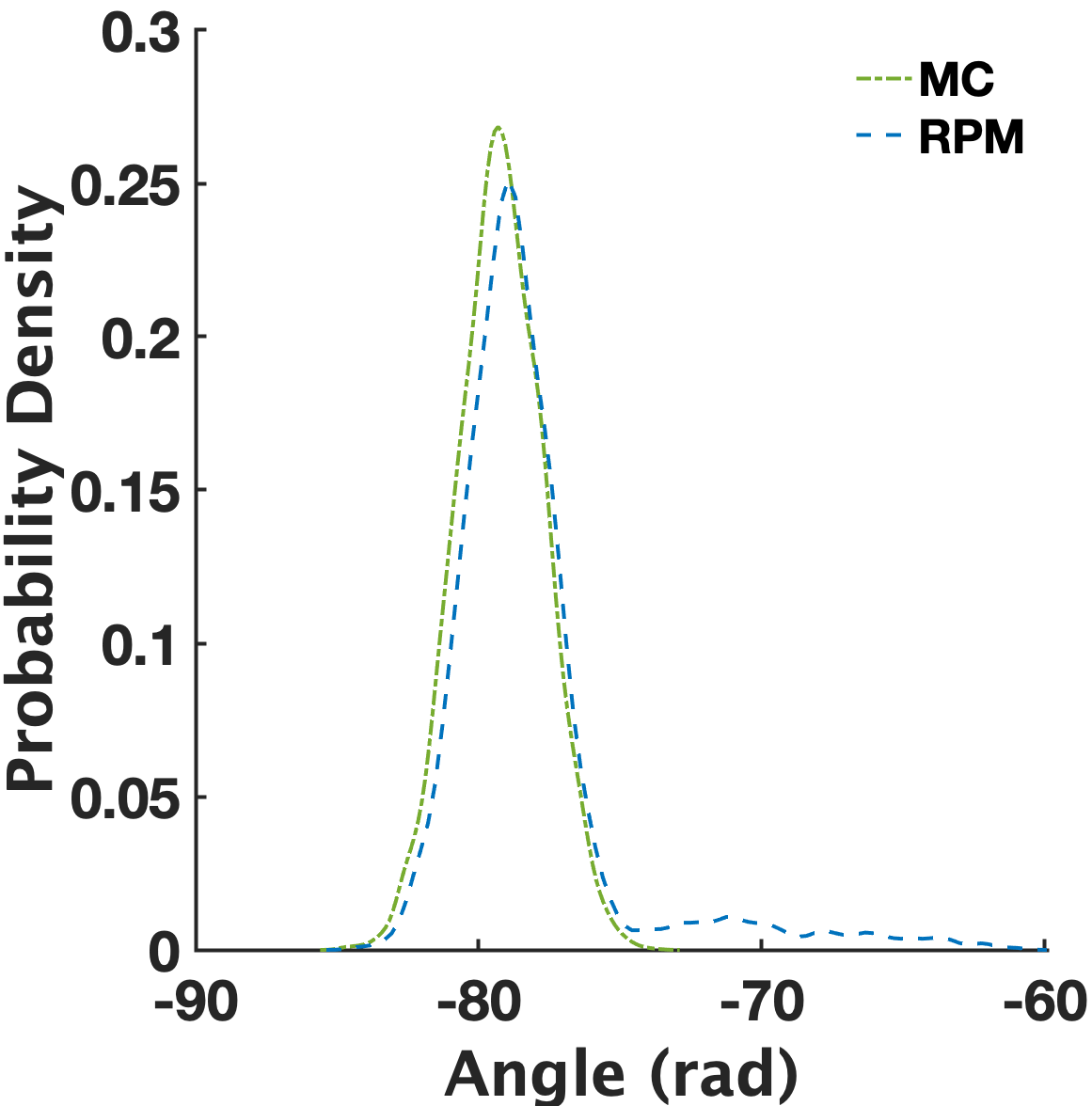}}}%
    \qquad
    \subfloat[\centering ]{{\includegraphics[height=4cm,width=4cm]{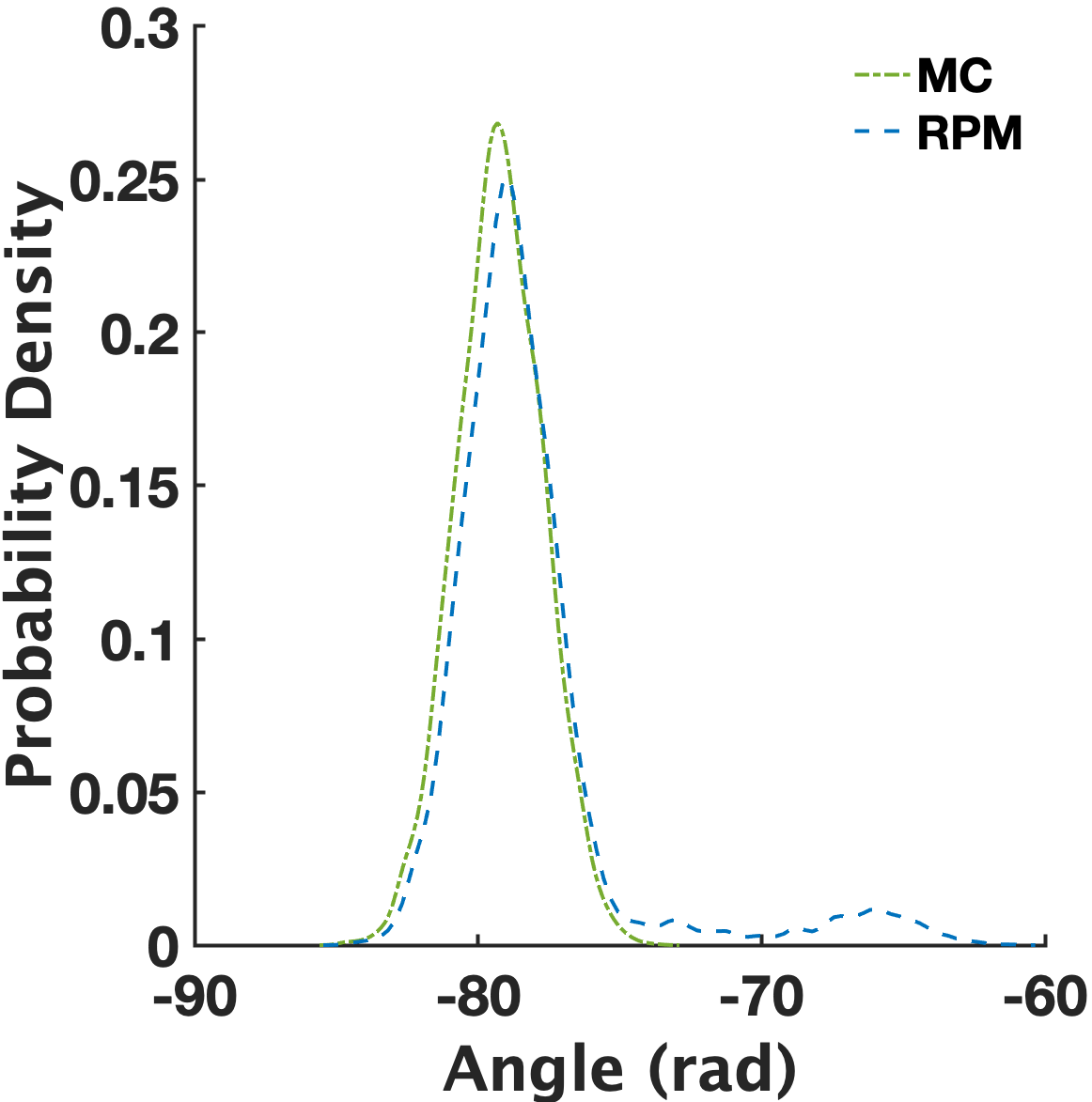}}}%
    \qquad
  \subfloat[\centering ]{{\includegraphics[height=4cm,width=4cm]{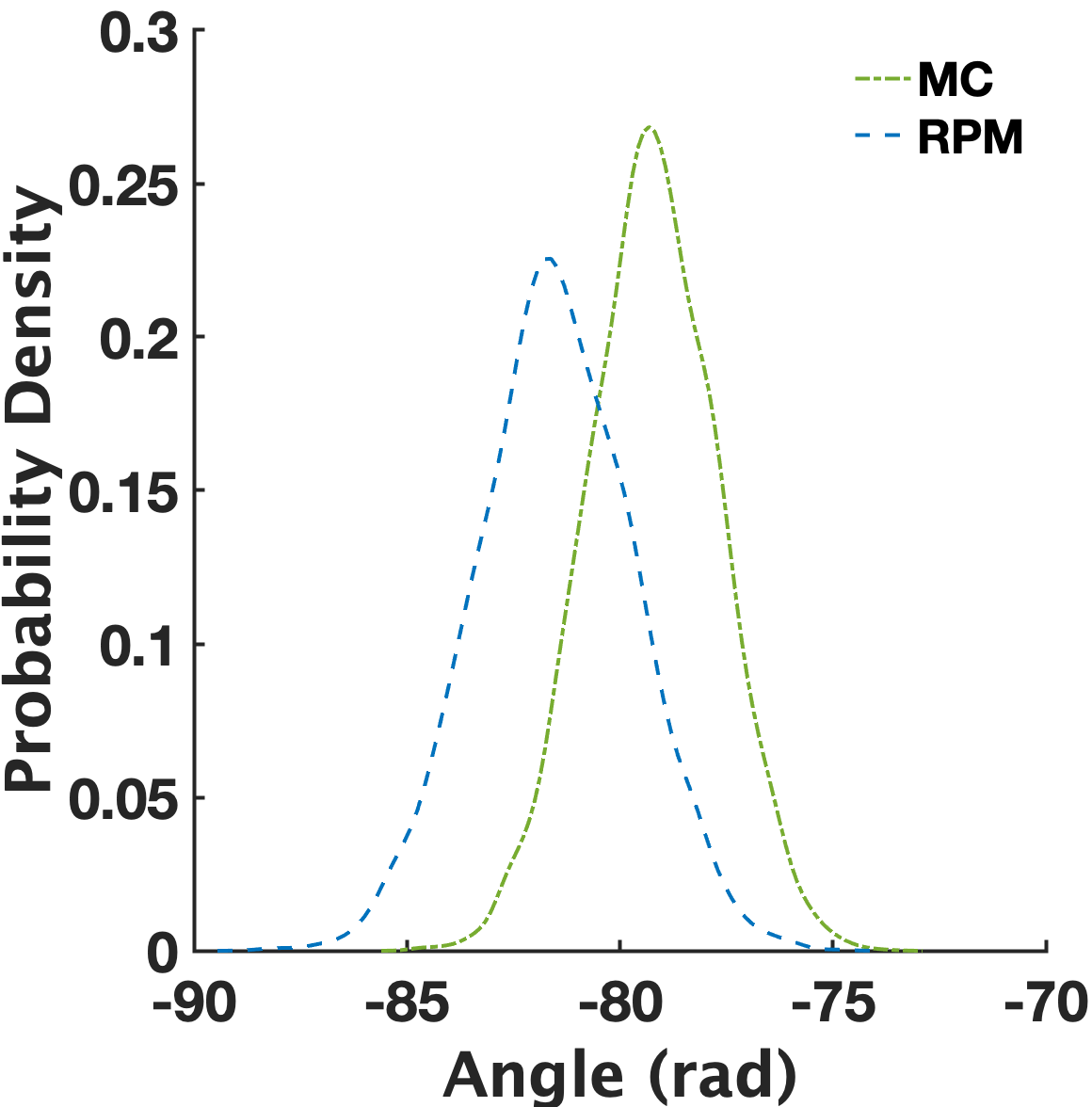}}}%
    \qquad
    \subfloat[\centering ]{{\includegraphics[height=4cm,width=4cm]{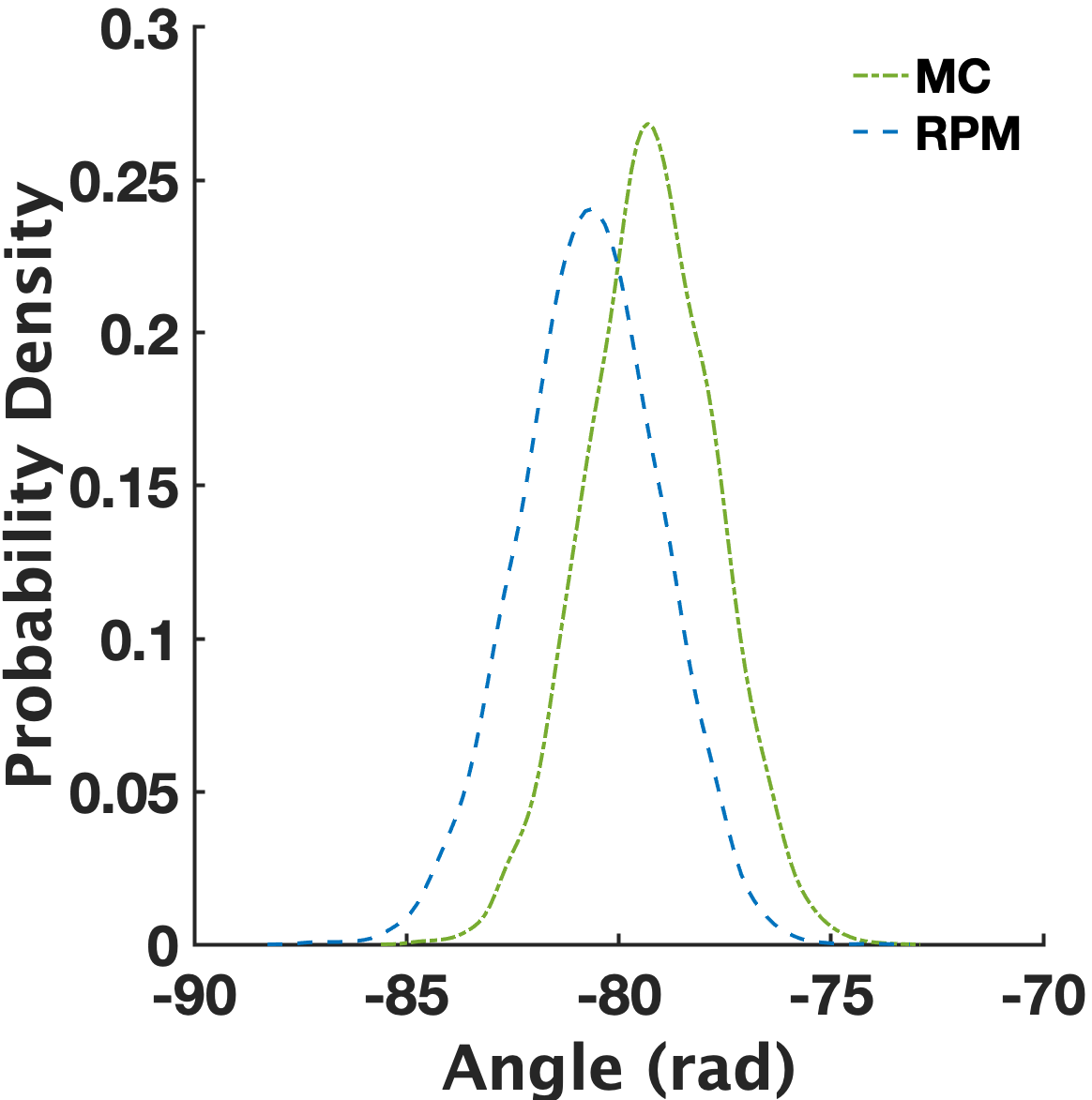}}}%
    \caption{ {Comparison between the GPM and the RPM probability density results for the voltage angle of Bus $2003.2$ in the $240-$bus network when (a) the training data set is added with $25\%$ of outliers ; (b) training data set is not added with outliers for linear basis; (c) the training data set is added with $25\%$ of outliers ; (d) training data set is not added with outliers for quadratic basis.}}
    \label{r_A_density}
\end{figure}


{Remark 1:
Note that, because of the lack of availability of the real measurements of the voltage phasors (output variables $\mathbf{y}$), they are obtained by running the power flow simulator using real measurements of active and reactive power injections (input variables $\mathbf{X}$).
}
\section{Conclusion and Future Work}
In this paper, we propose a robust process model to perform stochastic power flow calculations using time-series measurements of power injections and voltage phasors. The proposed model captures the natural stochastic dynamics introduced in the power grid by the RES and DGs. This is accomplished by training the RPM on recorded time series data set containing the measurements of active and reactive power injections from the RES and DGs and nodal voltage phasors. We demonstrate the RPM on the standard IEEE $33$-bus system and a real-world $240$-bus system. We show that the proposed methodology can handle $25\%$ of outliers i.e. bad leverage points and vertical outliers in the training data set using the root mean square errors and maximum absolute errors of the predicted values for voltage phasors.  

In future work, we will focus on extending the applications of the RPM for optimal power flow calculations. We will also investigate the performance of other robust estimators with high breakdown points for handling outliers in the training data of more than $25\%$. 

\newpage

\bibliography{reference} 

\begin{thebibliography}{10}

\bibitem{liacco1982role}
T.~D. Liacco, ``The role of state estimation in power system operation,'' {\em
  IFAC Proceedings Volumes}, vol.~15, no.~4, pp.~1531--1533, 1982.

\bibitem{ZhangNext-generationCenter}
P.~Zhang, F.~Li, N.~B. I. T. o.~S. Grid, and u.~2010, ``{Next-generation
  monitoring, analysis, and control for the future smart control center},''
  {\em ieeexplore.ieee.org}.

\bibitem{Allen2013AlgorithmSignals}
A.~Allen, S.~Santoso, and E.~Muljadi, ``{Algorithm for Screening Phasor
  Measurement Unit Data for Power System Events and Categories and Common
  Characteristics for Events Seen in Phasor Measurement Unit Relative
  Phase-Angle Differences and Frequency Signals},'' 2013.

\bibitem{LiuData-drivenLearning}
X.~Liu, X.~Zhang, L.~Chen, F.~X. J. o. M.~P. {\ldots}, and u.~2020,
  ``{Data-driven Transient Stability Assessment Model Considering Network
  Topology Changes via Mahalanobis Kernel Regression and Ensemble Learning},''
  {\em ieeexplore.ieee.org}.

\bibitem{McCamishAMeasurementsb}
B.~McCamish, R.~Meier, J.~Landford, R.~B. E. P.~S. {\ldots}, and u.~2016, ``{A
  backend framework for the efficient management of power system
  measurements},'' {\em Elsevier}.

\bibitem{Lin2018ComparisonGeneration}
C.~Lin, Z.~Bie, B.~Zhou, T.~Wang, and T.~Wang, ``Comparison of different
  methods in stochastic power flow with correlated wind power generation,''
  {\em IFAC-PapersOnLine}, vol.~51, pp.~67--72, 1 2018.

\bibitem{Xu2020ProbabilisticEmulator}
Y.~Xu, Z.~Hu, L.~Mili, M.~Korkali, and X.~Chen, ``Probabilistic power flow
  based on a gaussian process emulator,'' {\em IEEE Transactions on Power
  Systems}, vol.~35, pp.~3278--3281, 7 2020.

\bibitem{ren2015probabilistic}
Z.~Ren, W.~Li, R.~Billinton, and W.~Yan, ``Probabilistic power flow analysis
  based on the stochastic response surface method,'' {\em IEEE Transactions on
  Power Systems}, vol.~31, no.~3, pp.~2307--2315, 2016.

\bibitem{ni2016basis}
F.~Ni, P.~H. Nguyen, and J.~F. Cobben, ``Basis-adaptive sparse polynomial chaos
  expansion for probabilistic power flow,'' {\em IEEE Transactions on Power
  Systems}, vol.~32, no.~1, pp.~694--704, 2016.

\bibitem{wu2016probabilistic}
H.~Wu, Y.~Zhou, S.~Dong, and Y.~Song, ``Probabilistic load flow based on
  generalized polynomial chaos,'' {\em IEEE Transactions on Power Systems},
  vol.~32, no.~1, pp.~820--821, 2016.

\bibitem{wang2019data}
G.~Wang, H.~Xin, D.~Wu, P.~Ju, and X.~Jiang, ``Data-driven arbitrary polynomial
  chaos-based probabilistic load flow considering correlated uncertainties,''
  {\em IEEE Transactions on Power Systems}, vol.~34, no.~4, pp.~3274--3276,
  2019.

\bibitem{xu2019probabilistic}
Y.~Xu, L.~Mili, and J.~Zhao, ``Probabilistic power flow calculation and
  variance analysis based on hierarchical adaptive polynomial chaos-anova
  method,'' {\em IEEE Transactions on Power Systems}, vol.~34, no.~5,
  pp.~3316--3325, 2019.

\bibitem{laowanitwattana2018probabilistic}
J.~Laowanitwattana and S.~Uatrongjit, ``Probabilistic power flow analysis based
  on arbitrary polynomial chaos expansion for networks with uncertain renewable
  sources,'' {\em IEEJ transactions on Electrical and Electronic Engineering},
  vol.~13, no.~12, pp.~1754--1759, 2018.

\bibitem{laowanitwattana2021probabilistic}
J.~Laowanitwattana and S.~Uatrongjit, ``Probabilistic power flow analysis based
  on partial least square and arbitrary polynomial chaos expansion,'' {\em IEEE
  Transactions on Power Systems}, vol.~37, no.~2, pp.~1461--1470, 2021.

\bibitem{ye2022generalized}
K.~Ye, J.~Zhao, Y.~Zhang, X.~Liu, and H.~Zhang, ``A generalized computationally
  efficient copula-polynomial chaos framework for probabilistic power flow
  considering nonlinear correlations of pv injections,'' {\em International
  Journal of Electrical Power \& Energy Systems}, vol.~136, p.~107727, 2022.

\bibitem{yang2019fast}
Y.~Yang, Z.~Yang, J.~Yu, B.~Zhang, Y.~Zhang, and H.~Yu, ``Fast calculation of
  probabilistic power flow: A model-based deep learning approach,'' {\em IEEE
  Transactions on Smart Grid}, vol.~11, no.~3, pp.~2235--2244, 2019.

\bibitem{xiang2020probabilistic}
M.~Xiang, J.~Yu, Z.~Yang, Y.~Yang, H.~Yu, and H.~He, ``Probabilistic power flow
  with topology changes based on deep neural network,'' {\em International
  Journal of Electrical Power \& Energy Systems}, vol.~117, p.~105650, 2020.

\bibitem{wang2020probabilistic}
D.~Wang, K.~Zheng, Q.~Chen, G.~Luo, and X.~Zhang, ``Probabilistic power flow
  solution with graph convolutional network,'' in {\em 2020 IEEE PES Innovative
  Smart Grid Technologies Europe (ISGT-Europe)}, pp.~650--654, IEEE, 2020.

\bibitem{wu2021probabilistic}
H.~Wu, M.~Wang, Z.~Xu, and Y.~Jia, ``Probabilistic power flow of distribution
  system based on a graph-aware deep learning network,'' in {\em 2021 IEEE/IAS
  Industrial and Commercial Power System Asia (I\&CPS Asia)}, pp.~105--109,
  IEEE, 2021.

\bibitem{Ye2016IdentificationData}
X.~Ye, Z.~Lu, Y.~Qiao, Y.~Min, and M.~O'Malley, ``Identification and correction
  of outliers in wind farm time series power data,'' {\em IEEE Transactions on
  Power Systems}, vol.~31, pp.~4197--4205, 11 2016.

\bibitem{Li2020OutlierApplication}
G.~Li, Z.~Duan, L.~Liang, H.~Zhu, A.~Hu, Q.~Cui, B.~Chen, and W.~Hu, ``Outlier
  data mining method considering the output distribution characteristics for
  photovoltaic arrays and its application,'' {\em Energy Reports}, vol.~6,
  pp.~2345--2357, 11 2020.

\bibitem{wang2020data}
X.~Wang, X.~Wang, H.~Sheng, and X.~Lin, ``A data-driven sparse polynomial chaos
  expansion method to assess probabilistic total transfer capability for power
  systems with renewables,'' {\em IEEE Transactions on Power Systems}, vol.~36,
  no.~3, pp.~2573--2583, 2020.

\bibitem{xu2020data}
Y.~Xu, L.~Mili, M.~Korkali, K.~Karra, Z.~Zheng, and X.~Chen, ``A data-driven
  nonparametric approach for probabilistic load-margin assessment considering
  wind power penetration,'' {\em IEEE Transactions on Power Systems}, vol.~35,
  no.~6, pp.~4756--4768, 2020.

\bibitem{long2019image}
H.~Long, L.~Sang, Z.~Wu, and W.~Gu, ``Image-based abnormal data detection and
  cleaning algorithm via wind power curve,'' {\em IEEE Transactions on
  Sustainable Energy}, vol.~11, no.~2, pp.~938--946, 2019.

\bibitem{zheng2014raw}
L.~Zheng, W.~Hu, and Y.~Min, ``Raw wind data preprocessing: A data-mining
  approach,'' {\em IEEE Transactions on Sustainable Energy}, vol.~6, no.~1,
  pp.~11--19, 2014.

\bibitem{rousseeuw2005robust}
P.~J. Rousseeuw and A.~M. Leroy, {\em Robust regression and outlier detection}.
\newblock John wiley \& sons, 2005.

\bibitem{hampel1986robust}
F.~R. Hampel, E.~M. Ronchetti, P.~Rousseeuw, and W.~A. Stahel, {\em Robust
  statistics: the approach based on influence functions}.
\newblock Wiley-Interscience; New York, 1986.

\bibitem{Mill1996robustStatistics}
L.~Mill, M.~G. Cheniae, and N.~S. Vichare, ``Robust state estimation based on
  projection statistics,'' {\em IEEE Transactions on Power Systems}, vol.~11,
  no.~2, pp.~1118--1127, 1996.

\bibitem{donoho1992breakdown}
D.~L. Donoho and M.~Gasko, ``Breakdown properties of location estimates based
  on halfspace depth and projected outlyingness,'' {\em The Annals of
  Statistics}, pp.~1803--1827, 1992.

\bibitem{huber1992robust}
P.~J. Huber, ``Robust estimation of a location parameter,'' {\em Annals of
  Mathematical Statistics}, vol.~35, no.~1, pp.~73 -- 101, 1964.

\bibitem{maronna1995behavior}
R.~A. Maronna and V.~J. Yohai, ``The behavior of the stahel-donoho robust
  multivariate estimator,'' {\em Journal of the American Statistical
  Association}, vol.~90, no.~429, pp.~330--341, 1995.

\bibitem{gpmlbook}
C.~Rasmussen and C.~Williams, {\em Gaussian Processes for Machine Learning}.
\newblock Adaptive Computation and Machine Learning, Cambridge, MA, USA: MIT
  Press, Jan. 2006.

\bibitem{bu2019time}
F.~Bu, Y.~Yuan, Z.~Wang, K.~Dehghanpour, and A.~Kimber, ``A time-series
  distribution test system based on real utility data,'' in {\em 2019 North
  American Power Symposium (NAPS)}, pp.~1--6, IEEE, 2019.

\end{thebibliography}
\bibliographystyle{ieeetr}

 

% \end{IEEEbiography}



% \vfill

\end{document}